%% file: timedep.tex
\documentclass[12pt]{article}
\usepackage[sumlimits,intlimits,namelimits]{amsmath}
\usepackage{amssymb}
\usepackage[english]{babel}
\usepackage{graphicx,subfig,float}

\usepackage{cite} %Turns [A,B,C,D] into [A-D]

\usepackage[bordercolor=black,backgroundcolor=yellow,linecolor=black,
textsize=scriptsize,shadow]{todonotes}

\newcommand{\executeiffilenewer}[3]{%
\ifnum\pdfstrcmp{\pdffilemoddate{#1}}%
{\pdffilemoddate{#2}}>0%
{\immediate\write18{#3}}\fi%
}

\graphicspath{{.}{figures/}{figures/pdf_tex-files/}}

\global\parskip 6pt

\newcommand{\tr}{\text{tr}}
\newcommand{\sqg}{\sqrt{-g}}
\newcommand{\mO}{\mathcal{O}}
  
\newcommand{\pd}{\partial}
\newcommand{\eps}{\epsilon}
\newcommand{\totd}{\text{d}}

\renewcommand{\Re}{\text{Re\,}}
\renewcommand{\Im}{\text{Im\,}}

\newcommand{\tss}{\scriptstyle}

\usepackage{hyperref} % must be the last package

\begin{document}

\pagenumbering{Alph}
\begin{titlepage}
\thispagestyle{empty}
\hfill{MPP-2016-332}
\vspace*{.5cm}
\begin{center}
{\Large{{\bf Quantum Quenches in a Holographic Kondo Model}}} \\[.5ex]
\vspace{1cm}
\renewcommand{\thefootnote}{\fnsymbol{footnote}}
Johanna Erdmenger$^{1,2}$\footnote{jke@mpp.mpg.de},
Mario Flory$^{1,3}$\footnote{mflory@mpp.mpg.de, mflory@th.if.uj.edu.pl},
Max-Niklas Newrzella$^1$\footnote{maxnew@mpp.mpg.de},\\
Migael Strydom$^1$\footnote{migael.strydom@gmail.com} and 
Jackson M.\,S.\,Wu$^{4}$\footnote{jknwgm13@gmail.com}
\renewcommand{\thefootnote}{\arabic{footnote}}

\medskip

{\em $^1$ Max-Planck-Institut f\"ur Physik (Werner-Heisenberg-Institut),\\
F\"ohringer Ring 6, D-80805, Munich, 
Germany}
\\ \vspace{0.2 cm}
\textit{$^{2}$ Institut f\"ur Theoretische Physik und Astrophysik,
\\
Julius-Maximilians-Universit\"at W\"urzburg, Am Hubland, 97074 W\"urzburg}
\\ \vspace{0.2 cm}
\textit{$^{3}$ Institute of Physics, Jagiellonian University,
\\
\L{}ojasiewicza 11, 30-348 Krak\'ow, Poland}
\\ \vspace{0.2 cm}
\textit{$^{4}$Department of Physics and Astronomy,
University of Alabama,
\\
Tuscaloosa, AL 35487, USA}
\end{center}	

\begin{abstract}
\input{abstract.tex}
\end{abstract}

\end{titlepage}
\pagenumbering{arabic}

\tableofcontents
\newpage

\section{Introduction}	
\label{sec:intro}
\input{intro.tex}

\section{Setup}
\label{sec:setup}
\input{setup.tex}

\section{Results}
\label{sec:results}
\input{results.tex}

\section{Critical behaviour}
\label{sec:section}
\input{critical.tex}

\section{Summary and Outlook}
\label{sec:conclusion}
\input{conclusion.tex}

\appendix

\input{appendix.tex}

\bibliography{references}{}
\bibliographystyle{utphys}
 
\end{document}

%% file: abstract.tex
We study non-equilibrium dynamics and quantum quenches in a recent gauge/gravity duality model 
for a strongly coupled system interacting with a magnetic impurity with $SU(N)$ spin. 
At large $N$, it is convenient to write the impurity spin as a bilinear in Abrikosov fermions. 
The model describes an RG flow triggered by the marginally relevant Kondo operator.
There is a phase transition at a critical temperature, below which
an operator condenses which involves both an electron and an Abrikosov fermion field. This corresponds to a holographic   superconductor in AdS$_2$ and models the impurity screening. We quench 
the Kondo coupling either by a Gaussian pulse or by a hyperbolic tangent, the latter taking the system from the 
condensed to the uncondensed phase or vice-versa. We study the time dependence of the condensate induced by this quench. 
The timescale for equilibration is generically given by the leading quasinormal mode of the dual gravity model.
This mode also governs the formation of the screening cloud, which is obtained as the decrease of impurity degrees of freedom with time. In the condensed  phase, the leading quasinormal mode is imaginary and the relaxation of the condensate is over-damped.
For quenches whose final state is close to the critical point of the large $N$ phase transition, we study the critical slowing 
down and obtain the combination of critical exponents $z\nu=1$. When the final
state is exactly at the phase transition,  we find that the exponential ringing of the quasinormal modes 
is replaced by a power-law behaviour of the form $\sim t^{-a}\sin(b\log t)$. This indicates the emergence of a discrete scale invariance.

%% file: intro.tex
The AdS/CFT correspondence \cite{Maldacena:1997re,Gubser:1998bc,Witten:1998qj} 
and its generalisations to a more general 
gauge/gravity duality provide a new approach for studying strongly correlated systems.
One of the many applications of gauge/gravity duality is the study of
non-equilibrium strongly coupled systems. This applies in particular
to quantum quenches and to the subsequent relaxation to a new ground
state. While this idea was first studied in relation to heavy-ion
physics and the quark-gluon plasma~\cite{Balasubramanian:2011ur,Buchel:2012gw}, more recently quenches have been
considered for systems relevant to condensed matter as well~\cite{Bhaseen:2012gg}.

In this paper we use gauge/gravity duality to study quantum quenches for
strongly coupled systems interacting with a magnetic impurity. In
particular, we consider quenches in a recent
holographic model 
\cite{Erdmenger:2013dpa} that describes the RG flow triggered by an
impurity operator. This is a holographic variant of the well-known
Kondo model with certain distinct features owed to considering a model
for which a gravity dual may be constructed. As is 
standard in the AdS/CFT correspondence, the large $N$ limit needs to be taken to obtain a classical gravity dual. Thus 
in this model, the spin group of the impurity spin is $SU(N)$. The
spin is in a totally antisymmetric representation given by a Young tableau with $q$ boxes. In the large $N$ limit, it is convenient to introduce 
Abrikosov pseudo-fermions $\chi$ and to write the impurity field as a bilinear in these fermions,
i.e.~$S^a = \bar \chi T^a \chi$. For simplicity, the number of channels
(flavours) is taken to be $k=1$. At the UV fixed point of the RG flow, the strongly coupled 
theory is perturbed by the marginally relevant  operator $\kappa {\mathcal
O} {\mathcal O}^\dagger$ with ${\mathcal O} = \psi^\dagger \chi$. Here,
$\kappa$ is the Kondo coupling, $\psi$ is an electron and  $\chi$ an
Abrikosov fermion as introduced above. This operator triggers a flow
to an IR fixed point. $\kappa$ diverges at a temperature $T_K$, which defines the Kondo temperature.  
The IR fixed point is characterised by a non-trivial condensate 
$\langle {\mathcal O} \rangle$, i.e.~there is a critical temperature
$T_c$, slightly smaller than $T_K$, below which this condensate
forms. The condensation corresponds to screening of the impurity by
the  \textit{Kondo cloud}.

In the dual gravity model, according to \cite{Erdmenger:2013dpa}  this system is described by a Chern-Simons field 
in 2+1 dimensions dual to the electron current. Moreover, there is an Abelian
gauge field restricted to 1+1 dimensions, whose time component is dual to the charge density 
$\chi^\dagger \chi$, and a 1+1-dimensional complex scalar  dual to the operator 
${\mathcal O}$. In the probe limit, all of these fields are embedded in 
a BTZ black hole spacetime, i.e.~a black hole in $AdS_3$. The horizon of  
the black hole sets the temperature. The 1+1-dimensional subspace extends 
in the time direction and in the AdS radial direction.
The perturbation by $\kappa {\mathcal O} {\mathcal O}^\dagger$ is achieved
by considering the gravity dual of a `double trace' deformation as introduced 
in \cite{Witten:2001ua}. Solving the equations of motion, 
a second-order (mean field) phase transition is found at $T=T_c$, 
below which the scalar acquires a non-trivial condensate. Moreover, 
at low temperatures the charge density dual to the two-dimensional gauge field
decreases, such that the dimension of the spin representation is decreased. 
This corresponds to the screening of the impurity. Also, the electrons 
are subject to a phase shift which is obtained from the Wilson loop involving 
the Chern-Simons field in $AdS_3$. The resistivity is obtained from an analysis
of the leading irrelevant operator. Due to the large $N$ limit, 
the characteristic logarithmic behaviour at low temperatures is absent.
Rather, the resistivity has a polynomial dependence on temperature
with real exponent \cite{Erdmenger:2013dpa}.

In \cite{Erdmenger:2014xya,Erdmenger:2015spo,Erdmenger:2015xpq}, the model of \cite{Erdmenger:2013dpa} was extended to 
include the backreaction of the defect on the background geometry, which allows 
to the calculation of the entanglement entropy 
in particular. The two-impurity version of the model of \cite{Erdmenger:2013dpa} was studied in \cite{O'Bannon:2015gwa}. 
Recently, two-point functions for this model were calculated in
\cite{Erdmenger:2016vud,Erdmenger:2016jjg}, where it was found that
the spectral function displays a Fano resonance characteristic of
scattering between a continuum and a localised resonance. In the
screened phase where $\langle {\mathcal O} \rangle \neq 0 $, the
quasinormal modes (QNM) are purely imaginary and scale as $\omega \propto -i
\langle {\mathcal O} \rangle^2$. This is characteristic of the Kondo
resonance in the large $N$ limit.

In this paper, we restrict to the probe limit as in \cite{Erdmenger:2013dpa,O'Bannon:2015gwa}. We consider 
time-dependent configurations in which the time-dependence of the Kondo coupling 
is chosen as an input. 
In particular, we consider both Gaussian pulses in the Kondo coupling
$\kappa$ as well as $\tanh$-shaped transitions from the unscreened to
the screened phase and vice-versa. We consider fast quenches, in which
the quench time for the Kondo coupling is of order of the inverse
Kondo temperature.  

Our main result is that as
generally expected in the holographic approach, the response of the
system to quenches is
dominated by the QNM. These fix the equilibration time
for relaxation to the new ground state. In general, this relaxation time is independent of the original quench time. The QNMs ensure that the relaxation time is longer when the final state is closer to the phase transition. For relaxation
to the critical state at the phase transition, as obtained when
quenching the Kondo coupling to its critical value, the relaxation becomes polynomial rather than
exponential. In this case, there is a \textit{critical slowing down} and a damped
log-periodic behaviour which may be a sign of discrete scale invariance.

For quenches to
the screened phase we confirm that the leading QNM
behaves as $\omega \propto -i
\langle {\mathcal O} \rangle^2$ as seen in
\cite{Erdmenger:2016vud,Erdmenger:2016jjg}. The fact that the leading
QNM is imaginary implies that there are no oscillations
about the new ground state, and the relaxation is over-damped. For larger values of the condensate,
i.e. at very low temperatures, we see a different behaviour  
$\omega \propto -i \mathrm{ln}
\langle {\mathcal O} \rangle$, which corresponds to a deviation from
mean-field behaviour. The investigation of the zero-temperature behaviour will require a
refinement of the model corresponding to stabilisation of the IR fixed
point by a quartic contribution to the scalar potential on the gravity
side, which we leave for the future.

In the model considered, the number of degrees of freedom at the impurity site is represented by the charge density of the Abrikosov
fermions. This is related to the size of the spin representation. The charge density is holographically dual to the $AdS_2$ gauge field and
may be written as the flux of the $AdS_2$ gauge field through the boundary of $AdS_2$. Evaluating this flux at
the black hole horizon then gives a measure of the effective number of
impurity degrees of freedom. In the condensed phase, we observe that this flux is
reduced, which corresponds to a gravity dual realisation of
the screening. In this paper, we study the time evolution of the flux
after a Gaussian quench in the condensed phase and observe that it
decays exponentially, which corresponds to an exponential decrease of
the number of degrees of freedom after a quench. This models the time
dependence of the Kondo cloud formation at the impurity site.

Quantum quenches in the standard $SU(2)$ Kondo model were recently studied within condensed matter physics. These
investigations include
\cite{PhysRevLett.112.146804,2014PhRvB..90w5145K,PhysRevLett.113.180601,
PhysRevB.90.115101,2014arXiv1409.0646N}. 
In particular, \cite{PhysRevLett.112.146804} deals with the study of a quantum 
quench caused by the absorption of a photon by a quantum dot, while 
\cite{PhysRevB.90.115101} studies the universal behaviour of entanglement 
entropy after a quench of an impurity system. In \cite{2014PhRvB..90w5145K}, 
quenches in the 
pseudogap single-impurity Anderson model are investigated using numerical RG 
techniques. 
The system reaches an equilibrium state at late times. Furthermore, the 
spatiotemporal formation of the Kondo cloud after a quench was simulated in 
\cite{2014arXiv1409.0646N}, where an emergent lightcone structure is
observed. The non-equilibrium correlation functions are determined by
two different scales, initially by the lattice Fermi velocity and by
the Kondo temperature at late time. In our model considered here,
generically the Kondo temperature is the only scale.
Generally, due to the large $N$ limit for the spin symmetry required in our
holographic approach, comparison to the condensed matter results is
possible only concerning a limit number of aspects, in particular due
to the large $N$ phase transition which is not present in the $SU(2)$
case. The large $N$ limit implies that in the screened (condensed)
phase, there are no oscillations of the screening cloud and the
relaxation of the condensate is over-damped. On the other hand, in
$SU(2)$ Kondo quenches such oscillations are frequently present, as
seen for instance in \cite{PhysRevB.90.115101}.

The paper \cite{Bhaseen:2012gg} considers quenches in 3+1-dimensional holographic
superconductors with backreaction. Similarly to the present paper, for
some parameter regimes an over-damped behaviour is found. In contrast
to those results however, here we generically find over-damping
whenever the final state is in
condensed phase. This is due to the leading QNM being purely imaginary
in this phase, as expected from the presence of a Kondo resonance.
Moreover, \cite{Basu:2013soa} studies holographic quenches of a
double-trace operator in arbitrary dimensions, and the corresponding critical exponents
for quenches through the phase transition are obtained.

The structure of this paper is as follows: In section \ref{sec:setup}, we briefly describe the setup of the 
holographic Kondo model of \cite{Erdmenger:2013dpa}. Some additional details about the analytical and numerical 
treatment of the resulting equations of motion are relegated to appendices 
\ref{sec:ansatzEF} and \ref{sec:evolutionscheme}. We  then summarise 
our results in section \ref{sec:results}. We furthermore study quenches in the 
normal phase in section \ref{sec::normalQNM}, as well as in the condensed 
phase in section \ref{sec::condQNM}. In section \ref{sec::critExp} we study the phenomenon of critical slowing down near the phase transition. In section
\ref{sec::kappacrit}, this leads us to a study of the 
late-time behaviour of the system when the end state is exactly at the phase transition. 
We  end in section \ref{sec:conclusion} with a summary and an outlook.

%% file: setup.tex
\subsection{Action and equations of motion}

We consider the bottom-up model proposed in~\cite{Erdmenger:2013dpa}.
The bulk spacetime is that of a $(2+1)$-dimensional finite temperature BTZ black brane,
\begin{align}\label{eq:fullmetric}
ds^2 = G_{\mu\nu} \totd x^\mu \totd x^\nu = \frac{L^2}{z^2} \left( - 
h(z) \totd t^2 + 
\frac{\totd z^2}{h(z)} + \totd x^2 \right) \,, \quad h(z) = 1 - \frac{z^2}{z_H^2} \,,
\end{align}
where $L$ is the $AdS_3$ radius, and $z$ is the radial coordinate, with the 
boundary at $z=0$ and the 
horizon at $z=z_H$. The temperature of the dual field theory corresponds to the black brane's Hawking 
temperature, $T = 1/(2\pi z_H)$.
We apply the scaling symmetries available to set $z_H = 1$ and $L = 1$ for the rest of this work.

The model of~\cite{Erdmenger:2013dpa} has a non-dynamical co-dimension one 
hypersurface at $x=0$, which provides the gravity dual of the localised Kondo impurity of the 
boundary field theory. 
The action consists of an 
$AdS_3$ bulk contribution involving the Chern-Simons gauge field $A$, and an $AdS_2$ 
defect contribution involving a complex scalar $\Phi$ and a $U(1)$ gauge field 
$a$, \footnote{In \cite{O'Bannon:2015gwa}, a $U(2)$ gauge group was assumed for 
the gauge field $a$, corresponding to a two impurity Kondo-model.}
\begin{gather}
S = S_{CS} + S_{AdS_2} \,, 
\label{action}
\\
S_{CS} = -\frac{N}{4\pi}\int\tr\left( A \wedge dA + \frac{2}{3} A \wedge A 
\wedge A \right) \,, 
\label{AdS3action}
\\
S_{AdS_2} = -N\int \totd^2x \, \sqrt{-g} \left[
\frac{1}{4}\, f^{mn}f_{mn} + g^{mn} \left(D_m \Phi\right)^{\dagger} D_n \Phi 
+ V(\Phi^{\dagger} \Phi )
\right] \,,
\label{AdS2action}
\end{gather}
where $f_{mn} = \pd_m a_n - \pd_n a_m$ is the $U(1)$ field strength, $D_m$ is the gauge-covariant derivative,
\begin{equation}
D_m\Phi = (\partial_m + i A_m - i a_m)\Phi \,,
\end{equation}
and the $AdS_2$ metric $g$ is the pullback of the $AdS_3$ metric $G$ to the hypersurface by the immersion $x = 0$. 
The Roman indices $m,n$ run over $t,z$. For the remainder of this
paper, we restrict to a $U(1)$ flavour symmetry and thus the
Chern-Simons contribution \eqref{AdS3action}  to the action becomes Abelian.
The equations of motion are \cite{Erdmenger:2013dpa}
\begin{align}\label{eq:fulleom}
\eps^{n\mu\nu} F_{\mu\nu} &= \frac{4\pi}{N}\delta(x)J^n \,, \\
\partial_m\left(\sqg g^{mp}g^{nq}f_{pq}\right) &= J^{n} \,, \\
\partial_n J^{n} &= 0 \,, \\
\partial_m\left(\sqg g^{mn}\partial_n\phi\right) &= \sqg\Delta^{m}\Delta_{m}\phi
+\frac{1}{2}\sqg\frac{\partial V}{\partial \phi} \,,
\label{eq:fulleom2}
\end{align}
where we parametrised $\Phi = \phi\, e^{i\psi}$ and defined
\begin{equation}
J^n \equiv 2\sqrt{-g}g^{mn}\Delta_m\phi^2 \,, \qquad \Delta_m \equiv a_m - A_m
- \pd_m\psi,
\end{equation}
where greek indices run from 0 to 2 and latin indices from 0 to 1. $A_m$ are 
understood to be the components of the projection of $A$ to the hypersurface.
Upon gauging $A_z = 0$ and requiring regularity of the CS field at the 
horizon, $A_t(z_H) = 0$, only $A_x$ remains nontrivial.
Hence, the projection of the Chern-Simons field to the defect hypersurface 
vanishes which implies that $A$ decouples from the rest of the fields.
Due to this we are allowed to neglect the CS field when solving for the fields 
restricted to the defect. In principle, the CS field could be integrated from 
the solutions of $\Phi$ and $a$.

The field content of the model defined by equations \eqref{action} - 
\eqref{AdS2action} is to be interpreted in the light of the
holographic dictionary outlined in \cite{Erdmenger:2013dpa,O'Bannon:2015gwa}. 
Specifically, the Chern-Simons gauge field $A$ is holographically dual to the 
chiral current of conduction electons $\psi$ in the boundary theory. 
Similarly, the 
gauge field $a$, which is restricted to the $AdS_2$ subspace at $x=0$, is dual 
to the charge of the slave fermions $\chi$ that are restricted to the impurity 
at $x=0$ in the field theory picture. 
The scalar $\Phi$, charged under both the 
gauge groups of the fields $a$ and $A$, is then dual to the composite operator 
${\cal O} = \psi^\dagger \chi$, which indicates the coupling of the conduction 
band to the impurity. The potential term $V(\Phi^{\dagger} \Phi )$ in 
\eqref{AdS2action} is chosen as 
\begin{equation}
 V(\Phi^{\dagger} \Phi ) = M^2 \,\Phi^{\dagger} \Phi ,
 \label{eq:potential}
\end{equation}
with $M^2$ tuned to the 
Breitenlohner-Freedman bound to obtain the correct scaling dimensions.
At temperatures below a critical one, $T_c$, the scalar field exhibits 
an instability, which leads to its condensation and thus a nonvanishing expectation value,
$\langle\mO\rangle \neq 0$. This is interpreted as the formation of the Kondo cloud in the large-$N$ 
holographic model we consider here. We shall refer to the phase above $T_c$ where the scalar does not 
condense as the \textit{normal phase}, and the phase below $T_c$ where it
does as the \textit{condensed phase}. 
In the condensed phase, the impurity is screened.
Further details may be found in
\cite{Erdmenger:2013dpa,O'Bannon:2015gwa}.
Below, we review only those previous results that are relevant 
in the context of this paper. We present our new results on time dependence  in 
sections~\ref{sec:results} and~\ref{sec:section}. 

\subsection{Boundary behaviour and conditions on the Kondo coupling}

In the normal phase $T > T_c$, the solution for the gauge field is given by 
\begin{gather}
 a_t(t,z) = \frac{Q}{z} + \mu \,,
\end{gather}
where $Q$ denotes the electric flux at the boundary and $\mu$ 
the chemical potential. As explained in \cite{Erdmenger:2013dpa}, $Q$ defines 
the representation of the impurity spin and by following the same conventions, 
we set $Q=-\frac{1}{2}$. $|Q|$ is related to the number of boxes $q$ in the 
spin representation Young tableau by $|Q|=q/N$. In order to obtain regularity at 
the horizon, we must 
set $\mu = - Q$. 
To be able to map our bottom-up model to the Kondo model, we need to fix the 
scaling dimension of the scalar operator to be $\Delta_{\mathcal{O}} = 
\frac{1}{2}$, see \cite{Erdmenger:2013dpa}.
As the scalar field is restricted to live in an asymptotically $AdS_2$ space, the scaling dimension is given by
\begin{equation}
 \Delta_{\mathcal{O}} = \frac{d}{2} \pm \sqrt{\frac{d^2}{4} - Q^2 + M^2}\,,
\end{equation}
where we set $d=1$ \cite{Iqbal:2011aj}.
It can be seen that the correct scaling dimensions can only be obtained if we 
put the scalar exactly at the Breitenlohner-Freedman bound. With our choice of $Q$, 
this means setting $M = 0$.
The leading order behaviour of the scalar field near the bounday is then given 
by
\begin{equation}
 \Phi(t,z)= \phi_1(t,z)+ i \phi_2(t,z) \approx \sqrt{z} \,\left( \alpha(t) \log(z) + \beta(t)\right) + 
\ldots
\label{eq:boundary-exp}
\end{equation}
where $\alpha(t) = \alpha_1(t) + i \alpha_2(t)$ and $\beta(t) = \beta_1(t) + i 
\beta_2(t)$ are complex functions of time.
As was shown in \cite{Erdmenger:2013dpa,O'Bannon:2015gwa,Witten:2001ua}, the 
boundary condition for a boundary double trace operator $\sim \kappa 
\mathcal{O}\mathcal{O}^{\dagger}$ is given by requiring $\alpha = \kappa\beta$. 
Furthermore, it was demonstrated that an arbitrary energy scale 
$\Lambda$ has to be introduced due to the appearence of the logarithm in 
the boundary expansion.
The Kondo coupling $\kappa$ is running w.r.t.~rescalings of $\Lambda$ and 
eventually diverges at low temperatures at the \textit{Kondo temperature} 
$T_K = \Lambda\,e^{1/\kappa} / 2 \pi$,
which is invariant under these rescalings.
We avoided this energy scale by rescaling $z$ by $z_H$, which 
renders the radial coordinate $z$ dimensionless.
To avoid confusion, note that the notation of $\kappa$, $\beta$, $\alpha$ in 
\cite{Erdmenger:2013dpa} differs from ours by a subscript $(\cdot)_T$, which we 
left out for convenience.

We impose a time-dependent boundary condition on the Kondo coupling $\kappa(t) 
= \kappa_1(t) + i\,\kappa_2(t)$ which is a real function of time, 
i.e.~we set $\kappa_2(t) = 0$.
The relationships between the leading and subleading expansion coefficients at 
the boundary are hence given by
\begin{equation}
 \alpha_1(t) = \kappa_1(t) \, \beta_1(t) 
 \qquad \text{and} \qquad
 \alpha_2(t) = \kappa_1(t) \, \beta_2(t) \, .
 \label{eq:alpha-is-kappa-beta}
\end{equation}
At the same time, the electric flux $Q$ of the gauge field $a$ is required to 
stay constant, which is actually necessary to render the variational problem 
meaningful, c.f.~\cite{Erdmenger:2016jjg}.
Together with regularity at the event horizon, this fixes all boundary 
conditions for our system of partial differential equations.
After going through the holographic renormalisation procedure, which was carefully constructed and carried out in 
\cite{O'Bannon:2015gwa}, one finds~\footnote{In~\cite{Erdmenger:2013dpa}, the expectation value of the dual scalar operator
was identified as $\langle\mathcal{O}\rangle\propto\alpha$. See~\cite{O'Bannon:2015gwa} for a complete discussion of the difference.}
\begin{equation}
 \langle\mathcal{O}\rangle = -N \beta^{\dagger} \,.
 \label{eq:vev-beta}
\end{equation}
Since we are interested in the real-time dynamics of the scalar operator, $\beta$ will thus be the main
quantity we focus on below.

In equilibrium, the temperature measured in units of the Kondo temperature 
$T_K$ can be found from $\kappa_1$,
and it is the same relationship as that given in~\cite{Erdmenger:2013dpa}:
\begin{equation}
 \frac{T}{T_K} = \exp(-1/{\kappa_1}) \,.
 \label{eq:ToverTK}
\end{equation}
Furthermore, we have from~\cite{Erdmenger:2016jjg}
\begin{equation}
\log\frac{T_c}{T_K} = -2\Re\!\!\left[H\!\left(-\frac{1}{2} + i Q\right)\right] - \log 2 \,,
\end{equation}
where $H(z)$ is the harmonic number. For $|Q| = 1/2$, which is used throughout 
this paper, we have
\begin{equation}
\kappa_c = \kappa_1(T_c) = \left(2\Re\!\!\left[H\!\left(\frac{i - 1}{2} 
\right)\right] + \log 2\right)^{-1} \approx 8.9796 \,.
\label{eq:kappacrit}
\end{equation}

To study how the system evolves given a Kondo coupling that is changing in time, 
we investigate the time evolution of the expectation value of the scalar 
operator, $\langle\mathcal{O}(t)\rangle$. %which in the numerics is given by $\beta$.
We consider time-dependent profiles for the Kondo coupling with a form of either
a hyperbolic tangent,
\begin{equation}\label{eq:kappa-tanh}
\kappa_1(t) = \kappa_i + \frac{\Delta\kappa}{2}\big[\tanh\left(s(t - 
t_0)\right) + 1\big] \,,
\end{equation}
or a Gaussian,
\begin{equation}\label{eq:kappa-gauss}
\kappa_1(t) = \kappa_i + \Delta\kappa\,\exp\left(-s(t -t_0)^2\right) \,.
\end{equation}
We consider  the system to be initially prepared in an equilibrium state characterised 
by the Kondo coupling $\kappa_i$, which is then quenched to another state 
whose equilibrium is characterised by $\kappa_f$. 
For a hyperbolic tangent quench, $\kappa_f = \kappa_i + \Delta\kappa$, whereas for 
a Gaussian quench, $\kappa_f = \kappa_i$. In both cases, the
\textit{amplitude} parameter $\Delta\kappa$
controls how much $\kappa_1$ can change during the quench.
The \textit{steepness} parameter $s$ controls the speed of the quench, while 
the offset $t_0$ determines the midpoint of the quenching process. 
We shall refer to the set of parameters
$\{\kappa_i,\,\kappa_f,\,\Delta\kappa,\,s,\,t_0\}$ as the \textit{quench parameters}.

In view of the numerical analysis, we change to 
Eddington-Finkelstein coordinates and choose the radial gauge for the 
defect gauge field $a$.
The numerical solution of the evolution involves pseudospectral methods in the radial 
direction and an implicit evolution method in the time direction.
More details of the numerical implementation are given in  appendices 
\ref{sec:ansatzEF} and \ref{sec:evolutionscheme}.

%% file: results.tex
\subsection{Generic time evolution for phase transitions}
\label{sec::result-plots}

Here we present examples which show the generic form of the time-evolution of the scalar condensate
as the system is quenched from one phase to a different one. 
The input to the time-evolution problem is $\kappa_1(t)$, which we choose to have a
hyperbolic tangent profile as given in~\eqref{eq:kappa-tanh}. 
Similar quenches of a double-trace coupling were studied in~\cite{Basu:2013soa}, but in a different setting without
defect. Given the quench profile of the Kondo coupling, we solve the equations of motion 
to obtain the evolution of $\beta_{1,2}(t)$ and $\mu(t)$, and extract from it
information about the condensate $\langle \mathcal{O} (t)\rangle$. As we explain below,
QNMs can also be extracted from the time-evolution of the scalar condensate.

\begin{figure}[htbp]
\subfloat[
  Quench profile for $\kappa_1$ as given by eq.~\ref{eq:kappa-tanh} 
  with parameters $\kappa_i = 9$, $\kappa_f = 1$, $s = 1/10$, and $2\pi T\,t_0 = 50$.
  ]{
  \def\svgwidth{.45\textwidth}
\executeiffilenewer{fig1-a-n2c-kappa.svg}{fig1-a-n2c-kappa.pdf}%
{inkscape -z -D --file=fig1-a-n2c-kappa.svg %
--export-pdf=fig1-a-n2c-kappa.pdf --export-latex}%
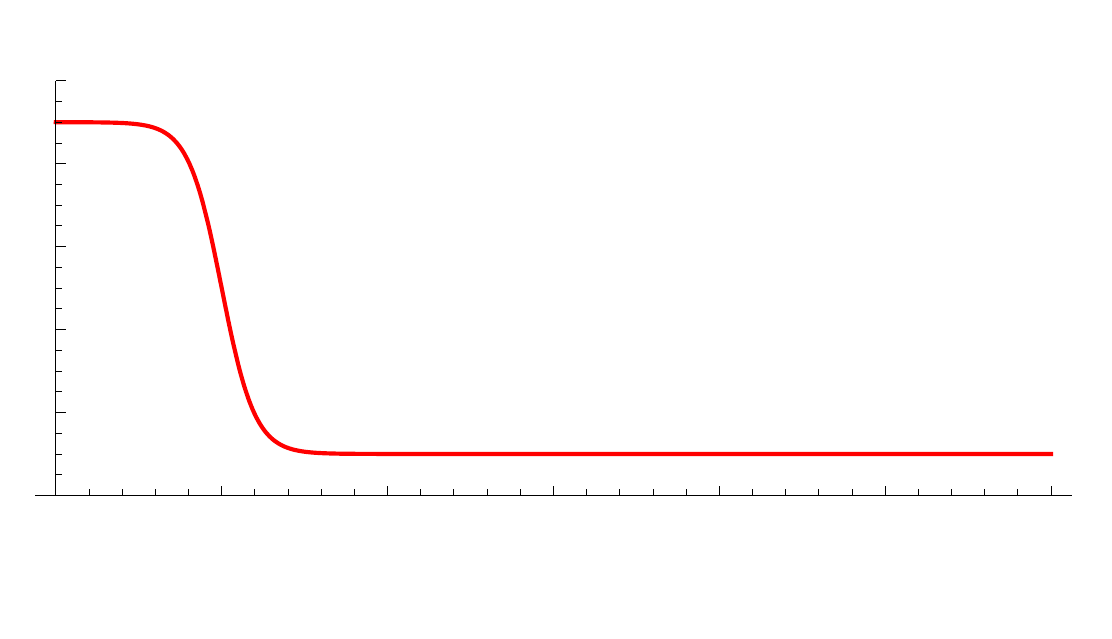%

}
\hfill
\subfloat[
  Time evolution of the absolute value of $\langle\mathcal{O}\rangle$.
  ]{
  \def\svgwidth{.45\textwidth}
\executeiffilenewer{fig1-d-n2c-abs-beta-linear.svg}{fig1-d-n2c-abs-beta-linear.pdf}%
{inkscape -z -D --file=fig1-d-n2c-abs-beta-linear.svg %
--export-pdf=fig1-d-n2c-abs-beta-linear.pdf --export-latex}%
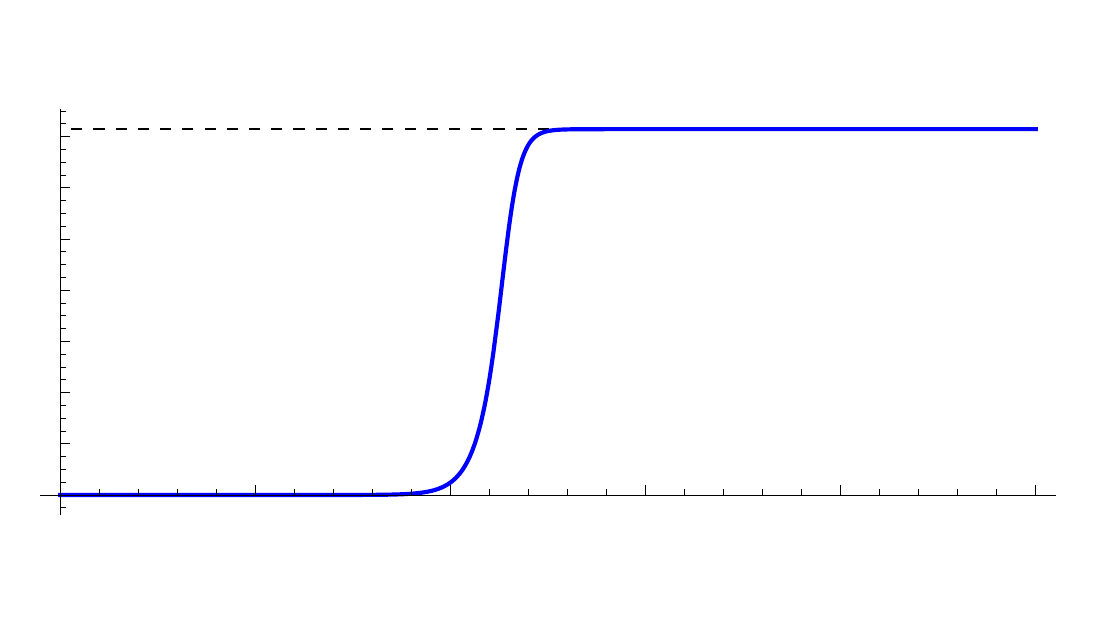%

  \label{fig:1b}
}
\hfill
\subfloat[
  Log plot of (b), indicating the exponential instability.
  ]{
  \def\svgwidth{.45\textwidth}
\executeiffilenewer{fig1-b-n2c-abs-beta.svg}{fig1-b-n2c-abs-beta.pdf}%
{inkscape -z -D --file=fig1-b-n2c-abs-beta.svg %
--export-pdf=fig1-b-n2c-abs-beta.pdf --export-latex}%
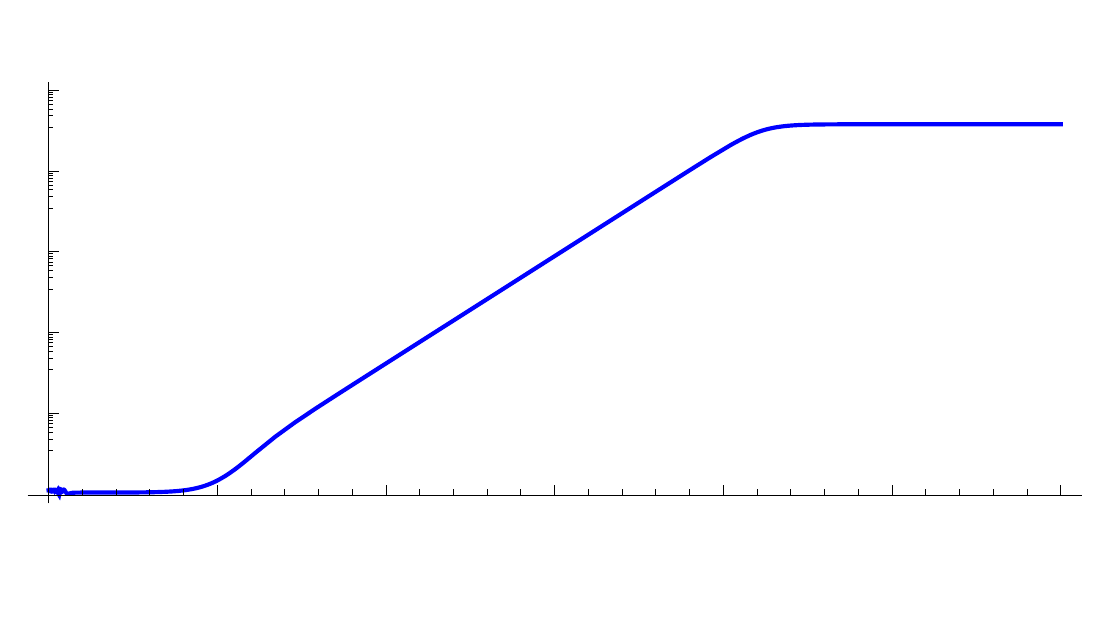%

}
\hfill
\subfloat[
  Log plot for the deviation at late times, indicating the QNM ringdown to the 
  equilibrium value for the new coupling $\kappa_f = \kappa_1^{nc}(\infty)$.
  ]{
  \def\svgwidth{.45\textwidth}
\executeiffilenewer{fig1-c-n2c-abs-beta-late.svg}{fig1-c-n2c-abs-beta-late.pdf}%
{inkscape -z -D --file=fig1-c-n2c-abs-beta-late.svg %
--export-pdf=fig1-c-n2c-abs-beta-late.pdf --export-latex}%
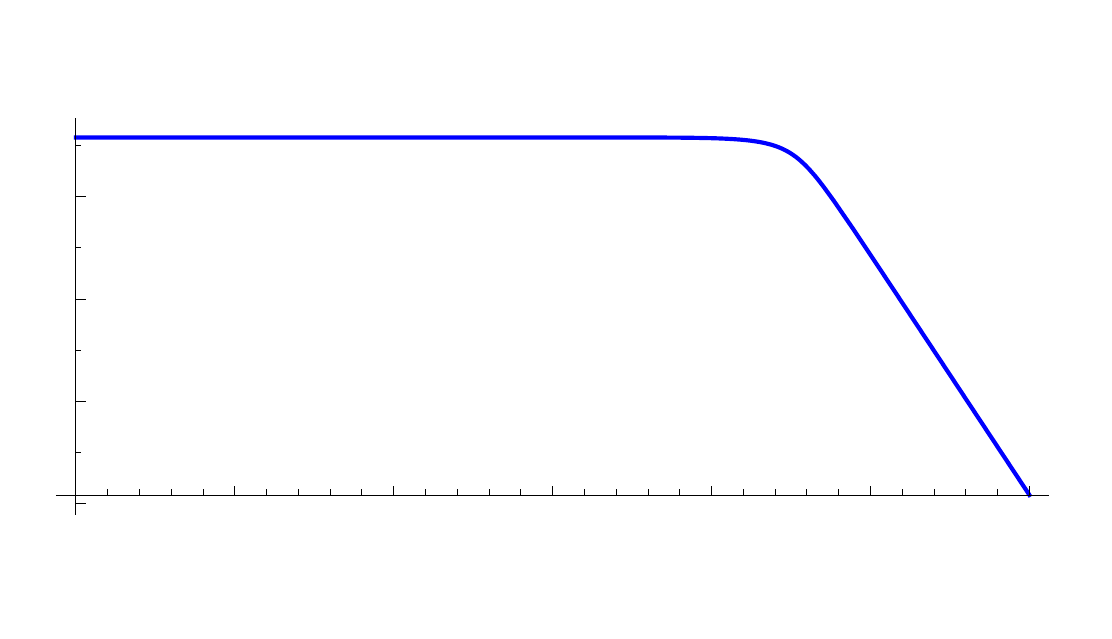%

}
\caption{Typical numerical evolution of the scalar operator (b) for a quench  
from the normal to the condensed phase (a). Note the different time scales 
involved due to the instability mode (c) and the QNM ringdown to the final 
equilibrium (d).
}
\label{fig:generic-norm2cond-amp}
\end{figure}

\begin{figure}[htbp]
\subfloat[]{
  \def\svgwidth{.45\textwidth}
\executeiffilenewer{fig2-a-n2c-re-and-im-linear.svg}{fig2-a-n2c-re-and-im-linear.pdf}%
{inkscape -z -D --file=fig2-a-n2c-re-and-im-linear.svg %
--export-pdf=fig2-a-n2c-re-and-im-linear.pdf --export-latex}%
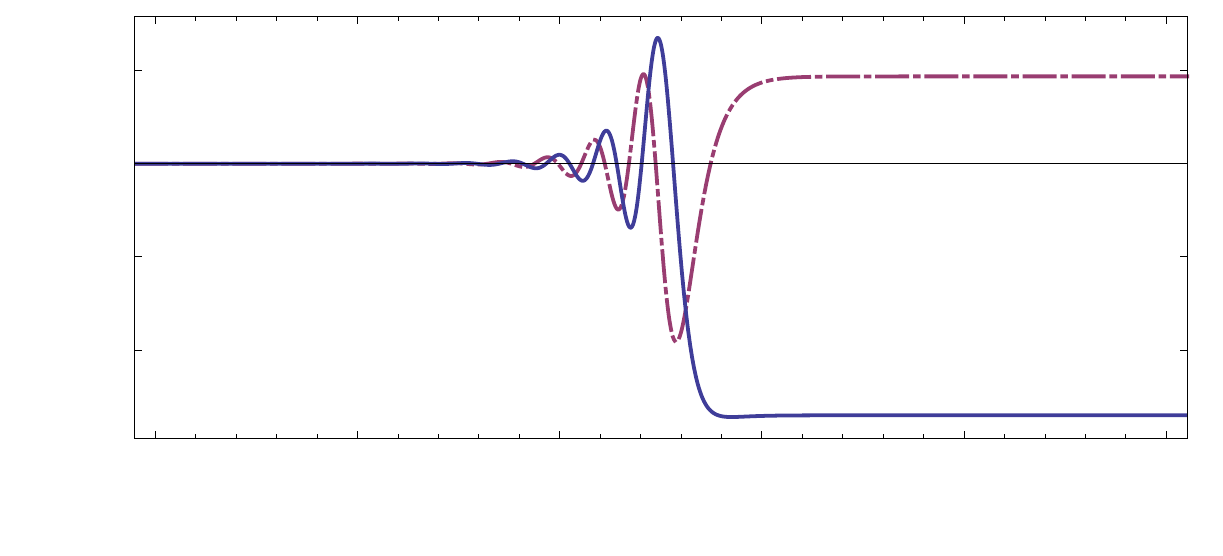%

}
\hfill
\subfloat[]{
  \def\svgwidth{.45\textwidth}
\executeiffilenewer{fig2-b-n2c-re-log.svg}{fig2-b-n2c-re-log.pdf}%
{inkscape -z -D --file=fig2-b-n2c-re-log.svg %
--export-pdf=fig2-b-n2c-re-log.pdf --export-latex}%
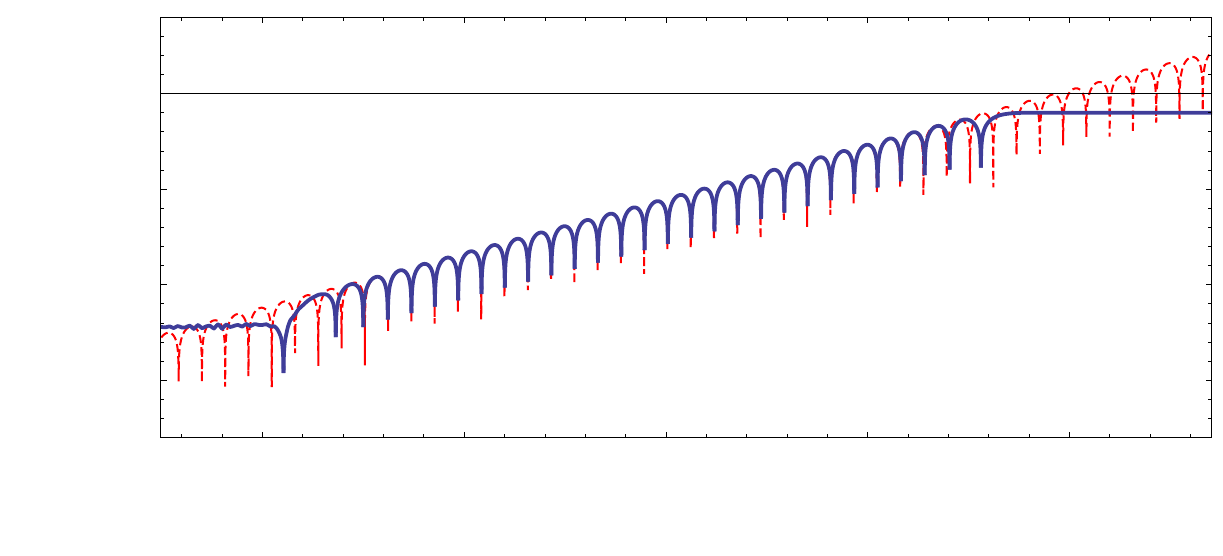%

}
\caption{Time evolution of the real and imaginary part 
of $\langle\mathcal{O}\rangle$ (a), and the normalised absolute value of 
$\Re\langle\mathcal{O}\rangle$ (b) for the normal-to-condensed quench, 
$\kappa_1^{nc}(t)$. 
The red curve in (b) is a numerical fit to the unstable QNM behaviour 
of the form given by Eq.~\eqref{eq:fit}, with $\omega_I > 0$.}
\label{fig:generic-norm2cond-re-and-im}
\end{figure}

\begin{figure}[htbp]
\centering
\def\svgwidth{.99\textwidth}
\executeiffilenewer{fig3-n2c-Delta_t.svg}{fig3-n2c-Delta_t.pdf}%
{inkscape -z -D --file=fig3-n2c-Delta_t.svg %
--export-pdf=fig3-n2c-Delta_t.pdf --export-latex}%
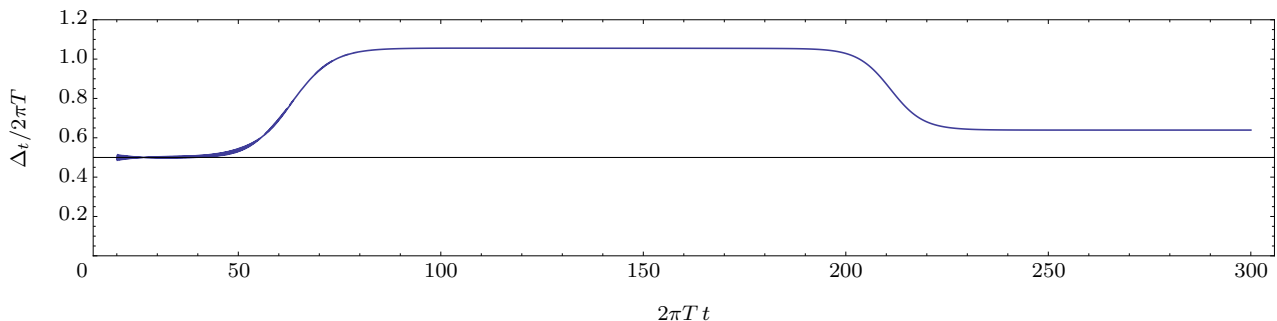%

\caption{Time evolution of $\Delta_t = \mu - \pd_t \psi_0$
for the normal-to-condensed quench $\kappa_1^{nc}(t)$.}
\label{fig:generic-norm2cond-Delta_t}
\end{figure}

Figs.~\ref{fig:generic-norm2cond-amp} to~\ref{fig:generic-norm2cond-Delta_t} 
show the time evolution of 
the scalar operator expectation value $\langle\mathcal{O}(t)\rangle$
for a quench
from the normal phase ($T > T_c$, $\kappa_1 > \kappa_c$) to the condensed phase 
($T < T_c$, $\kappa_1 < \kappa_c$). 
The profile for this ``normal-to-condensed'' quench, $\kappa_1^{nc}(t)$, 
is shown in fig.~\ref{fig:generic-norm2cond-amp} (a). 
Fig.~\ref{fig:generic-norm2cond-amp} (b) shows the absolute value of 
$\langle\mathcal{O}(t)\rangle$, 
and we see first a clear exponential rise (fig.~\ref{fig:generic-norm2cond-amp} 
(c)), and then an exponential decay to a constant value 
(fig.~\ref{fig:generic-norm2cond-amp} (d)).~\footnote{In the normal phase, the 
scalar 
condensate vanishes, $\langle\mathcal{O}\rangle\propto\beta^\dagger = 0$. However 
numerically, zero is only represented up to machine precision, i.e.~a finite 
quantity is regarded as zero if its magnitude is less than the machine 
precision. To avoid numerical artefacts at machine precision and to have a 
firmer control of the numerical accuracy, we set $\beta_1(t=0) = \beta_2(t=0) = 10^{-10}$, 
which explains the initial plateau.}
Fig.~\ref{fig:generic-norm2cond-re-and-im} shows finer details of the time evolution for the 
normal-to-condensed quench, $\kappa_1^{nc}(t)$. We see that $\langle\mathcal{O}(t)\rangle$ oscillates 
and then settles exponentially to a non-zero value dictated by $\kappa_f$ in the condensed phase. 
Note that appreciable changes in $\langle\mathcal{O}(t)\rangle$ do not begin until well after the end 
of the quench. 

We model the time evolution of the scalar field using a QNM behaviour of the form 
\begin{equation}
f(t) = a \, e^{-i \omega t} + b \,, \quad \omega = \omega_R + i\omega_I \,,
\label{eq:fit}
\end{equation}
to fit our results, 
with $\omega$ is the complex QNM frequency. The fit is depicted by the
red curve in fig.~\ref{fig:generic-norm2cond-re-and-im}, which agrees
very well with the full numerical result given by the blue curve. The initial 
behaviour just after the quench is described by a QNM with $\omega_R \neq 0$, leading to an oscillation profile, and with
$\omega_I > 0$, leading to an exponential rise indicating an
instability. Physically, this comes about since
the quench is driving the system out of the normal phase, and
instabilities must occur prior to the formation of a scalar condensate
leading to the new stable ground state. 
Just after the quench, the dominant QNM  is thus associated with the instability of the
normal phase. We will make this more precise in the next subsection.

On the other hand, the relaxation at late times is described by a QNM with $\omega_I < 0$ and $\omega_R = 0$, i.e. a pure 
exponential decay. Such behaviour is expected as the system reach near its final equilibrium state, and is controlled
by the QNM of the final state, see e.g.~\cite{Bantilan:2012vu}. Note that this exponential decay 
is governed by the lowest-lying QNM, i.e.~one with the smallest
$|\omega_I|$. Higher QNMs would produce much 
faster decays than that from the lowest QNM, which would be exponentially suppressed in comparison, 
and not be visible at late times. 
Note also that since the time-evolution is described by the QNMs, its time scales
are given by the appropriate $\omega_I^{-1}$ in the appropriate
regime. These scales are all long compared to the 
time scale of the quench profile.

Fig.~\ref{fig:generic-norm2cond-Delta_t} shows the time evolution of the gauge 
invariant quantity 
$\Delta_t = \mu - \pd_t \psi_0$
\footnote{
The quantity $\psi_0$ is the leading order 
expansion coefficient of the phase of the scalar field, and is given by 
$\psi_0 = \arctan \left(\beta_2(t) / \beta_1(t) \right)$.
}
for the normal-to-condensed quench  $\kappa_1^{nc}(t)$. 
It starts out at $\approx 1/2$ since in the normal phase
$\mu(0) = 1/2$ and the phase rotation $\partial_t\psi_0$ is approximately zero 
if the coupling parameter $\kappa_i = 9$ is close to its critical value. 
As the system is quenched, $\Delta_t$ rises to an intermediate plateau.
This is due to the fact that an instability mode of the normal phase
is turned on, whose 
imaginary part is just the phase rotation velocity.
Eventually at $2\pi Tt \approx 200$, the scalar field becomes macroscopic in 
size, see fig.~\ref{fig:1b}. 
This causes backreaction on $\Delta_t$, which then drops to a new asymptotic constant 
value for $2\pi Tt > 200$.

\begin{figure}[htbp]
\subfloat[]{
  \def\svgwidth{.47\textwidth}
\executeiffilenewer{fig4-a-c2n-kappa.svg}{fig4-a-c2n-kappa.pdf}%
{inkscape -z -D --file=fig4-a-c2n-kappa.svg %
--export-pdf=fig4-a-c2n-kappa.pdf --export-latex}%
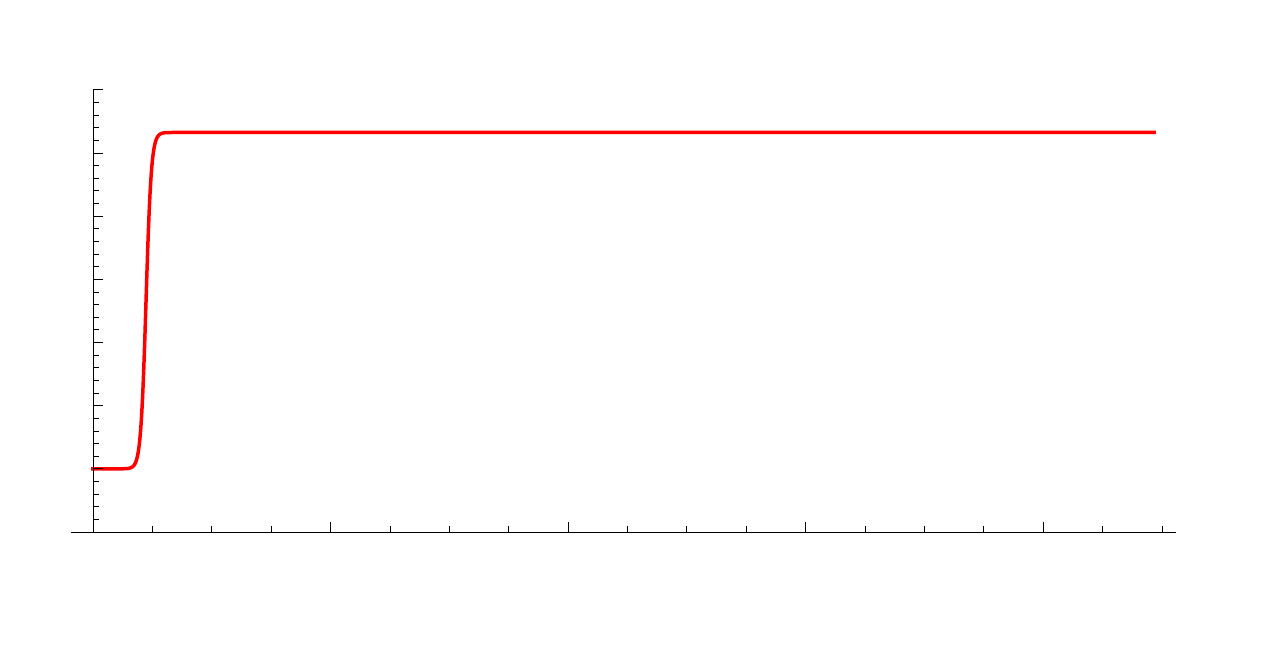%

  \label{fig:generic-cond2norm-abs-beta-a}
}
\hfill
\subfloat[]{
  \def\svgwidth{.47\textwidth}
\executeiffilenewer{fig4-b-c2n-abs-beta.svg}{fig4-b-c2n-abs-beta.pdf}%
{inkscape -z -D --file=fig4-b-c2n-abs-beta.svg %
--export-pdf=fig4-b-c2n-abs-beta.pdf --export-latex}%
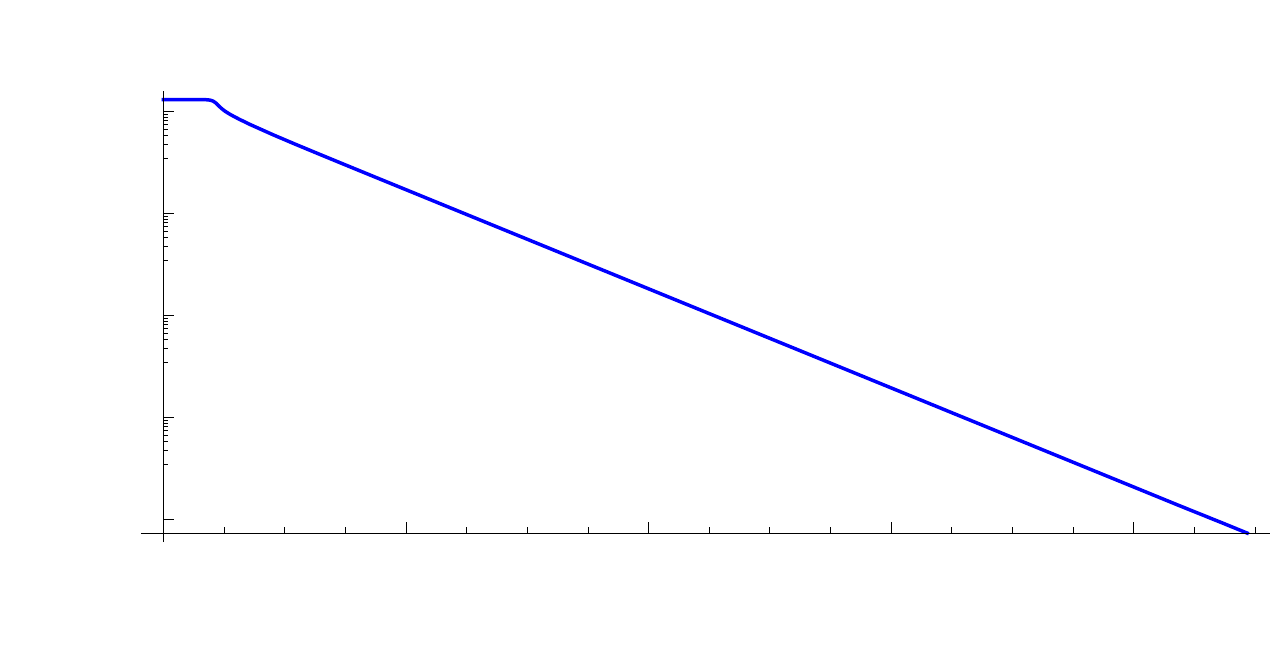%

  \label{fig:generic-cond2norm-abs-beta-b}
}
\caption{The quench profile (left) and the time evolution of the absolute value of 
$\langle\mathcal{O}\rangle$ (right) for a quench from the condensed to the normal phase. 
The quench parameters are $\kappa_i = 8$, $\kappa_f \approx 10.7$, $s \approx 
0.022$ and $2\pi T \,t_0 \approx 447.$}
\label{fig:generic-cond2norm-abs-beta}
\end{figure}

\begin{figure}[htbp]
\subfloat{
  \def\svgwidth{.45\textwidth}
\executeiffilenewer{fig5-a-c2n-re-and-im-linear.svg}{fig5-a-c2n-re-and-im-linear.pdf}%
{inkscape -z -D --file=fig5-a-c2n-re-and-im-linear.svg %
--export-pdf=fig5-a-c2n-re-and-im-linear.pdf --export-latex}%
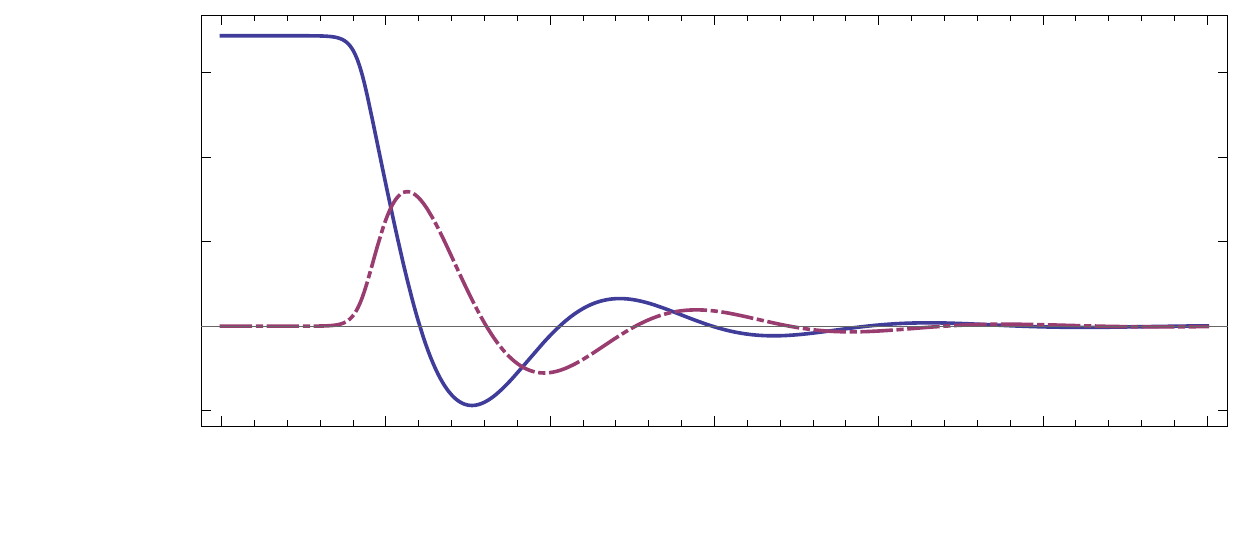%

}
\hfill
\subfloat{
  \def\svgwidth{.45\textwidth}
\executeiffilenewer{fig5-b-c2n-log.svg}{fig5-b-c2n-log.pdf}%
{inkscape -z -D --file=fig5-b-c2n-log.svg %
--export-pdf=fig5-b-c2n-log.pdf --export-latex}%
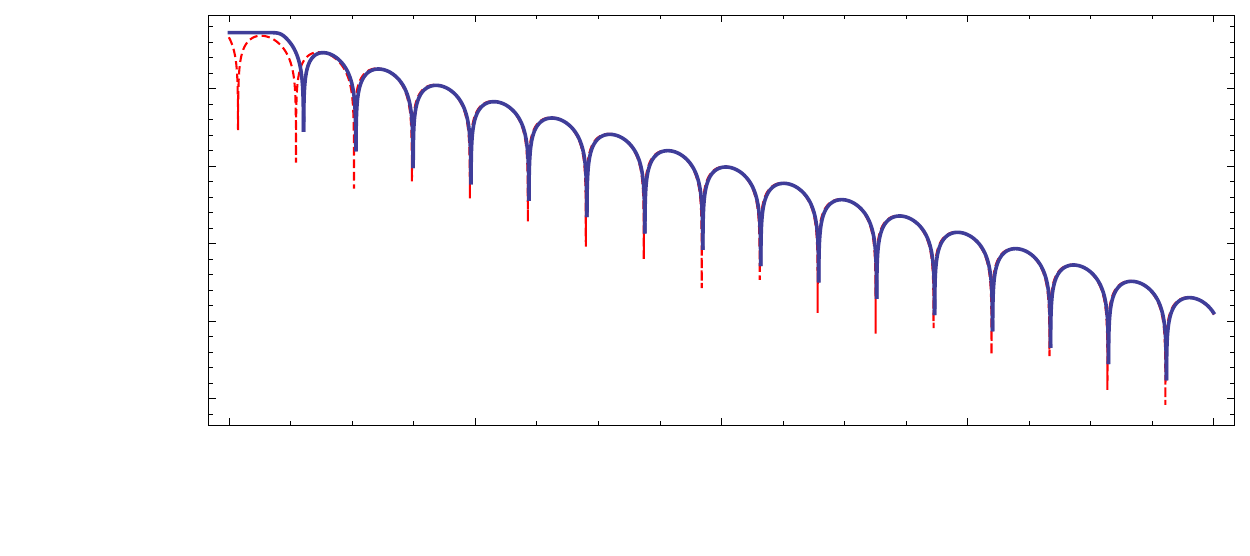%

}
\caption{Time evolution of the real and imaginary part of $\langle\mathcal{O}\rangle$ (left),
and the absolute value of $\Re\langle\mathcal{O}\rangle$ (right) for the condensed-to-normal 
quench, $\kappa_1^{cn}(t)$. The red curve in the right plot is a numerical fit to the lowest
QNM ringdown of the form given by Eq.~\eqref{eq:fit}.}
\label{fig:generic-cond2norm-re-and-im}
\end{figure}

\begin{figure}[htbp]
\centering
\def\svgwidth{.99\textwidth}
\executeiffilenewer{fig6-c2n-Delta_t.svg}{fig6-c2n-Delta_t.pdf}%
{inkscape -z -D --file=fig6-c2n-Delta_t.svg %
--export-pdf=fig6-c2n-Delta_t.pdf --export-latex}%
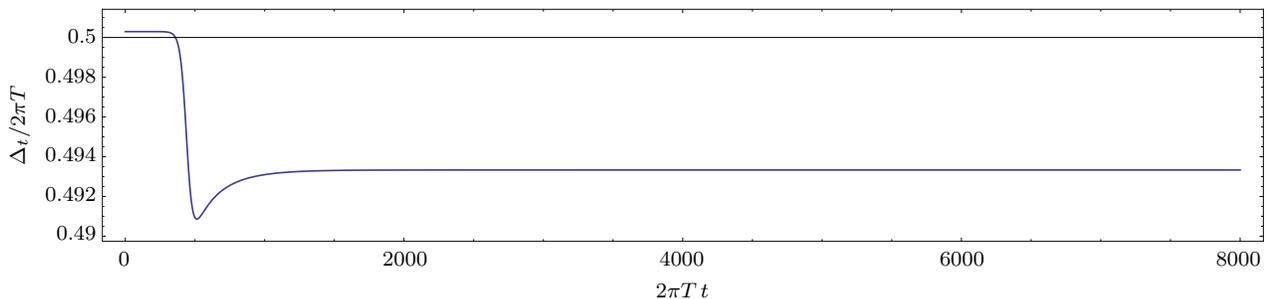%

\caption{Time evolution of the gauge invariant quantity, $\Delta_t = \mu - \pd_t 
\psi_0$.} 
\label{fig:generic-cond2norm-Delta_t}
\end{figure}

Let us now turn to a quench from the condensed to the normal phase, as
shown in
Fig.~\ref{fig:generic-cond2norm-abs-beta}
to~\ref{fig:generic-cond2norm-Delta_t}. The profile for the condensed-to-normal 
quench, $\kappa_1^{cn}(t)$, is shown in
fig.~\ref{fig:generic-cond2norm-abs-beta-a}.
Fig.~\ref{fig:generic-cond2norm-abs-beta-b} shows the absolute value 
of $\langle\mathcal{O}(t)\rangle$, 
and we see now a clear exponential decay to zero right after the
quench, as expected for the scalar in the normal phase.

Fig.~\ref{fig:generic-cond2norm-re-and-im} shows finer details of the time evolution for the condensed-to-normal 
quench, $\kappa_1^{cn}(t)$. We see that  $\langle\mathcal{O}(t)\rangle$ oscillates and then settles exponentially to 
zero in the normal phase. Again, the time evolution is very well described by the 
QNM behaviour of the form given by Eq.~\eqref{eq:fit}. This time,
the governing QNM is the lowest-lying in the normal phase with $\omega_I < 0$. 
Note the period of the oscillations is much longer (by one to two orders of magnitude) compared to the 
normal-to-condensed quench since here, $|\omega_I|$  is much smaller.

Interestingly, the QNM behaviour takes over almost immediately after the quench, 
rather than at late times. 
There does not appear to be a nonlinear regime between the end of the quench and 
the start of the ringdown. 
This appears to be a universal feature of holography, where strongly-coupled 
systems are modelled by dual 
gravitational dynamics~\cite{Ishii:2015gia,Bantilan:2012vu}.

Fig.~\ref{fig:generic-cond2norm-Delta_t} shows the time evolution of the gauge invariant quantity
$\Delta_t = \mu - \pd_t \psi_0$ for the condensed-to-normal quench $\kappa_1^{cn}(t)$. Compared
to the normal-to-condensed quench, there is no plateau or basin structure seen between the initial
and the final equilibrium values of $\Delta_t$.
The value of $\Delta_t$ does not 
asymptote to $\mu_{c} = 1/2$ (for $Q = 1/2$) due to the ongoing 
phase rotation of the scalar field, whose velocity asymptotes to a constant 
which matches the offset in \ref{fig:generic-cond2norm-Delta_t}. This offset 
arises from the real part of the lowest QNM of the normal phase.

%%%%%%%%%%%%%%%%%%%%%%%%%%%%%%%%%%%%%%%%%%%%%%%%%%%%%%%%%%%%%%%%%%%%%%%%%%%%%%%%
%%%%%%%%%%%%%%%%%%%%%%%%%%%%%%%%%%%%%%%%%%%%%%%%%%%%%%%%%%%%%%%%%%%%%%%%%%%%%%%%

\subsection{QNMs in normal phase}
\label{sec::normalQNM}

By the holographic correspondence, the QNMs of the scalar 
fluctuations are related to the poles of the two-point function of the dual scalar 
operator $\mathcal{O}$. 
In the normal phase, these poles are analyically given 
by~\cite{Erdmenger:2016vud}
\begin{equation}\label{kappa}
\frac{1}{\kappa_{T}(\omega)} = -\log\frac{T}{T_K}
= H\!\left(-\frac{1}{2} + i Q - i\frac{\omega}{2\pi T}\right) +  
  H\!\left(-\frac{1}{2} - i Q\right) + \log 2 \,.
\end{equation}
For complex frequencies, $\kappa_T(\omega)$ 
is a complex function. As the temperature is lowered, poles in the lower half of 
the complex $\omega$-plane move up towards the origin, arriving there at the 
critical temperature, $T_c$. They
 cross into the upper half plane below $T_c$, signalling an 
instability~\cite{Erdmenger:2016vud}. This is qualitatively similar to the 
behaviour of the lowest-lying QNMs in holographic superconductors 
\cite{Amado:2009ts,Bhaseen:2012gg}. The critical coupling at 
which phase transition occurs is thus $\kappa_T(0) = \kappa_c$, the same critical
value as that given in Eq.~\eqref{eq:kappacrit}

\begin{figure}[htbp]
 \subfloat[][
 Contour plot of $|\kappa|$ in the complex frequency plane. 
 White space denotes cropping for values $|\kappa|>2$.
 ]{
  \def\svgwidth{.3\textwidth}
\executeiffilenewer{fig7-a-contour-overview.svg}{fig7-a-contour-overview.pdf}%
{inkscape -z -D --file=fig7-a-contour-overview.svg %
--export-pdf=fig7-a-contour-overview.pdf --export-latex}%
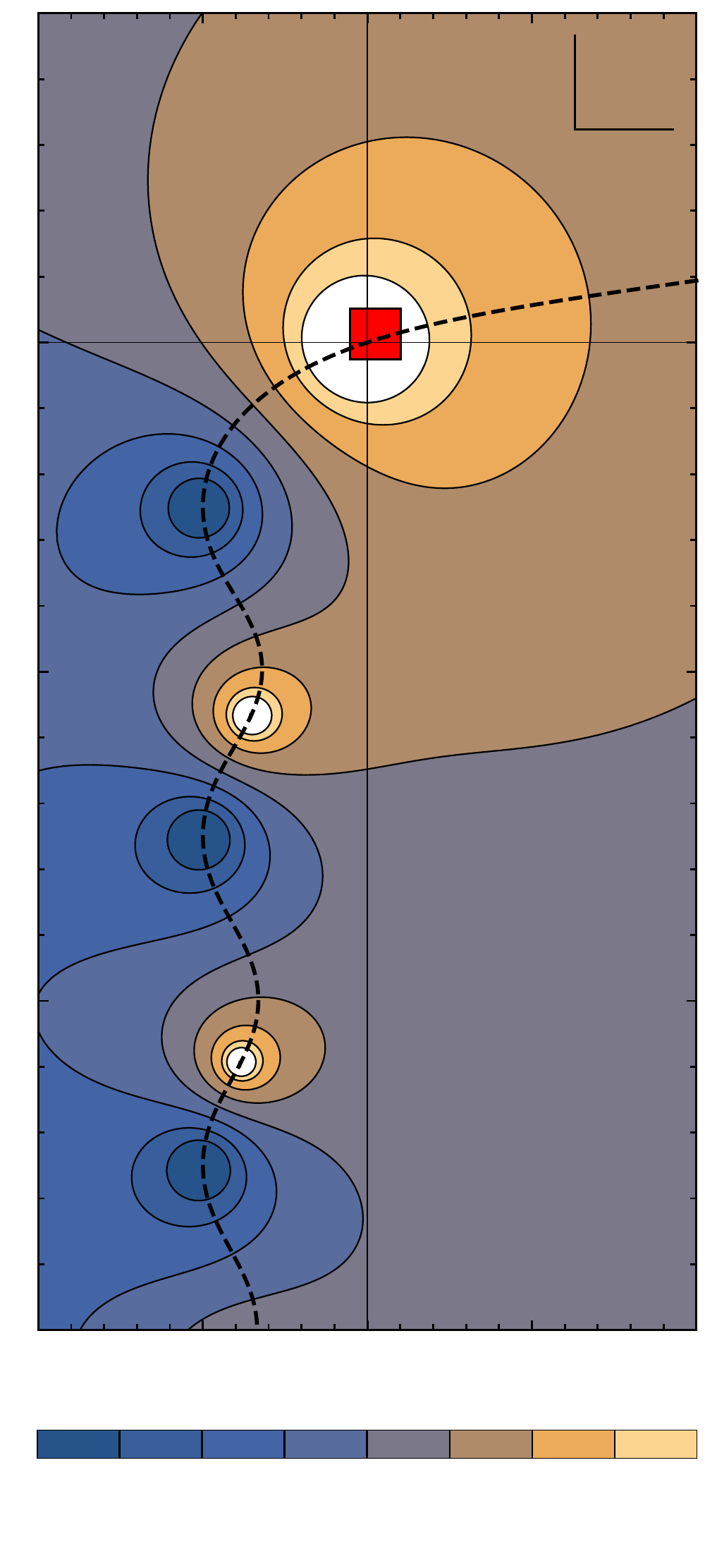%

 }
 \hfill
 \subfloat[][
  Blowup of the red square in (a). 
  Red dots indicate QNMs extracted from numerical fitting at different temperatures.
  White space denotes cropping for values $|\kappa|>100$.
 ]{
  \def\svgwidth{.6\textwidth}
\executeiffilenewer{fig7-b-contour-detail.svg}{fig7-b-contour-detail.pdf}%
{inkscape -z -D --file=fig7-b-contour-detail.svg %
--export-pdf=fig7-b-contour-detail.pdf --export-latex}%
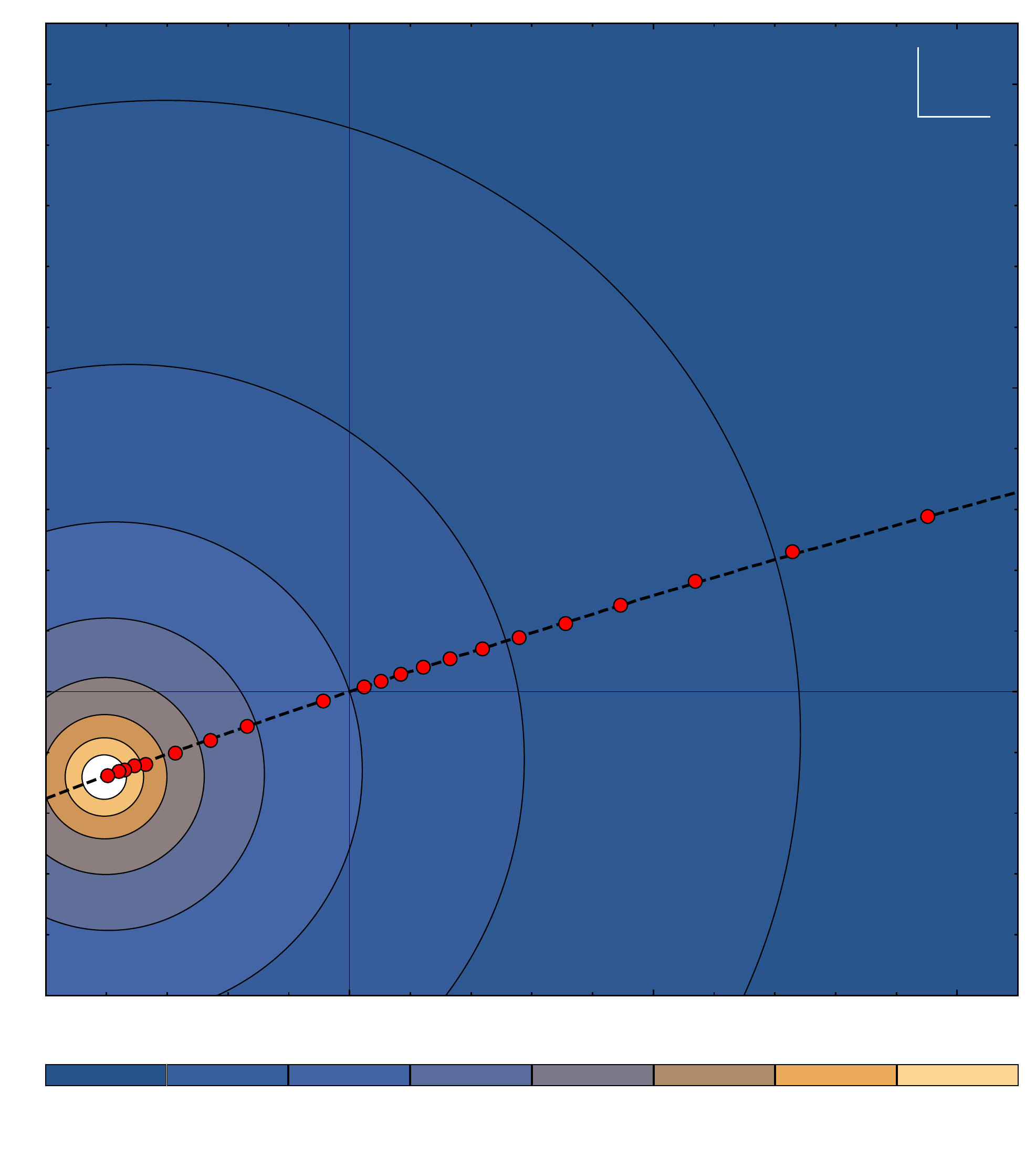%

 }
 \caption{Contour plot of $|\kappa(\omega)|$ over the complex $\omega$-plane. 
A blow-up around the origin is shown in (b). Along the dashed
curve  
$\kappa(\omega)$ is real. The contours denote constant values of 
$|\kappa|$, with the colour scale indicating the value. 
The contour $\kappa(\omega) = \kappa_{c}$ intersects the dashed line at the 
origin in (b). The red dots are lowest lying QNM found from fitting the time 
evolution of the scalar operator after the quench
to the QNM behaviour defined in eq.~\eqref{eq:fit}.
}
\label{fig:compPlane}
\end{figure}
Fig.~\ref{fig:compPlane} shows a contour plot of the magnitude of the Kondo coupling, $|\kappa|$, on the complex 
$\omega$ plane. The dashed curve traces out a path on which $\kappa$ is real. 
Following this curve, we have $\kappa > \kappa_c$ ($T > T_c$) in the lower 
half plane, $\kappa < \kappa_c$ ($T < T_c$) in the upper half plane, and 
$\kappa = \kappa_c$ at the origin as expected.
The right of fig.~\ref{fig:compPlane} shows a blow-up of the region in the 
contour plot 
around the origin marked by the red square. The QNMs found numerically from 
fitting the time evolution of the scalar condensate
to the QNM behaviour given by Eq.~\eqref{eq:fit} are marked by the red dots. 
Those coming from fitting the exponential rise in the normal-to-condensed quench
give QNMs in the upper half plane, while those from the exponential decay 
in the condensed-to-normal quench give QNMs in the lower half plane.
We see that they fall perfectly on the dashed curved analytically given by 
Eq.~\eqref{kappa}, indicating that it is indeed the normal phase QNM associated
with instability that governs the exponential rise seen in fig.~\ref{fig:generic-norm2cond-amp},
and the lowest QNM in the normal phase that governs the relaxation seen in 
fig.~\ref{fig:generic-cond2norm-abs-beta}.

Note that close to $T_c$, QNMs in the normal phase can be found analytically~\cite{Erdmenger:2016jjg}. In 
particular, the lowest QNM is given by
\begin{equation}
\frac{\omega}{2\pi T_c} = \frac{-i}{\psi'\left(\frac{1}{2} + i Q\right)}\left(\frac{T}{T_c} - 1\right) \,,
\end{equation}
which implies a relaxation time scale
\begin{equation}
\tau \sim \omega^{-1} \sim \frac{1}{T_K}\left(\frac{T}{T_c} - 1\right)^{-1} \,, \quad T \gtrsim T_c \,, 
\end{equation}
and we have used the fact that $T_c \sim T_K$ for $|Q| = 1/2$.

%%%%%%%%%%%%%%%%%%%%%%%%%%%%%%%%%%%%%%%%%%%%%%%%%%%%%%%%%%%%%%%%%%%%%%%%%%%%%%%%
%%%%%%%%%%%%%%%%%%%%%%%%%%%%%%%%%%%%%%%%%%%%%%%%%%%%%%%%%%%%%%%%%%%%%%%%%%%%%%%%

\subsection{QNMs in condensed phase}
\label{sec::condQNM}

In the condensed phase, the two-point function of the scalar operator, as well 
as the associated QNMs, can be found by solving a coupled system of fluctuation 
equations. Close to $T_c$ when the scalar condensate is small, the QNMs 
can also be found semi-analyically, see \cite{Erdmenger:2016jjg}. 
Here, we obtain QNMs in the condensed phase in yet another way by fitting the relaxation 
of $\langle\mathcal{O}\rangle$ after Gaussian quenches, and we find excellent 
agreement to those found in \cite{Erdmenger:2016jjg}.

\begin{figure}[htbp]
 \subfloat[
 Normalised to $2 \pi T$ and shown in complex plane.
 ]{
  \def\svgwidth{.46\textwidth}
\executeiffilenewer{fig8-QNM-cond.svg}{fig8-QNM-cond.pdf}%
{inkscape -z -D --file=fig8-QNM-cond.svg %
--export-pdf=fig8-QNM-cond.pdf --export-latex}%
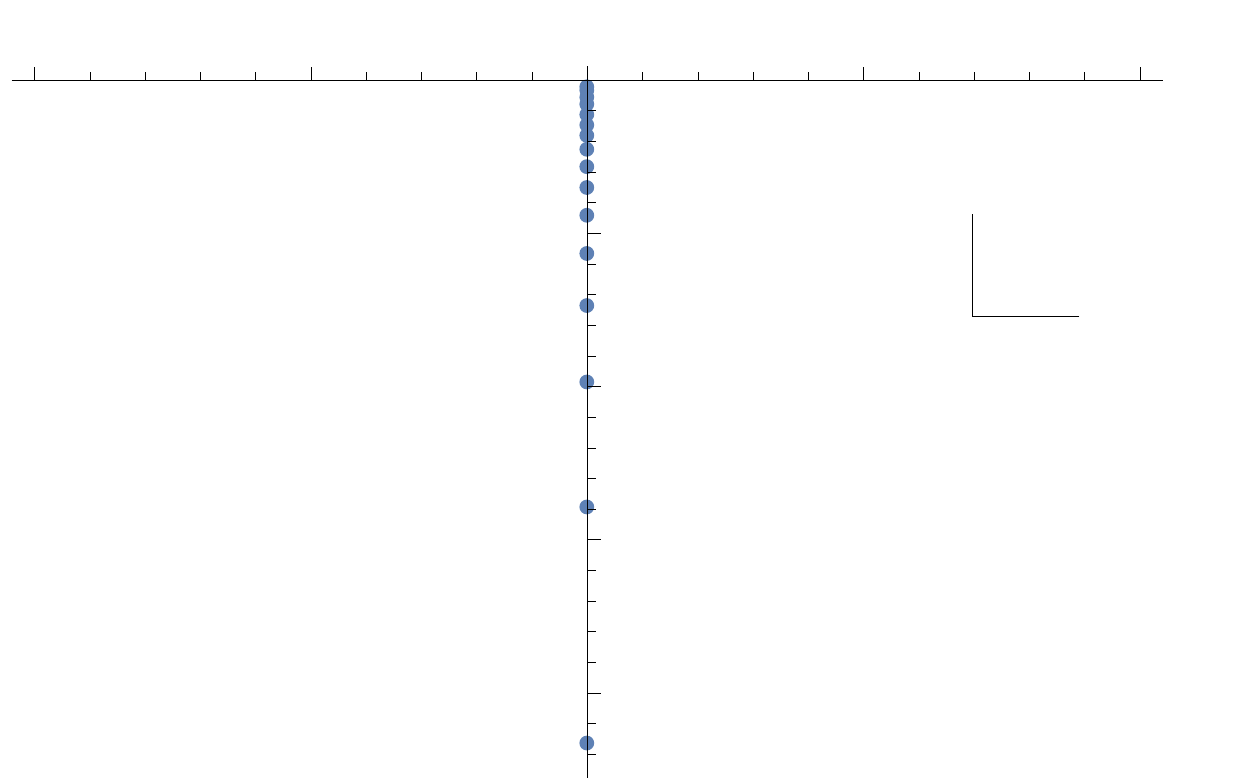%

  \label{fig:QNMs-cond-2a}
 }
 \hfill
 \subfloat[
 Normalised to $2 \pi T_c$ and plotted vs.~$T/T_c$.
 ]{
  \def\svgwidth{.49\textwidth}
\executeiffilenewer{fig9-QNM-cond-phase-normalised.svg}{fig9-QNM-cond-phase-normalised.pdf}%
{inkscape -z -D --file=fig9-QNM-cond-phase-normalised.svg %
--export-pdf=fig9-QNM-cond-phase-normalised.pdf --export-latex}%
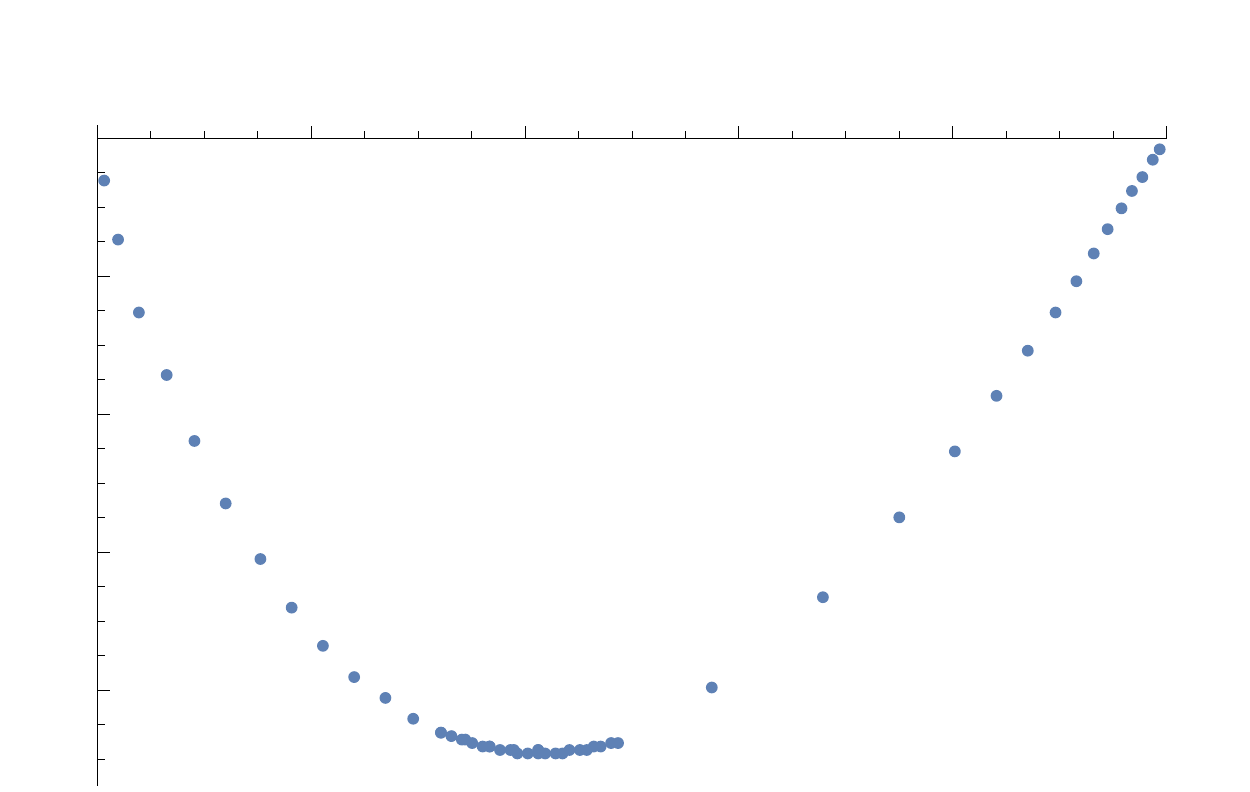%

  \label{fig:QNMs-cond-2b}
 }
 \caption{
 The lowest QNMs extracted from fitting the relaxation behaviour after Gaussian 
 quenches in the condensed phase characterised by 
 $0.008 \leq T/T_c \leq 1$.
}
\label{fig:QNMs-cond}
\end{figure}

Fig.~\ref{fig:QNMs-cond-2a} shows the lowest QNMs, i.e.~QNMs closest to the 
real axis, found from fitting the relaxation at late times after a quench from 
states initially in the condensed phase at various $T < T_c$.
We see they are all purely imaginary, and move down the imaginary axis as the
temperature is lowered, all in agreement with \cite{Erdmenger:2016jjg}. This 
means that the relaxation
is purely an exponential decay without oscillation, in contrast to
what is observed 
in~\cite{Bhaseen:2012gg}, where there
are other regimes in addition to the over-damped one we see here.
Fig.~\ref{fig:QNMs-cond-2b} exhibits the temperature dependence of the lowest 
QNM in the condensed phase, when the temperature is normalised by $T_c$. We see 
that this purely 
imaginary mode first moves down the imaginary axis as temperature decreases 
below $T_c$ and then turnes back up towards the origin. 
This shows that the system experiences \textit{critical slowing down} both 
as $T\rightarrow T_c$ and as $T\rightarrow 0$. See section \ref{sec::critExp} 
for further discussion of this phenomenon in the $T\rightarrow T_c$ 
limit.

\begin{figure}[htbp]
\centering
 \subfloat[
 Behaviour for $T \lesssim T_c$. 
 The dashed line is a linear fit for the region $1 > T/T_c > 0.998$ passing through the origin.
 ]{
  \def\svgwidth{.47\textwidth}
\executeiffilenewer{fig10-omega-vs-alpha-square-linear-NEW.svg}{fig10-omega-vs-alpha-square-linear-NEW.pdf}%
{inkscape -z -D --file=fig10-omega-vs-alpha-square-linear-NEW.svg %
--export-pdf=fig10-omega-vs-alpha-square-linear-NEW.pdf --export-latex}%
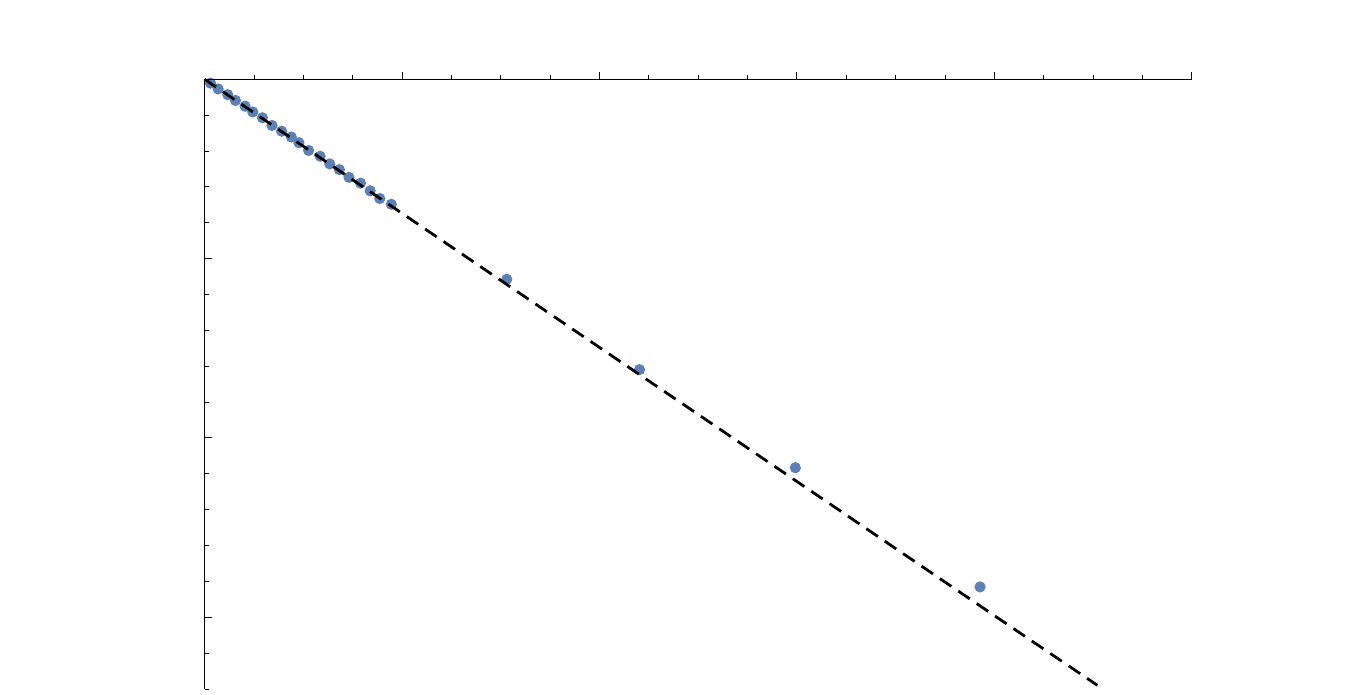%

  \label{fig:omega-vs-alpha-lin}
 }
 \hfill
 \subfloat[
 Full temperature range from $T/T_c = 1$ down to $T/T_c \approx 0.008$. The dashed 
line is a fit for the region $T/T_c < 0.33$.
 ]{
  \def\svgwidth{.43\textwidth}
\executeiffilenewer{fig10-omega-vs-alpha-square-log-linear.svg}{fig10-omega-vs-alpha-square-log-linear.pdf}%
{inkscape -z -D --file=fig10-omega-vs-alpha-square-log-linear.svg %
--export-pdf=fig10-omega-vs-alpha-square-log-linear.pdf --export-latex}%
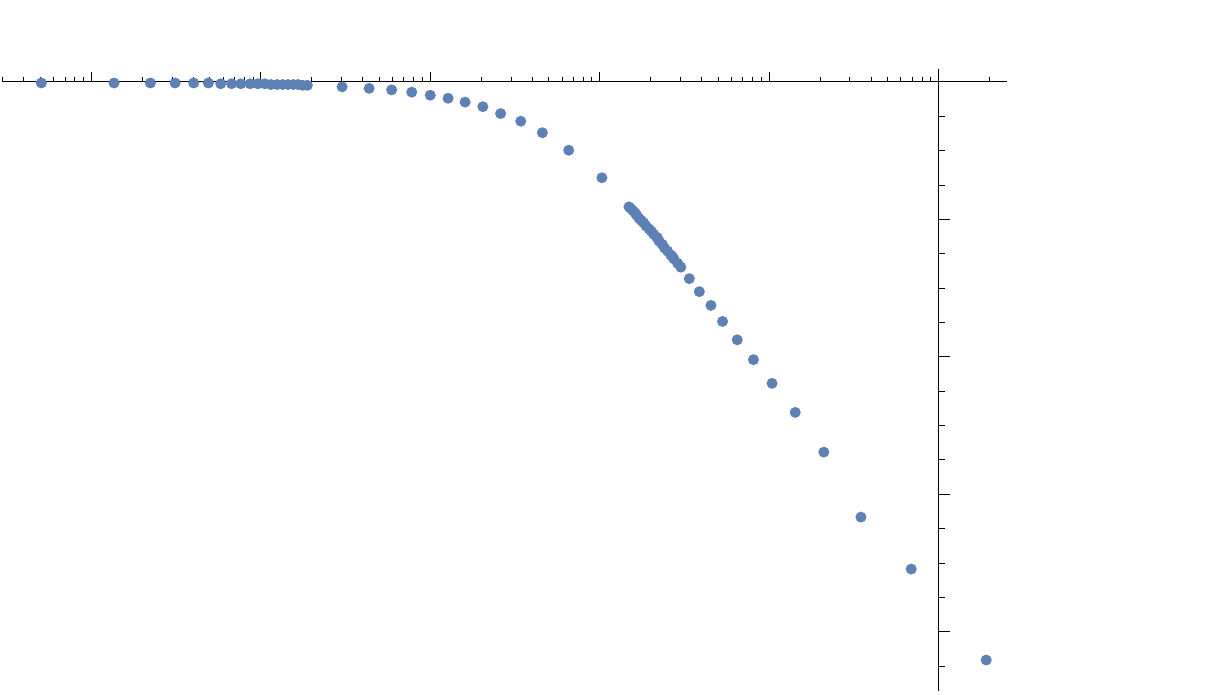%

  \label{fig:omega-vs-alpha-log-lin}
 }
 \caption{Functional dependence of the lowest QNM on
   $\kappa_1^2\langle\mathcal{O}\rangle^2$ in the condensed phase.}
\label{fig:omega-vs-alpha}
\end{figure}
Fig.~\ref{fig:omega-vs-alpha} shows how the lowest QNM vary as a function of $\kappa_1^2\langle\mathcal{O}\rangle^2$.
Close to $T_c$, we find that the lowest QNM has a linear dependence on 
$\kappa_1^2\langle\mathcal{O}\rangle^2/N^2$,
\begin{equation}\label{eq:relation-omega-beta}
\omega_I = b\,\kappa_1^2\langle\mathcal{O}\rangle^2/N^2 \,, \quad b \approx -18.7 \,,
\end{equation}
where $b$ is found from a numerical fit.
This value of $b$ is very close to that in~\cite{Erdmenger:2016jjg}. The difference is about $6\%$, and is due to the 
slight differences in the data being fitted. Deviation from this linear relation began to be visible for $T > 0.988 T_c$.
Note that from~\cite{Erdmenger:2013dpa}, $\langle\mathcal{O}\rangle\propto(T_c - T)^{1/2}$ for $T \lesssim T_c$.
Given $T/T_K = e^{-1/\kappa_1}$, this implies a relaxation time scale 
\begin{equation}
\tau\sim\omega^{-1} \sim \frac{b}{T_K}\log^2\frac{T}{T_K}\left(1 - \frac{T}{T_c}\right)^{-1} \,, \quad T \lesssim T_c \,.
\end{equation}

Interestingly, at low temperatures, this linear dependence turns into a 
logarithmic one,
\begin{equation}\label{eq:omegalog}
\frac{\omega_I}{2\pi T} = b\,
\log\left(\frac{\kappa_1^2\,\langle\mathcal{O}\rangle^2}{2\pi T N^2}\right) \,,
\quad b \approx - 0.14 \, .
\end{equation}
This log behaviour indicates a deviation from mean-field behaviour. To 
understand this further, it is necessary to study the IR behaviour around zero 
temperature. This however requires the stabilisation of the scalar at the IR 
fixed point 
by adding a quartic term to the scalar potential in \eqref{eq:potential}, which 
is left for further study. Note that this is also interesting in the
context of finding the zero-temperature impurity entropy at the defect  \cite{Erdmenger:2015spo}.

%%%%%%%%%%%%%%%%%%%%%%%%%%%%%%%%%%%%%%%%%%%%%%%%%%%%%%%%%%%%%%%%%%%%%%%%%%%%%%%%
%%%%%%%%%%%%%%%%%%%%%%%%%%%%%%%%%%%%%%%%%%%%%%%%%%%%%%%%%%%%%%%%%%%%%%%%%%%%%%%%

\subsection{Evolution of the screening of the impurity}
\label{sec::horizonQNM}

One particular goal of the analysis presented is to obtain the time 
evolution of the screening due to the formation of the Kondo cloud. 
According to \cite{Erdmenger:2013dpa}, the electric flux at the 
asymptotic boundary and the event horizon provides a measure of the
number of impurity degrees of freedom in the UV and the IR, respectively. 
Its decrease thus corresponds to the screening of the impurity in the IR. 
This is due to the fact that the flux involves
$a_t$, which is dual to the charge density determining the size of the
$SU(N)$ spin representation.
The electric flux at the event horizon is a quantity that may be traced in 
Eddington-Finkelstein coordinates. Although there is no
straightforward map of horizon dynamics to the boundary, the decay
constant of the horizon flux still encodes information about the
decrease of impurity degrees of freedom.

\begin{figure}[H]
\subfloat[][
 $\kappa_1 = 1$, i.e.~$T \approx 0.41\,T_c$.
 ]{
  \def\svgwidth{.45\textwidth}
\executeiffilenewer{fig11-a-DOF-at-hor.svg}{fig11-a-DOF-at-hor.pdf}%
{inkscape -z -D --file=fig11-a-DOF-at-hor.svg %
--export-pdf=fig11-a-DOF-at-hor.pdf --export-latex}%
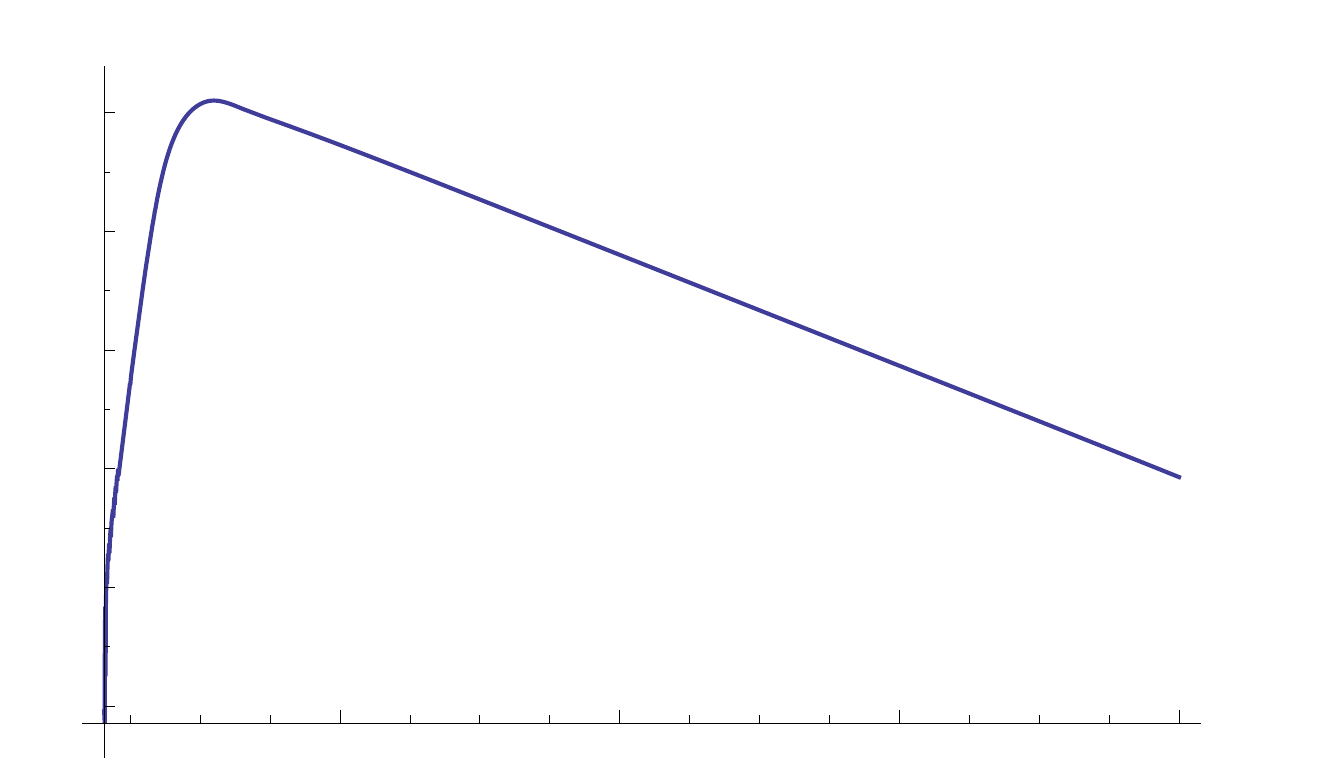%

  \label{fig:DOF-a}
}
 \hfill
\subfloat[][
 $\kappa_1 = 8.5$, i.e.~$T \approx 0.99\,T_c$.
 ]{
  \def\svgwidth{.45\textwidth}
\executeiffilenewer{fig11-b-DOF-at-hor-2.svg}{fig11-b-DOF-at-hor-2.pdf}%
{inkscape -z -D --file=fig11-b-DOF-at-hor-2.svg %
--export-pdf=fig11-b-DOF-at-hor-2.pdf --export-latex}%
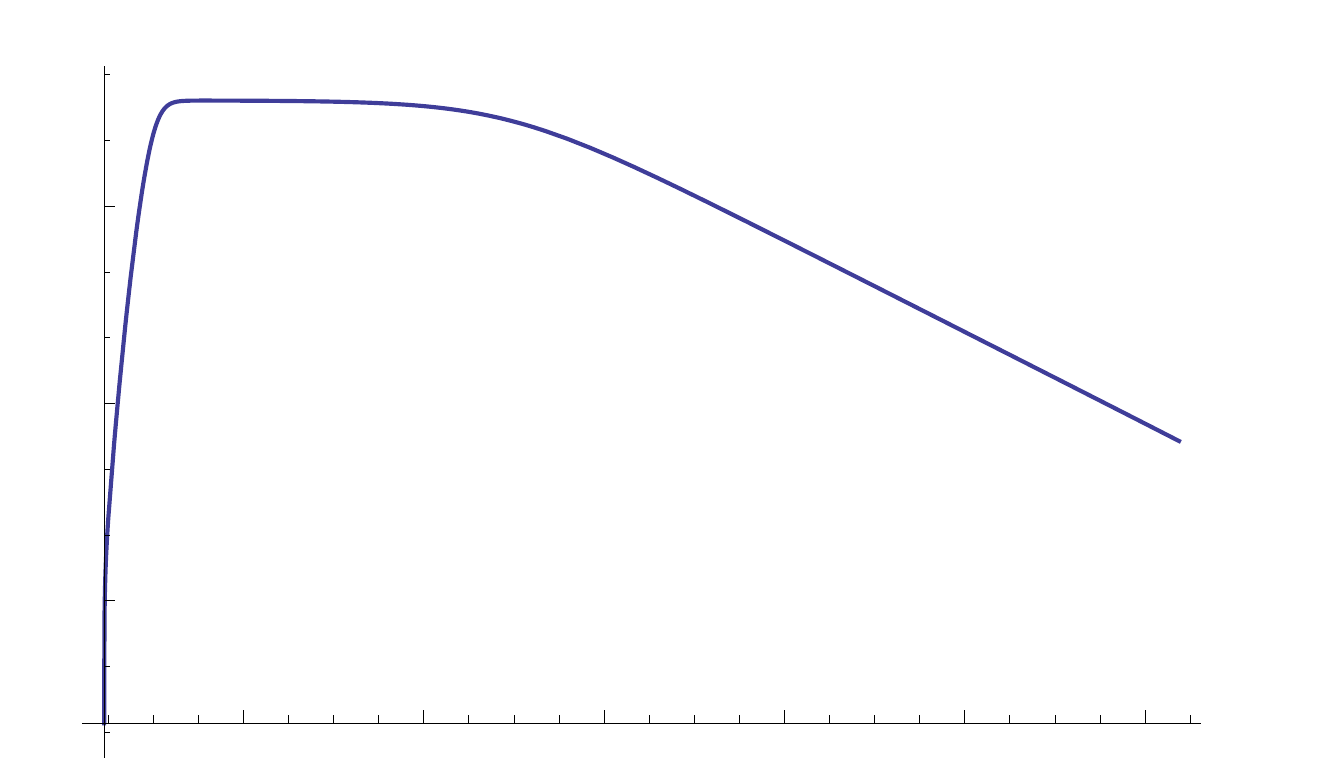%

  \label{fig:DOF-b}
}
\caption{A log plot of $D$ as defined in \eqref{eq:D} for different Gaussian 
quenches, showing the late-time  exponential decay of the effective IR degrees of freedom due to 
screening.
Signs of critical slowing down can be seen in 
(b) as the system is close to the critical temperature $T_c$.}
\label{fig:horizon-av}
\end{figure}

In our choice of coordinates and gauge fixing (see appendix 
\ref{sec:ansatzEF}), $a'_v(v,y)$ is proportional to the electric flux.
We define a new variable $D$ by
\begin{equation}
  D = \frac{a'_v(v,1)}{a'_v(0,1)}-1 \, ,
  \label{eq:D}
\end{equation}
which starts out at zero  at $t=0$ by construction and becomes nontrivial during a 
generic quench.
In figure \ref{fig:horizon-av}, we show the evolution of $D$ for a Gaussian 
quench around $\kappa_1 = 1$ and $\kappa_1 = 8.5$, respectively. We
observe an initial rise due to the Gaussian quench which takes the
system to a state with smaller condensate, and subsequently an
exponential decay which corresponds to the reduction of impurity
degrees of freedom due to screening. For the Kondo coupling $\kappa_1
= 8.5$ shown in fig.~\ref{fig:DOF-b}, which is
closer to the phase transition, there is a plateau at intermediate
times which is a sign of the onset of critical slowing-down near the
phase transition.
Fitting the evolution at late times to an exponential decay reveals that it is 
governed by a QNM which, not suprisingly, coincides with the complex
frequency obtained by analysing the time dependence of boundary quantities.
Of course, it would be interesting to examine further observables
that contain information about the evolution of the screening, in
particular the evolution of the Kondo cloud in the spatial direction
ambient to the defect (i.e.~in $1+1$ dimensions). This
will allow a comparison to the Kondo cloud evolution obtained
e.g.~in~\cite{2014arXiv1409.0646N}.
This may be obtained holographically
with methods proposed in \cite{Erdmenger:2015spo}, however for time-dependent 
couplings. It requires to include the backreaction of the geometry 
to the field content on the defect hypersurface. This is left for future 
research.

%%%%%%%%%%%%%%%%%%%%%%%%%%%%%%%%%%%%%%%%%%%%%%%%%%%%%%%%%%%%%%%%%%%%%%%%%%%%%%%%
%%%%%%%%%%%%%%%%%%%%%%%%%%%%%%%%%%%%%%%%%%%%%%%%%%%%%%%%%%%%%%%%%%%%%%%%%%%%%%%%

%% file: fig1-a-n2c-kappa.pdf_tex
%% Creator: Inkscape inkscape 0.48.4, www.inkscape.org
%% Accompanies image file 'fig1-a-n2c-kappa.pdf' (pdf, eps, ps)
\begingroup%
  \makeatletter%
  \providecommand\color[2][]{%
    \errmessage{(Inkscape) Color is used for the text in Inkscape, but the package 'color.sty' is not loaded}%
    \renewcommand\color[2][]{}%
  }%
  \providecommand\transparent[1]{%
    \errmessage{(Inkscape) Transparency is used (non-zero) for the text in Inkscape, but the package 'transparent.sty' is not loaded}%
    \renewcommand\transparent[1]{}%
  }%
  \providecommand\rotatebox[2]{#2}%
  \ifx\svgwidth\undefined%
    \setlength{\unitlength}{320bp}%
    \ifx\svgscale\undefined%
      \relax%
    \else%
      \setlength{\unitlength}{\unitlength * \real{\svgscale}}%
    \fi%
  \else%
    \setlength{\unitlength}{\svgwidth}%
  \fi%
  \global\let\svgwidth\undefined%
  \global\let\svgscale\undefined%
  \makeatother%
  \begin{picture}(1,0.575)%
    \put(0,0){\includegraphics[width=\unitlength]{fig1-a-n2c-kappa.pdf}}%
    \put(0.05189071,0.52708267){\color[rgb]{0,0,0}\makebox(0,0)[b]{\smash{$\tss{\kappa_1^{nc}}$}}}%
    \put(0.02783167,0.44921833){\color[rgb]{0,0,0}\makebox(0,0)[rb]{\smash{$\tss{9}$}}}%
    \put(0.02783167,0.33947248){\color[rgb]{0,0,0}\makebox(0,0)[rb]{\smash{$\tss{6}$}}}%
    \put(0.02783167,0.22972083){\color[rgb]{0,0,0}\makebox(0,0)[rb]{\smash{$\tss{3}$}}}%
    \put(0.02783167,0.08725075){\color[rgb]{0,0,0}\makebox(0,0)[rb]{\smash{$\tss{0}$}}}%
    \put(0.64913914,0.08725075){\color[rgb]{0,0,0}\makebox(0,0)[b]{\smash{$\tss{200}$}}}%
    \put(0.9482116,0.08725075){\color[rgb]{0,0,0}\makebox(0,0)[b]{\smash{$\tss{300}$}}}%
    \put(0.49834454,0.01414947){\color[rgb]{0,0,0}\makebox(0,0)[b]{\smash{$\tss{2 \pi T \,t}$}}}%
    \put(0.34750578,0.08725075){\color[rgb]{0,0,0}\makebox(0,0)[b]{\smash{$\tss{100}$}}}%
  \end{picture}%
\endgroup%

%% file: fig1-d-n2c-abs-beta-linear.pdf_tex
%% Creator: Inkscape inkscape 0.48.4, www.inkscape.org
%% Accompanies image file 'fig1-d-n2c-abs-beta-linear.pdf' (pdf, eps, ps)
\begingroup%
  \makeatletter%
  \providecommand\color[2][]{%
    \errmessage{(Inkscape) Color is used for the text in Inkscape, but the package 'color.sty' is not loaded}%
    \renewcommand\color[2][]{}%
  }%
  \providecommand\transparent[1]{%
    \errmessage{(Inkscape) Transparency is used (non-zero) for the text in Inkscape, but the package 'transparent.sty' is not loaded}%
    \renewcommand\transparent[1]{}%
  }%
  \providecommand\rotatebox[2]{#2}%
  \ifx\svgwidth\undefined%
    \setlength{\unitlength}{320bp}%
    \ifx\svgscale\undefined%
      \relax%
    \else%
      \setlength{\unitlength}{\unitlength * \real{\svgscale}}%
    \fi%
  \else%
    \setlength{\unitlength}{\svgwidth}%
  \fi%
  \global\let\svgwidth\undefined%
  \global\let\svgscale\undefined%
  \makeatother%
  \begin{picture}(1,0.575)%
    \put(0,0){\includegraphics[width=\unitlength]{fig1-d-n2c-abs-beta-linear.pdf}}%
    \put(0.23105764,0.0806016){\color[rgb]{0,0,0}\makebox(0,0)[b]{\smash{$\tss{100}$}}}%
    \put(0.40704918,0.08638687){\color[rgb]{0,0,0}\makebox(0,0)[b]{\smash{$\tss{200}$}}}%
    \put(0.58464856,0.0907785){\color[rgb]{0,0,0}\makebox(0,0)[b]{\smash{$\tss{300}$}}}%
    \put(0.75711675,0.08986838){\color[rgb]{0,0,0}\makebox(0,0)[b]{\smash{$\tss{400}$}}}%
    \put(0.93309526,0.08807268){\color[rgb]{0,0,0}\makebox(0,0)[b]{\smash{$\tss{500}$}}}%
    \put(0.51286674,0.00536995){\color[rgb]{0,0,0}\makebox(0,0)[b]{\smash{$\tss{2 \pi T \,t}$}}}%
    \put(0.05316753,0.51175737){\color[rgb]{0,0,0}\makebox(0,0)[b]{\smash{$\tss{\frac{|\langle\mathcal{O}\rangle|}{N\sqrt{2 \pi \,T}}}$}}}%
    \put(0.03259032,0.12039528){\color[rgb]{0,0,0}\makebox(0,0)[rb]{\smash{$\tss{0}$}}}%
    \put(0.03259032,0.34934704){\color[rgb]{0,0,0}\makebox(0,0)[rb]{\smash{$\tss{0.1}$}}}%
  \end{picture}%
\endgroup%

%% file: fig1-b-n2c-abs-beta.pdf_tex
%% Creator: Inkscape inkscape 0.91, www.inkscape.org
%% Accompanies image file 'fig1-b-n2c-abs-beta.pdf' (pdf, eps, ps)
\begingroup%
  \makeatletter%
  \providecommand\color[2][]{%
    \errmessage{(Inkscape) Color is used for the text in Inkscape, but the package 'color.sty' is not loaded}%
    \renewcommand\color[2][]{}%
  }%
  \providecommand\transparent[1]{%
    \errmessage{(Inkscape) Transparency is used (non-zero) for the text in Inkscape, but the package 'transparent.sty' is not loaded}%
    \renewcommand\transparent[1]{}%
  }%
  \providecommand\rotatebox[2]{#2}%
  \ifx\svgwidth\undefined%
    \setlength{\unitlength}{320bp}%
    \ifx\svgscale\undefined%
      \relax%
    \else%
      \setlength{\unitlength}{\unitlength * \real{\svgscale}}%
    \fi%
  \else%
    \setlength{\unitlength}{\svgwidth}%
  \fi%
  \global\let\svgwidth\undefined%
  \global\let\svgscale\undefined%
  \makeatother%
  \begin{picture}(1,0.575)%
    \put(0,0){\includegraphics[width=\unitlength,page=1]{fig1-b-n2c-abs-beta.pdf}}%
    \put(0.49942845,0.01414947){\color[rgb]{0,0,0}\makebox(0,0)[b]{\smash{$\tss{2 \pi T \,t}$}}}%
    \put(0.03595532,0.18585877){\color[rgb]{0,0,0}\makebox(0,0)[rb]{\smash{$\tss{10^{-8}}$}}}%
    \put(0.03595532,0.2589029){\color[rgb]{0,0,0}\makebox(0,0)[rb]{\smash{$\tss{10^{-6}}$}}}%
    \put(0.03595532,0.33072559){\color[rgb]{0,0,0}\makebox(0,0)[rb]{\smash{$\tss{10^{-4}}$}}}%
    \put(0.03595532,0.40415675){\color[rgb]{0,0,0}\makebox(0,0)[rb]{\smash{$\tss{10^{-2}}$}}}%
    \put(0.03595532,0.47624481){\color[rgb]{0,0,0}\makebox(0,0)[rb]{\smash{$\tss{1}$}}}%
    \put(0.04508714,0.53442428){\color[rgb]{0,0,0}\makebox(0,0)[b]{\smash{$\tss{\frac{|\langle\mathcal{O}\rangle|}{N\sqrt{2 \pi \,T}}}$}}}%
    \put(0.34802751,0.08849968){\color[rgb]{0,0,0}\makebox(0,0)[b]{\smash{$\tss{100}$}}}%
    \put(0.64967011,0.08849968){\color[rgb]{0,0,0}\makebox(0,0)[b]{\smash{$\tss{200}$}}}%
    \put(0.95302018,0.08849968){\color[rgb]{0,0,0}\makebox(0,0)[b]{\smash{$\tss{300}$}}}%
  \end{picture}%
\endgroup%

%% file: fig1-c-n2c-abs-beta-late.pdf_tex
%% Creator: Inkscape inkscape 0.91, www.inkscape.org
%% Accompanies image file 'fig1-c-n2c-abs-beta-late.pdf' (pdf, eps, ps)
\begingroup%
  \makeatletter%
  \providecommand\color[2][]{%
    \errmessage{(Inkscape) Color is used for the text in Inkscape, but the package 'color.sty' is not loaded}%
    \renewcommand\color[2][]{}%
  }%
  \providecommand\transparent[1]{%
    \errmessage{(Inkscape) Transparency is used (non-zero) for the text in Inkscape, but the package 'transparent.sty' is not loaded}%
    \renewcommand\transparent[1]{}%
  }%
  \providecommand\rotatebox[2]{#2}%
  \ifx\svgwidth\undefined%
    \setlength{\unitlength}{320bp}%
    \ifx\svgscale\undefined%
      \relax%
    \else%
      \setlength{\unitlength}{\unitlength * \real{\svgscale}}%
    \fi%
  \else%
    \setlength{\unitlength}{\svgwidth}%
  \fi%
  \global\let\svgwidth\undefined%
  \global\let\svgscale\undefined%
  \makeatother%
  \begin{picture}(1,0.575)%
    \put(0,0){\includegraphics[width=\unitlength,page=1]{fig1-c-n2c-abs-beta-late.pdf}}%
    \put(0.06803971,0.51324425){\color[rgb]{0,0,0}\makebox(0,0)[b]{\smash{$\tss{\frac{|\langle\mathcal{O}(t)\rangle - \langle\mathcal{O(\infty)}\rangle|}{N\sqrt{2 \pi \,T}}}$}}}%
    \put(0.52776136,0.01794367){\color[rgb]{0,0,0}\makebox(0,0)[b]{\smash{$\tss{2 \pi T \,t}$}}}%
    \put(0.35335925,0.07476574){\color[rgb]{0,0,0}\makebox(0,0)[b]{\smash{$\tss{100}$}}}%
    \put(0.63888845,0.06798113){\color[rgb]{0,0,0}\makebox(0,0)[b]{\smash{$\tss{200}$}}}%
    \put(0.92782125,0.08494264){\color[rgb]{0,0,0}\makebox(0,0)[b]{\smash{$\tss{300}$}}}%
    \put(0.05060378,0.20742996){\color[rgb]{0,0,0}\makebox(0,0)[rb]{\smash{$\tss{10^{-6}}$}}}%
    \put(0.05779994,0.29964175){\color[rgb]{0,0,0}\makebox(0,0)[rb]{\smash{$\tss{10^{-4}}$}}}%
    \put(0.05420187,0.38506651){\color[rgb]{0,0,0}\makebox(0,0)[rb]{\smash{$\tss{10^{-2}}$}}}%
  \end{picture}%
\endgroup%

%% file: fig2-a-n2c-re-and-im-linear.pdf_tex
%% Creator: Inkscape inkscape 0.91, www.inkscape.org
%% Accompanies image file 'fig2-a-n2c-re-and-im-linear.pdf' (pdf, eps, ps)
\begingroup%
  \makeatletter%
  \providecommand\color[2][]{%
    \errmessage{(Inkscape) Color is used for the text in Inkscape, but the package 'color.sty' is not loaded}%
    \renewcommand\color[2][]{}%
  }%
  \providecommand\transparent[1]{%
    \errmessage{(Inkscape) Transparency is used (non-zero) for the text in Inkscape, but the package 'transparent.sty' is not loaded}%
    \renewcommand\transparent[1]{}%
  }%
  \providecommand\rotatebox[2]{#2}%
  \ifx\svgwidth\undefined%
    \setlength{\unitlength}{352bp}%
    \ifx\svgscale\undefined%
      \relax%
    \else%
      \setlength{\unitlength}{\unitlength * \real{\svgscale}}%
    \fi%
  \else%
    \setlength{\unitlength}{\svgwidth}%
  \fi%
  \global\let\svgwidth\undefined%
  \global\let\svgscale\undefined%
  \makeatother%
  \begin{picture}(1,0.45454545)%
    \put(-0.39440909,1.11635059){\color[rgb]{0,0,0}\makebox(0,0)[lt]{\begin{minipage}{1.70549349\unitlength}\raggedleft \end{minipage}}}%
    \put(0,0){\includegraphics[width=\unitlength,page=1]{fig2-a-n2c-re-and-im-linear.pdf}}%
    \put(0.09523166,0.31415054){\color[rgb]{0,0,0}\makebox(0,0)[rb]{\smash{$\tss{0}$}}}%
    \put(0.09523166,0.39033918){\color[rgb]{0,0,0}\makebox(0,0)[rb]{\smash{$\tss{0.05}$}}}%
    \put(0.09523166,0.23796944){\color[rgb]{0,0,0}\makebox(0,0)[rb]{\smash{$\tss{-0.05}$}}}%
    \put(0.09523166,0.16145187){\color[rgb]{0,0,0}\makebox(0,0)[rb]{\smash{$\tss{-0.1}$}}}%
    \put(0.12654584,0.05538767){\color[rgb]{0,0,0}\makebox(0,0)[b]{\smash{$\tss{100}$}}}%
    \put(0.29233149,0.05538767){\color[rgb]{0,0,0}\makebox(0,0)[b]{\smash{$\tss{150}$}}}%
    \put(0.45882762,0.05538767){\color[rgb]{0,0,0}\makebox(0,0)[b]{\smash{$\tss{200}$}}}%
    \put(0.62342284,0.05538767){\color[rgb]{0,0,0}\makebox(0,0)[b]{\smash{$\tss{250}$}}}%
    \put(0.78943753,0.05538767){\color[rgb]{0,0,0}\makebox(0,0)[b]{\smash{$\tss{300}$}}}%
    \put(0.95299966,0.05538767){\color[rgb]{0,0,0}\makebox(0,0)[b]{\smash{$\tss{350}$}}}%
    \put(0.96312297,0.13610255){\color[rgb]{0.24705882,0.23921569,0.6}\makebox(0,0)[rb]{\smash{$\tss{\Re\langle\mathcal{O}\rangle/N\sqrt{2\pi T}}$}}}%
    \put(0.96312297,0.34829859){\color[rgb]{0.6,0.23921569,0.44313725}\makebox(0,0)[rb]{\smash{$\tss{\Im\langle\mathcal{O}\rangle/N\sqrt{2\pi T}}$}}}%
    \put(0.54080582,0.01177042){\color[rgb]{0,0,0}\makebox(0,0)[b]{\smash{$\tss{2 \pi T \,t}$}}}%
  \end{picture}%
\endgroup%

%% file: fig2-b-n2c-re-log.pdf_tex
%% Creator: Inkscape inkscape 0.91, www.inkscape.org
%% Accompanies image file 'fig2-b-n2c-re-log.pdf' (pdf, eps, ps)
\begingroup%
  \makeatletter%
  \providecommand\color[2][]{%
    \errmessage{(Inkscape) Color is used for the text in Inkscape, but the package 'color.sty' is not loaded}%
    \renewcommand\color[2][]{}%
  }%
  \providecommand\transparent[1]{%
    \errmessage{(Inkscape) Transparency is used (non-zero) for the text in Inkscape, but the package 'transparent.sty' is not loaded}%
    \renewcommand\transparent[1]{}%
  }%
  \providecommand\rotatebox[2]{#2}%
  \ifx\svgwidth\undefined%
    \setlength{\unitlength}{352bp}%
    \ifx\svgscale\undefined%
      \relax%
    \else%
      \setlength{\unitlength}{\unitlength * \real{\svgscale}}%
    \fi%
  \else%
    \setlength{\unitlength}{\svgwidth}%
  \fi%
  \global\let\svgwidth\undefined%
  \global\let\svgscale\undefined%
  \makeatother%
  \begin{picture}(1,0.45454545)%
    \put(-0.39440909,1.11635059){\color[rgb]{0,0,0}\makebox(0,0)[lt]{\begin{minipage}{1.70549349\unitlength}\raggedleft \end{minipage}}}%
    \put(0,0){\includegraphics[width=\unitlength,page=1]{fig2-b-n2c-re-log.pdf}}%
    \put(0.12205682,0.37286502){\color[rgb]{0,0,0}\makebox(0,0)[rb]{\smash{$\tss{1}$}}}%
    \put(0.12205682,0.29548884){\color[rgb]{0,0,0}\makebox(0,0)[rb]{\smash{$\tss{10^{-10}}$}}}%
    \put(0.12205682,0.21730268){\color[rgb]{0,0,0}\makebox(0,0)[rb]{\smash{$\tss{10^{-20}}$}}}%
    \put(0.12205682,0.13930397){\color[rgb]{0,0,0}\makebox(0,0)[rb]{\smash{$\tss{10^{-30}}$}}}%
    \put(0.21442187,0.05538767){\color[rgb]{0,0,0}\makebox(0,0)[b]{\smash{$\tss{50}$}}}%
    \put(0.3795414,0.05538767){\color[rgb]{0,0,0}\makebox(0,0)[b]{\smash{$\tss{100}$}}}%
    \put(0.54519039,0.05538767){\color[rgb]{0,0,0}\makebox(0,0)[b]{\smash{$\tss{150}$}}}%
    \put(0.70955322,0.05538767){\color[rgb]{0,0,0}\makebox(0,0)[b]{\smash{$\tss{200}$}}}%
    \put(0.8753842,0.05538767){\color[rgb]{0,0,0}\makebox(0,0)[b]{\smash{$\tss{250}$}}}%
    \put(0.06316376,0.42978805){\color[rgb]{0,0,0}\makebox(0,0)[b]{\smash{$\tss{\frac{\left|\Re\langle\mathcal{O}\rangle\right|}{N\sqrt{2\pi T}}}$}}}%
    \put(0.56155482,0.01177042){\color[rgb]{0,0,0}\makebox(0,0)[b]{\smash{$\tss{2 \pi T \,t}$}}}%
  \end{picture}%
\endgroup%

%% file: fig3-n2c-Delta_t.pdf_tex
%% Creator: Inkscape inkscape 0.48.4, www.inkscape.org
%% Accompanies image file 'fig3-n2c-Delta_t.pdf' (pdf, eps, ps)
\begingroup%
  \makeatletter%
  \providecommand\color[2][]{%
    \errmessage{(Inkscape) Color is used for the text in Inkscape, but the package 'color.sty' is not loaded}%
    \renewcommand\color[2][]{}%
  }%
  \providecommand\transparent[1]{%
    \errmessage{(Inkscape) Transparency is used (non-zero) for the text in Inkscape, but the package 'transparent.sty' is not loaded}%
    \renewcommand\transparent[1]{}%
  }%
  \providecommand\rotatebox[2]{#2}%
  \ifx\svgwidth\undefined%
    \setlength{\unitlength}{800bp}%
    \ifx\svgscale\undefined%
      \relax%
    \else%
      \setlength{\unitlength}{\unitlength * \real{\svgscale}}%
    \fi%
  \else%
    \setlength{\unitlength}{\svgwidth}%
  \fi%
  \global\let\svgwidth\undefined%
  \global\let\svgscale\undefined%
  \makeatother%
  \begin{picture}(1,0.25)%
    \put(0,0){\includegraphics[width=\unitlength]{fig3-n2c-Delta_t.pdf}}%
    \put(0.05826973,0.24183472){\color[rgb]{0,0,0}\makebox(0,0)[rb]{\smash{$\tss{1.2}$}}}%
    \put(0.05826973,0.21064477){\color[rgb]{0,0,0}\makebox(0,0)[rb]{\smash{$\tss{1.0}$}}}%
    \put(0.05826973,0.17891877){\color[rgb]{0,0,0}\makebox(0,0)[rb]{\smash{$\tss{0.8}$}}}%
    \put(0.05826973,0.14826044){\color[rgb]{0,0,0}\makebox(0,0)[rb]{\smash{$\tss{0.6}$}}}%
    \put(0.05826973,0.11683949){\color[rgb]{0,0,0}\makebox(0,0)[rb]{\smash{$\tss{0.4}$}}}%
    \put(0.05826973,0.08694382){\color[rgb]{0,0,0}\makebox(0,0)[rb]{\smash{$\tss{0.2}$}}}%
    \put(0.05826973,0.04180609){\color[rgb]{0,0,0}\makebox(0,0)[rb]{\smash{$\tss{0}$}}}%
    \put(0.17802217,0.04180609){\color[rgb]{0,0,0}\makebox(0,0)[b]{\smash{$\tss{50}$}}}%
    \put(0.33740453,0.04180609){\color[rgb]{0,0,0}\makebox(0,0)[b]{\smash{$\tss{100}$}}}%
    \put(0.49748293,0.04180609){\color[rgb]{0,0,0}\makebox(0,0)[b]{\smash{$\tss{150}$}}}%
    \put(0.65733825,0.04180609){\color[rgb]{0,0,0}\makebox(0,0)[b]{\smash{$\tss{200}$}}}%
    \put(0.81759428,0.04180609){\color[rgb]{0,0,0}\makebox(0,0)[b]{\smash{$\tss{250}$}}}%
    \put(0.9776157,0.04180609){\color[rgb]{0,0,0}\makebox(0,0)[b]{\smash{$\tss{300}$}}}%
    \put(0.52923516,0.00788888){\color[rgb]{0,0,0}\makebox(0,0)[b]{\smash{$\tss{2 \pi T \,t}$}}}%
    \put(0.0091163,0.1590042){\color[rgb]{0,0,0}\rotatebox{90}{\makebox(0,0)[b]{\smash{$\tss{\Delta_t / 2 \pi T}$}}}}%
  \end{picture}%
\endgroup%

%% file: fig4-a-c2n-kappa.pdf_tex
%% Creator: Inkscape inkscape 0.48.4, www.inkscape.org
%% Accompanies image file 'fig4-a-c2n-kappa.pdf' (pdf, eps, ps)
\begingroup%
  \makeatletter%
  \providecommand\color[2][]{%
    \errmessage{(Inkscape) Color is used for the text in Inkscape, but the package 'color.sty' is not loaded}%
    \renewcommand\color[2][]{}%
  }%
  \providecommand\transparent[1]{%
    \errmessage{(Inkscape) Transparency is used (non-zero) for the text in Inkscape, but the package 'transparent.sty' is not loaded}%
    \renewcommand\transparent[1]{}%
  }%
  \providecommand\rotatebox[2]{#2}%
  \ifx\svgwidth\undefined%
    \setlength{\unitlength}{368bp}%
    \ifx\svgscale\undefined%
      \relax%
    \else%
      \setlength{\unitlength}{\unitlength * \real{\svgscale}}%
    \fi%
  \else%
    \setlength{\unitlength}{\svgwidth}%
  \fi%
  \global\let\svgwidth\undefined%
  \global\let\svgscale\undefined%
  \makeatother%
  \begin{picture}(1,0.52173913)%
    \put(0,0){\includegraphics[width=\unitlength]{fig4-a-c2n-kappa.pdf}}%
    \put(0.48818733,0.01475517){\color[rgb]{0,0,0}\makebox(0,0)[b]{\smash{$\tss{2 \pi T \,t}$}}}%
    \put(0.81666723,0.06401779){\color[rgb]{0,0,0}\makebox(0,0)[b]{\smash{$\tss{8000}$}}}%
    \put(0.63132767,0.06401779){\color[rgb]{0,0,0}\makebox(0,0)[b]{\smash{$\tss{6000}$}}}%
    \put(0.44429427,0.06401779){\color[rgb]{0,0,0}\makebox(0,0)[b]{\smash{$\tss{4000}$}}}%
    \put(0.25822561,0.06401779){\color[rgb]{0,0,0}\makebox(0,0)[b]{\smash{$\tss{2000}$}}}%
    \put(0.05662066,0.06401779){\color[rgb]{0,0,0}\makebox(0,0)[rb]{\smash{$\tss{0}$}}}%
    \put(0.05662066,0.14771725){\color[rgb]{0,0,0}\makebox(0,0)[rb]{\smash{$\tss{8}$}}}%
    \put(0.05662066,0.24533736){\color[rgb]{0,0,0}\makebox(0,0)[rb]{\smash{$\tss{9}$}}}%
    \put(0.05662066,0.34221495){\color[rgb]{0,0,0}\makebox(0,0)[rb]{\smash{$\tss{10}$}}}%
    \put(0.05662066,0.43642049){\color[rgb]{0,0,0}\makebox(0,0)[rb]{\smash{$\tss{11}$}}}%
    \put(0.07387651,0.48621091){\color[rgb]{0,0,0}\makebox(0,0)[b]{\smash{$\tss{\kappa_1^{cn}}$}}}%
  \end{picture}%
\endgroup%

%% file: fig4-b-c2n-abs-beta.pdf_tex
%% Creator: Inkscape inkscape 0.91, www.inkscape.org
%% Accompanies image file 'fig4-b-c2n-abs-beta.pdf' (pdf, eps, ps)
\begingroup%
  \makeatletter%
  \providecommand\color[2][]{%
    \errmessage{(Inkscape) Color is used for the text in Inkscape, but the package 'color.sty' is not loaded}%
    \renewcommand\color[2][]{}%
  }%
  \providecommand\transparent[1]{%
    \errmessage{(Inkscape) Transparency is used (non-zero) for the text in Inkscape, but the package 'transparent.sty' is not loaded}%
    \renewcommand\transparent[1]{}%
  }%
  \providecommand\rotatebox[2]{#2}%
  \ifx\svgwidth\undefined%
    \setlength{\unitlength}{368bp}%
    \ifx\svgscale\undefined%
      \relax%
    \else%
      \setlength{\unitlength}{\unitlength * \real{\svgscale}}%
    \fi%
  \else%
    \setlength{\unitlength}{\svgwidth}%
  \fi%
  \global\let\svgwidth\undefined%
  \global\let\svgscale\undefined%
  \makeatother%
  \begin{picture}(1,0.52173913)%
    \put(1.18104917,-0.07223208){\color[rgb]{0,0,0}\makebox(0,0)[lt]{\begin{minipage}{1.30114\unitlength}\raggedleft \end{minipage}}}%
    \put(0,0){\includegraphics[width=\unitlength,page=1]{fig4-b-c2n-abs-beta.pdf}}%
    \put(0.88692557,0.06401779){\color[rgb]{0,0,0}\makebox(0,0)[b]{\smash{$\tss{8000}$}}}%
    \put(0.69819508,0.06401779){\color[rgb]{0,0,0}\makebox(0,0)[b]{\smash{$\tss{6000}$}}}%
    \put(0.50890085,0.06401779){\color[rgb]{0,0,0}\makebox(0,0)[b]{\smash{$\tss{4000}$}}}%
    \put(0.32170184,0.06401779){\color[rgb]{0,0,0}\makebox(0,0)[b]{\smash{$\tss{2000}$}}}%
    \put(0.12009691,0.06401779){\color[rgb]{0,0,0}\makebox(0,0)[rb]{\smash{$\tss{0}$}}}%
    \put(0.12009691,0.18693267){\color[rgb]{0,0,0}\makebox(0,0)[rb]{\smash{$\tss{10^{-9}}$}}}%
    \put(0.12009691,0.26718635){\color[rgb]{0,0,0}\makebox(0,0)[rb]{\smash{$\tss{10^{-7}}$}}}%
    \put(0.12009691,0.34659225){\color[rgb]{0,0,0}\makebox(0,0)[rb]{\smash{$\tss{10^{-5}}$}}}%
    \put(0.12009691,0.42854144){\color[rgb]{0,0,0}\makebox(0,0)[rb]{\smash{$\tss{10^{-3}}$}}}%
    \put(0.55213995,0.01475517){\color[rgb]{0,0,0}\makebox(0,0)[b]{\smash{$\tss{2 \pi T \,t}$}}}%
    \put(-0.61289527,0.65734865){\color[rgb]{0,0,0}\makebox(0,0)[lt]{\begin{minipage}{3.57486489\unitlength}\raggedleft \end{minipage}}}%
    \put(0.12763542,0.48621091){\color[rgb]{0,0,0}\makebox(0,0)[b]{\smash{$\tss{\left|\langle\mathcal{O}\rangle\right|/N\sqrt{2\pi T}}$}}}%
  \end{picture}%
\endgroup%

%% file: fig5-a-c2n-re-and-im-linear.pdf_tex
%% Creator: Inkscape inkscape 0.91, www.inkscape.org
%% Accompanies image file 'fig5-a-c2n-re-and-im-linear.pdf' (pdf, eps, ps)
\begingroup%
  \makeatletter%
  \providecommand\color[2][]{%
    \errmessage{(Inkscape) Color is used for the text in Inkscape, but the package 'color.sty' is not loaded}%
    \renewcommand\color[2][]{}%
  }%
  \providecommand\transparent[1]{%
    \errmessage{(Inkscape) Transparency is used (non-zero) for the text in Inkscape, but the package 'transparent.sty' is not loaded}%
    \renewcommand\transparent[1]{}%
  }%
  \providecommand\rotatebox[2]{#2}%
  \ifx\svgwidth\undefined%
    \setlength{\unitlength}{360bp}%
    \ifx\svgscale\undefined%
      \relax%
    \else%
      \setlength{\unitlength}{\unitlength * \real{\svgscale}}%
    \fi%
  \else%
    \setlength{\unitlength}{\svgwidth}%
  \fi%
  \global\let\svgwidth\undefined%
  \global\let\svgscale\undefined%
  \makeatother%
  \begin{picture}(1,0.44444444)%
    \put(0,0){\includegraphics[width=\unitlength,page=1]{fig5-a-c2n-re-and-im-linear.pdf}}%
    \put(0.14163128,0.1763857){\color[rgb]{0,0,0}\makebox(0,0)[rb]{\smash{$\tss{0}$}}}%
    \put(0.14163128,0.24440546){\color[rgb]{0,0,0}\makebox(0,0)[rb]{\smash{$\tss{0.0005}$}}}%
    \put(0.14163128,0.31242524){\color[rgb]{0,0,0}\makebox(0,0)[rb]{\smash{$\tss{0.001}$}}}%
    \put(0.14163128,0.380445){\color[rgb]{0,0,0}\makebox(0,0)[rb]{\smash{$\tss{0.0015}$}}}%
    \put(0.14163128,0.10836588){\color[rgb]{0,0,0}\makebox(0,0)[rb]{\smash{$\tss{-0.0005}$}}}%
    \put(0.14150663,0.06288996){\color[rgb]{0,0,0}\makebox(0,0)[rb]{\smash{$\tss{0}$}}}%
    \put(0.43687961,0.06288996){\color[rgb]{0,0,0}\makebox(0,0)[rb]{\smash{$\tss{1000}$}}}%
    \put(0.7074638,0.06288996){\color[rgb]{0,0,0}\makebox(0,0)[rb]{\smash{$\tss{2000}$}}}%
    \put(0.9660885,0.06288996){\color[rgb]{0,0,0}\makebox(0,0)[rb]{\smash{$\tss{3000}$}}}%
    \put(0.57172382,0.01532125){\color[rgb]{0,0,0}\makebox(0,0)[b]{\smash{$\tss{2 \pi T \,t}$}}}%
    \put(0.30118979,0.38214811){\color[rgb]{0.24705882,0.23921569,0.6}\makebox(0,0)[lb]{\smash{$\tss{\Re\langle\mathcal{O}\rangle/N\sqrt{2\pi T}}$}}}%
    \put(0.35226988,0.29258853){\color[rgb]{0.6,0.23921569,0.44313725}\makebox(0,0)[lb]{\smash{$\tss{\Im\langle\mathcal{O}\rangle/N\sqrt{2\pi T}}$}}}%
  \end{picture}%
\endgroup%

%% file: fig5-b-c2n-log.pdf_tex
%% Creator: Inkscape inkscape 0.91, www.inkscape.org
%% Accompanies image file 'fig5-b-c2n-log.pdf' (pdf, eps, ps)
\begingroup%
  \makeatletter%
  \providecommand\color[2][]{%
    \errmessage{(Inkscape) Color is used for the text in Inkscape, but the package 'color.sty' is not loaded}%
    \renewcommand\color[2][]{}%
  }%
  \providecommand\transparent[1]{%
    \errmessage{(Inkscape) Transparency is used (non-zero) for the text in Inkscape, but the package 'transparent.sty' is not loaded}%
    \renewcommand\transparent[1]{}%
  }%
  \providecommand\rotatebox[2]{#2}%
  \ifx\svgwidth\undefined%
    \setlength{\unitlength}{360bp}%
    \ifx\svgscale\undefined%
      \relax%
    \else%
      \setlength{\unitlength}{\unitlength * \real{\svgscale}}%
    \fi%
  \else%
    \setlength{\unitlength}{\svgwidth}%
  \fi%
  \global\let\svgwidth\undefined%
  \global\let\svgscale\undefined%
  \makeatother%
  \begin{picture}(1,0.44444444)%
    \put(0,0){\includegraphics[width=\unitlength,page=1]{fig5-b-c2n-log.pdf}}%
    \put(0.1523827,0.06288996){\color[rgb]{0,0,0}\makebox(0,0)[rb]{\smash{$\tss{0}$}}}%
    \put(0.37859823,0.06288996){\color[rgb]{0,0,0}\makebox(0,0)[rb]{\smash{$\tss{2000}$}}}%
    \put(0.57838681,0.06288996){\color[rgb]{0,0,0}\makebox(0,0)[rb]{\smash{$\tss{4000}$}}}%
    \put(0.7760612,0.06288996){\color[rgb]{0,0,0}\makebox(0,0)[rb]{\smash{$\tss{6000}$}}}%
    \put(0.9726785,0.06288996){\color[rgb]{0,0,0}\makebox(0,0)[rb]{\smash{$\tss{8000}$}}}%
    \put(0.1523827,0.36278742){\color[rgb]{0,0,0}\makebox(0,0)[rb]{\smash{$\tss{10^{-10}}$}}}%
    \put(0.1523827,0.24069439){\color[rgb]{0,0,0}\makebox(0,0)[rb]{\smash{$\tss{10^{-20}}$}}}%
    \put(0.1523827,0.11860135){\color[rgb]{0,0,0}\makebox(0,0)[rb]{\smash{$\tss{10^{-30}}$}}}%
    \put(0.57731208,0.01532125){\color[rgb]{0,0,0}\makebox(0,0)[b]{\smash{$\tss{2 \pi T \,t}$}}}%
    \put(0.0225683,0.26838809){\color[rgb]{0,0,0}\rotatebox{90}{\makebox(0,0)[b]{\smash{$\tss{\left|\Re\langle\mathcal{O}\rangle\right|/N\sqrt{2\pi T}}$}}}}%
  \end{picture}%
\endgroup%

%% file: fig6-c2n-Delta_t.pdf_tex
%% Creator: Inkscape inkscape 0.48.4, www.inkscape.org
%% Accompanies image file 'fig6-c2n-Delta_t.pdf' (pdf, eps, ps)
\begingroup%
  \makeatletter%
  \providecommand\color[2][]{%
    \errmessage{(Inkscape) Color is used for the text in Inkscape, but the package 'color.sty' is not loaded}%
    \renewcommand\color[2][]{}%
  }%
  \providecommand\transparent[1]{%
    \errmessage{(Inkscape) Transparency is used (non-zero) for the text in Inkscape, but the package 'transparent.sty' is not loaded}%
    \renewcommand\transparent[1]{}%
  }%
  \providecommand\rotatebox[2]{#2}%
  \ifx\svgwidth\undefined%
    \setlength{\unitlength}{800bp}%
    \ifx\svgscale\undefined%
      \relax%
    \else%
      \setlength{\unitlength}{\unitlength * \real{\svgscale}}%
    \fi%
  \else%
    \setlength{\unitlength}{\svgwidth}%
  \fi%
  \global\let\svgwidth\undefined%
  \global\let\svgscale\undefined%
  \makeatother%
  \begin{picture}(1,0.25)%
    \put(0,0){\includegraphics[width=\unitlength]{fig6-c2n-Delta_t.pdf}}%
    \put(0.01345812,0.14722569){\color[rgb]{0,0,0}\rotatebox{90}{\makebox(0,0)[b]{\smash{$\tss{\Delta_t / 2 \pi T}$}}}}%
    \put(0.53283913,0.01099079){\color[rgb]{0,0,0}\makebox(0,0)[b]{\smash{$\tss{2 \pi T \,t}$}}}%
    \put(0.06991135,0.2121522){\color[rgb]{0,0,0}\makebox(0,0)[rb]{\smash{$\tss{0.5}$}}}%
    \put(0.06991135,0.18101383){\color[rgb]{0,0,0}\makebox(0,0)[rb]{\smash{$\tss{0.498}$}}}%
    \put(0.06991135,0.14987549){\color[rgb]{0,0,0}\makebox(0,0)[rb]{\smash{$\tss{0.496}$}}}%
    \put(0.06991135,0.11873714){\color[rgb]{0,0,0}\makebox(0,0)[rb]{\smash{$\tss{0.494}$}}}%
    \put(0.06991135,0.08759878){\color[rgb]{0,0,0}\makebox(0,0)[rb]{\smash{$\tss{0.492}$}}}%
    \put(0.06991135,0.05646042){\color[rgb]{0,0,0}\makebox(0,0)[rb]{\smash{$\tss{0.49}$}}}%
    \put(0.09183125,0.03648239){\color[rgb]{0,0,0}\makebox(0,0)[b]{\smash{$\tss{0}$}}}%
    \put(0.31383934,0.03648239){\color[rgb]{0,0,0}\makebox(0,0)[b]{\smash{$\tss{2000}$}}}%
    \put(0.53248569,0.03648239){\color[rgb]{0,0,0}\makebox(0,0)[b]{\smash{$\tss{4000}$}}}%
    \put(0.75385841,0.03648239){\color[rgb]{0,0,0}\makebox(0,0)[b]{\smash{$\tss{6000}$}}}%
    \put(0.97210869,0.03648239){\color[rgb]{0,0,0}\makebox(0,0)[b]{\smash{$\tss{8000}$}}}%
  \end{picture}%
\endgroup%

%% file: fig7-a-contour-overview.pdf_tex
%% Creator: Inkscape inkscape 0.48.4, www.inkscape.org
%% Accompanies image file 'fig7-a-contour-overview.pdf' (pdf, eps, ps)
\begingroup%
  \makeatletter%
  \providecommand\color[2][]{%
    \errmessage{(Inkscape) Color is used for the text in Inkscape, but the package 'color.sty' is not loaded}%
    \renewcommand\color[2][]{}%
  }%
  \providecommand\transparent[1]{%
    \errmessage{(Inkscape) Transparency is used (non-zero) for the text in Inkscape, but the package 'transparent.sty' is not loaded}%
    \renewcommand\transparent[1]{}%
  }%
  \providecommand\rotatebox[2]{#2}%
  \ifx\svgwidth\undefined%
    \setlength{\unitlength}{288bp}%
    \ifx\svgscale\undefined%
      \relax%
    \else%
      \setlength{\unitlength}{\unitlength * \real{\svgscale}}%
    \fi%
  \else%
    \setlength{\unitlength}{\svgwidth}%
  \fi%
  \global\let\svgwidth\undefined%
  \global\let\svgscale\undefined%
  \makeatother%
  \begin{picture}(1,2.22222222)%
    \put(0,0){\includegraphics[width=\unitlength]{fig7-a-contour-overview.pdf}}%
    \put(0.89358213,2.100317){\color[rgb]{0,0,0}\makebox(0,0)[b]{\smash{$\frac{\omega}{2\,\pi\,T}$}}}%
    \put(0.02790312,2.19122399){\color[rgb]{0,0,0}\makebox(0,0)[rb]{\smash{$\tss{1}$}}}%
    \put(0.02790312,1.72407712){\color[rgb]{0,0,0}\makebox(0,0)[rb]{\smash{$\tss{0}$}}}%
    \put(0.02790312,1.26114093){\color[rgb]{0,0,0}\makebox(0,0)[rb]{\smash{$\tss{-1}$}}}%
    \put(0.02790312,0.79236333){\color[rgb]{0,0,0}\makebox(0,0)[rb]{\smash{$\tss{-2}$}}}%
    \put(0.02790312,0.33380824){\color[rgb]{0,0,0}\makebox(0,0)[rb]{\smash{$\tss{-3}$}}}%
    \put(0.05384943,0.27295926){\color[rgb]{0,0,0}\makebox(0,0)[b]{\smash{$\tss{-1}$}}}%
    \put(0.28829058,0.27295926){\color[rgb]{0,0,0}\makebox(0,0)[b]{\smash{$\tss{-0.5}$}}}%
    \put(0.52166605,0.27295926){\color[rgb]{0,0,0}\makebox(0,0)[b]{\smash{$\tss{0}$}}}%
    \put(0.75504155,0.27295926){\color[rgb]{0,0,0}\makebox(0,0)[b]{\smash{$\tss{0.5}$}}}%
    \put(0.98565176,0.27295926){\color[rgb]{0,0,0}\makebox(0,0)[b]{\smash{$\tss{1}$}}}%
    \put(0.98729653,0.08850535){\color[rgb]{0,0,0}\makebox(0,0)[b]{\smash{$\tss{2}$}}}%
    \put(0.0528658,0.08850535){\color[rgb]{0,0,0}\makebox(0,0)[b]{\smash{$\tss{0}$}}}%
    \put(0.28647349,0.08850535){\color[rgb]{0,0,0}\makebox(0,0)[b]{\smash{$\tss{0.14}$}}}%
    \put(0.52008115,0.08850535){\color[rgb]{0,0,0}\makebox(0,0)[b]{\smash{$\tss{0.35}$}}}%
    \put(0.75368887,0.08850535){\color[rgb]{0,0,0}\makebox(0,0)[b]{\smash{$\tss{0.89}$}}}%
    \put(0.02790312,0.16223684){\color[rgb]{0,0,0}\makebox(0,0)[rb]{\smash{$\tss{|\kappa|}$}}}%
  \end{picture}%
\endgroup%

%% file: fig7-b-contour-detail.pdf_tex
%% Creator: Inkscape inkscape 0.48.4, www.inkscape.org
%% Accompanies image file 'fig7-b-contour-detail.pdf' (pdf, eps, ps)
\begingroup%
  \makeatletter%
  \providecommand\color[2][]{%
    \errmessage{(Inkscape) Color is used for the text in Inkscape, but the package 'color.sty' is not loaded}%
    \renewcommand\color[2][]{}%
  }%
  \providecommand\transparent[1]{%
    \errmessage{(Inkscape) Transparency is used (non-zero) for the text in Inkscape, but the package 'transparent.sty' is not loaded}%
    \renewcommand\transparent[1]{}%
  }%
  \providecommand\rotatebox[2]{#2}%
  \ifx\svgwidth\undefined%
    \setlength{\unitlength}{568bp}%
    \ifx\svgscale\undefined%
      \relax%
    \else%
      \setlength{\unitlength}{\unitlength * \real{\svgscale}}%
    \fi%
  \else%
    \setlength{\unitlength}{\svgwidth}%
  \fi%
  \global\let\svgwidth\undefined%
  \global\let\svgscale\undefined%
  \makeatother%
  \begin{picture}(1,1.12676056)%
    \put(0,0){\includegraphics[width=\unitlength]{fig7-b-contour-detail.pdf}}%
    \put(0.03143928,1.03964032){\color[rgb]{0,0,0}\makebox(0,0)[rb]{\smash{$\tss{0.1}$}}}%
    \put(0.03143928,0.74715874){\color[rgb]{0,0,0}\makebox(0,0)[rb]{\smash{$\tss{0.05}$}}}%
    \put(0.03143928,0.45211395){\color[rgb]{0,0,0}\makebox(0,0)[rb]{\smash{$\tss{0}$}}}%
    \put(0.03143928,0.15830266){\color[rgb]{0,0,0}\makebox(0,0)[rb]{\smash{$\tss{-0.05}$}}}%
    \put(0.04413338,0.13190401){\color[rgb]{0,0,0}\makebox(0,0)[b]{\smash{$\tss{-0.05}$}}}%
    \put(0.33930057,0.13190401){\color[rgb]{0,0,0}\makebox(0,0)[b]{\smash{$\tss{0}$}}}%
    \put(0.62974742,0.13190401){\color[rgb]{0,0,0}\makebox(0,0)[b]{\smash{$\tss{0.05}$}}}%
    \put(0.92619132,0.13190401){\color[rgb]{0,0,0}\makebox(0,0)[b]{\smash{$\tss{0.1}$}}}%
    \put(0.92535939,1.0447374){\color[rgb]{1,1,1}\makebox(0,0)[b]{\smash{$\frac{\omega}{2\,\pi\,T}$}}}%
    \put(0.0314782,0.08226035){\color[rgb]{0,0,0}\makebox(0,0)[rb]{\smash{$\tss{|\kappa|}$}}}%
    \put(0.9798988,0.04487624){\color[rgb]{0,0,0}\makebox(0,0)[b]{\smash{$\tss{100}$}}}%
    \put(0.15885944,0.04487624){\color[rgb]{0,0,0}\makebox(0,0)[b]{\smash{$\tss{3.6}$}}}%
    \put(0.27695629,0.04487624){\color[rgb]{0,0,0}\makebox(0,0)[b]{\smash{$\tss{5.7}$}}}%
    \put(0.51019395,0.04487624){\color[rgb]{0,0,0}\makebox(0,0)[b]{\smash{$\tss{14.2}$}}}%
    \put(0.74382347,0.04487624){\color[rgb]{0,0,0}\makebox(0,0)[b]{\smash{$\tss{35.7}$}}}%
    \put(0.39718746,0.04487624){\color[rgb]{0,0,0}\makebox(0,0)[b]{\smash{$\tss{\kappa_{c}}$}}}%
    \put(0.62535649,0.04487624){\color[rgb]{0,0,0}\makebox(0,0)[b]{\smash{$\tss{22.6}$}}}%
    \put(0.87183536,0.04487624){\color[rgb]{0,0,0}\makebox(0,0)[b]{\smash{$\tss{56.7}$}}}%
  \end{picture}%
\endgroup%

%% file: fig8-QNM-cond.pdf_tex
%% Creator: Inkscape inkscape 0.48.4, www.inkscape.org
%% Accompanies image file 'fig8-QNM-cond.pdf' (pdf, eps, ps)
\begingroup%
  \makeatletter%
  \providecommand\color[2][]{%
    \errmessage{(Inkscape) Color is used for the text in Inkscape, but the package 'color.sty' is not loaded}%
    \renewcommand\color[2][]{}%
  }%
  \providecommand\transparent[1]{%
    \errmessage{(Inkscape) Transparency is used (non-zero) for the text in Inkscape, but the package 'transparent.sty' is not loaded}%
    \renewcommand\transparent[1]{}%
  }%
  \providecommand\rotatebox[2]{#2}%
  \ifx\svgwidth\undefined%
    \setlength{\unitlength}{360bp}%
    \ifx\svgscale\undefined%
      \relax%
    \else%
      \setlength{\unitlength}{\unitlength * \real{\svgscale}}%
    \fi%
  \else%
    \setlength{\unitlength}{\svgwidth}%
  \fi%
  \global\let\svgwidth\undefined%
  \global\let\svgscale\undefined%
  \makeatother%
  \begin{picture}(1,0.625)%
    \put(0,0){\includegraphics[width=\unitlength]{fig8-QNM-cond.pdf}}%
    \put(0.82089971,0.40514904){\color[rgb]{0,0,0}\makebox(0,0)[b]{\smash{$\frac{\omega}{2\,\pi\,T}$}}}%
    \put(0.69210896,0.53081631){\color[rgb]{0,0,0}\makebox(0,0)[b]{\smash{$\tss{0.5}$}}}%
    \put(0.91387363,0.53081631){\color[rgb]{0,0,0}\makebox(0,0)[b]{\smash{$\tss{1.0}$}}}%
    \put(0.2493579,0.53081631){\color[rgb]{0,0,0}\makebox(0,0)[b]{\smash{$\tss{-0.5}$}}}%
    \put(0.0274507,0.53081631){\color[rgb]{0,0,0}\makebox(0,0)[b]{\smash{$\tss{-1.0}$}}}%
    \put(0.45736821,0.43624827){\color[rgb]{0,0,0}\makebox(0,0)[rb]{\smash{$\tss{-0.05}$}}}%
    \put(0.45736821,0.31234228){\color[rgb]{0,0,0}\makebox(0,0)[rb]{\smash{$\tss{-0.1}$}}}%
    \put(0.45736821,0.18843632){\color[rgb]{0,0,0}\makebox(0,0)[rb]{\smash{$\tss{-0.15}$}}}%
    \put(0.45736821,0.06453034){\color[rgb]{0,0,0}\makebox(0,0)[rb]{\smash{$\tss{-0.2}$}}}%
  \end{picture}%
\endgroup%

%% file: fig9-QNM-cond-phase-normalised.pdf_tex
%% Creator: Inkscape inkscape 0.48.4, www.inkscape.org
%% Accompanies image file 'fig9-QNM-cond-phase-normalised.pdf' (pdf, eps, ps)
\begingroup%
  \makeatletter%
  \providecommand\color[2][]{%
    \errmessage{(Inkscape) Color is used for the text in Inkscape, but the package 'color.sty' is not loaded}%
    \renewcommand\color[2][]{}%
  }%
  \providecommand\transparent[1]{%
    \errmessage{(Inkscape) Transparency is used (non-zero) for the text in Inkscape, but the package 'transparent.sty' is not loaded}%
    \renewcommand\transparent[1]{}%
  }%
  \providecommand\rotatebox[2]{#2}%
  \ifx\svgwidth\undefined%
    \setlength{\unitlength}{360bp}%
    \ifx\svgscale\undefined%
      \relax%
    \else%
      \setlength{\unitlength}{\unitlength * \real{\svgscale}}%
    \fi%
  \else%
    \setlength{\unitlength}{\svgwidth}%
  \fi%
  \global\let\svgwidth\undefined%
  \global\let\svgscale\undefined%
  \makeatother%
  \begin{picture}(1,0.63333333)%
    \put(0,0){\includegraphics[width=\unitlength]{fig9-QNM-cond-phase-normalised.pdf}}%
    \put(0.07559659,0.58256412){\color[rgb]{0,0,0}\makebox(0,0)[b]{\smash{$\frac{\omega_I}{2 \pi T_c}$}}}%
    \put(0.96054823,0.5169682){\color[rgb]{0,0,0}\makebox(0,0)[lb]{\smash{$\frac{T}{T_c}$}}}%
    \put(0.25075395,0.55030484){\color[rgb]{0,0,0}\makebox(0,0)[b]{\smash{$\tss{0.2}$}}}%
    \put(0.42148387,0.55030484){\color[rgb]{0,0,0}\makebox(0,0)[b]{\smash{$\tss{0.4}$}}}%
    \put(0.59221382,0.55030484){\color[rgb]{0,0,0}\makebox(0,0)[b]{\smash{$\tss{0.6}$}}}%
    \put(0.76294386,0.55030484){\color[rgb]{0,0,0}\makebox(0,0)[b]{\smash{$\tss{0.8}$}}}%
    \put(0.06104664,0.51426222){\color[rgb]{0,0,0}\makebox(0,0)[rb]{\smash{$\tss{0}$}}}%
    \put(0.06104664,0.40443844){\color[rgb]{0,0,0}\makebox(0,0)[rb]{\smash{$\tss{-0.02}$}}}%
    \put(0.06104664,0.29461469){\color[rgb]{0,0,0}\makebox(0,0)[rb]{\smash{$\tss{-0.04}$}}}%
    \put(0.06104664,0.18479091){\color[rgb]{0,0,0}\makebox(0,0)[rb]{\smash{$\tss{-0.06}$}}}%
    \put(0.06104664,0.07496715){\color[rgb]{0,0,0}\makebox(0,0)[rb]{\smash{$\tss{-0.08}$}}}%
    \put(0.93367377,0.55030484){\color[rgb]{0,0,0}\makebox(0,0)[b]{\smash{$\tss{1.0}$}}}%
  \end{picture}%
\endgroup%

%% file: fig10-omega-vs-alpha-square-linear-NEW.pdf_tex
%% Creator: Inkscape inkscape 0.48.4, www.inkscape.org
%% Accompanies image file 'fig10-omega-vs-alpha-square-linear-NEW.pdf' (pdf, eps, ps)
\begingroup%
  \makeatletter%
  \providecommand\color[2][]{%
    \errmessage{(Inkscape) Color is used for the text in Inkscape, but the package 'color.sty' is not loaded}%
    \renewcommand\color[2][]{}%
  }%
  \providecommand\transparent[1]{%
    \errmessage{(Inkscape) Transparency is used (non-zero) for the text in Inkscape, but the package 'transparent.sty' is not loaded}%
    \renewcommand\transparent[1]{}%
  }%
  \providecommand\rotatebox[2]{#2}%
  \ifx\svgwidth\undefined%
    \setlength{\unitlength}{392bp}%
    \ifx\svgscale\undefined%
      \relax%
    \else%
      \setlength{\unitlength}{\unitlength * \real{\svgscale}}%
    \fi%
  \else%
    \setlength{\unitlength}{\svgwidth}%
  \fi%
  \global\let\svgwidth\undefined%
  \global\let\svgscale\undefined%
  \makeatother%
  \begin{picture}(1,0.51020408)%
    \put(0,0){\includegraphics[width=\unitlength]{fig10-omega-vs-alpha-square-linear-NEW.pdf}}%
    \put(0.15089052,0.48760395){\color[rgb]{0,0,0}\makebox(0,0)[b]{\smash{$\tss{\frac{\omega_I}{2 \pi T}}$}}}%
    \put(0.89532776,0.43917367){\color[rgb]{0,0,0}\makebox(0,0)[lb]{\smash{$\tss{\frac{\kappa_1^{\,2}\,\langle \mathcal{O} \rangle^2}{2 \pi T}}$}}}%
    \put(0.29728666,0.47755876){\color[rgb]{0,0,0}\makebox(0,0)[b]{\smash{$\tss{2\cdot 10^{-4}}$}}}%
    \put(0.44205901,0.47755876){\color[rgb]{0,0,0}\makebox(0,0)[b]{\smash{$\tss{4\cdot 10^{-4}}$}}}%
    \put(0.58683137,0.47755876){\color[rgb]{0,0,0}\makebox(0,0)[b]{\smash{$\tss{6\cdot 10^{-4}}$}}}%
    \put(0.73160366,0.47755876){\color[rgb]{0,0,0}\makebox(0,0)[b]{\smash{$\tss{8\cdot 10^{-4}}$}}}%
    \put(0.87637603,0.47755876){\color[rgb]{0,0,0}\makebox(0,0)[b]{\smash{$\tss{10^{-3}}$}}}%
    \put(0.13694407,0.43984095){\color[rgb]{0,0,0}\makebox(0,0)[rb]{\smash{$\tss{0}$}}}%
    \put(0.13694407,0.31555501){\color[rgb]{0,0,0}\makebox(0,0)[rb]{\smash{$\tss{-5\cdot 10^{-3}}$}}}%
    \put(0.13694407,0.18421884){\color[rgb]{0,0,0}\makebox(0,0)[rb]{\smash{$\tss{-10^{-2}}$}}}%
    \put(0.13694407,0.05288266){\color[rgb]{0,0,0}\makebox(0,0)[rb]{\smash{$\tss{-1.5\cdot 10^{-2}}$}}}%
  \end{picture}%
\endgroup%

%% file: fig10-omega-vs-alpha-square-log-linear.pdf_tex
%% Creator: Inkscape inkscape 0.91, www.inkscape.org
%% Accompanies image file 'fig10-omega-vs-alpha-square-log-linear.pdf' (pdf, eps, ps)
\begingroup%
  \makeatletter%
  \providecommand\color[2][]{%
    \errmessage{(Inkscape) Color is used for the text in Inkscape, but the package 'color.sty' is not loaded}%
    \renewcommand\color[2][]{}%
  }%
  \providecommand\transparent[1]{%
    \errmessage{(Inkscape) Transparency is used (non-zero) for the text in Inkscape, but the package 'transparent.sty' is not loaded}%
    \renewcommand\transparent[1]{}%
  }%
  \providecommand\rotatebox[2]{#2}%
  \ifx\svgwidth\undefined%
    \setlength{\unitlength}{352bp}%
    \ifx\svgscale\undefined%
      \relax%
    \else%
      \setlength{\unitlength}{\unitlength * \real{\svgscale}}%
    \fi%
  \else%
    \setlength{\unitlength}{\svgwidth}%
  \fi%
  \global\let\svgwidth\undefined%
  \global\let\svgscale\undefined%
  \makeatother%
  \begin{picture}(1,0.56818182)%
    \put(0,0){\includegraphics[width=\unitlength,page=1]{fig10-omega-vs-alpha-square-log-linear.pdf}}%
    \put(0.79507776,0.38182692){\color[rgb]{0,0,0}\makebox(0,0)[lb]{\smash{$\tss{-0.2}$}}}%
    \put(0.79507776,0.26992083){\color[rgb]{0,0,0}\makebox(0,0)[lb]{\smash{$\tss{-0.4}$}}}%
    \put(0.79507776,0.15801479){\color[rgb]{0,0,0}\makebox(0,0)[lb]{\smash{$\tss{-0.6}$}}}%
    \put(0.79507776,0.04610872){\color[rgb]{0,0,0}\makebox(0,0)[lb]{\smash{$\tss{-0.8}$}}}%
    \put(0.84897388,0.48913492){\color[rgb]{0,0,0}\makebox(0,0)[lb]{\smash{$\tss{\frac{\kappa_1^{\,2}\,\langle \mathcal{O} \rangle^2}{2 \pi T}}$}}}%
    \put(0.76688246,0.53251895){\color[rgb]{0,0,0}\makebox(0,0)[b]{\smash{$\tss{\frac{\omega_I}{2 \pi T}}$}}}%
    \put(0.07594695,0.52164299){\color[rgb]{0,0,0}\makebox(0,0)[b]{\smash{$\tss{10^{-5}}$}}}%
    \put(0.21415808,0.52164299){\color[rgb]{0,0,0}\makebox(0,0)[b]{\smash{$\tss{10^{-4}}$}}}%
    \put(0.35236924,0.52164299){\color[rgb]{0,0,0}\makebox(0,0)[b]{\smash{$\tss{10^{-3}}$}}}%
    \put(0.49058039,0.52164299){\color[rgb]{0,0,0}\makebox(0,0)[b]{\smash{$\tss{10^{-2}}$}}}%
    \put(0.62879155,0.52164299){\color[rgb]{0,0,0}\makebox(0,0)[b]{\smash{$\tss{10^{-1}}$}}}%
    \put(0,0){\includegraphics[width=\unitlength,page=2]{fig10-omega-vs-alpha-square-log-linear.pdf}}%
  \end{picture}%
\endgroup%

%% file: fig11-a-DOF-at-hor.pdf_tex
%% Creator: Inkscape inkscape 0.91, www.inkscape.org
%% Accompanies image file 'fig11-a-DOF-at-hor.pdf' (pdf, eps, ps)
\begingroup%
  \makeatletter%
  \providecommand\color[2][]{%
    \errmessage{(Inkscape) Color is used for the text in Inkscape, but the package 'color.sty' is not loaded}%
    \renewcommand\color[2][]{}%
  }%
  \providecommand\transparent[1]{%
    \errmessage{(Inkscape) Transparency is used (non-zero) for the text in Inkscape, but the package 'transparent.sty' is not loaded}%
    \renewcommand\transparent[1]{}%
  }%
  \providecommand\rotatebox[2]{#2}%
  \ifx\svgwidth\undefined%
    \setlength{\unitlength}{384bp}%
    \ifx\svgscale\undefined%
      \relax%
    \else%
      \setlength{\unitlength}{\unitlength * \real{\svgscale}}%
    \fi%
  \else%
    \setlength{\unitlength}{\svgwidth}%
  \fi%
  \global\let\svgwidth\undefined%
  \global\let\svgscale\undefined%
  \makeatother%
  \begin{picture}(1,0.5859375)%
    \put(0,0){\includegraphics[width=\unitlength,page=1]{fig11-a-DOF-at-hor.pdf}}%
    \put(0.25483282,0.00478981){\color[rgb]{0,0,0}\makebox(0,0)[b]{\smash{$\tss{20}$}}}%
    \put(0.46494381,0.00478981){\color[rgb]{0,0,0}\makebox(0,0)[b]{\smash{$\tss{40}$}}}%
    \put(0.67505479,0.00478981){\color[rgb]{0,0,0}\makebox(0,0)[b]{\smash{$\tss{60}$}}}%
    \put(0.88516585,0.00478981){\color[rgb]{0,0,0}\makebox(0,0)[b]{\smash{$\tss{80}$}}}%
    \put(0.06593703,0.49288495){\color[rgb]{0,0,0}\makebox(0,0)[rb]{\smash{$\tss{10^{-2}}$}}}%
    \put(0.06593703,0.40379409){\color[rgb]{0,0,0}\makebox(0,0)[rb]{\smash{$\tss{10^{-4}}$}}}%
    \put(0.06593703,0.31470325){\color[rgb]{0,0,0}\makebox(0,0)[rb]{\smash{$\tss{10^{-6}}$}}}%
    \put(0.06593703,0.2256124){\color[rgb]{0,0,0}\makebox(0,0)[rb]{\smash{$\tss{10^{-8}}$}}}%
    \put(0.06593703,0.13652155){\color[rgb]{0,0,0}\makebox(0,0)[rb]{\smash{$\tss{10^{-10}}$}}}%
    \put(0.07730612,0.55410043){\color[rgb]{0,0,0}\makebox(0,0)[b]{\smash{$\tss{D}$}}}%
    \put(0.9258639,0.04573373){\color[rgb]{0,0,0}\makebox(0,0)[lb]{\smash{$\tss{2\pi T\,t}$}}}%
  \end{picture}%
\endgroup%

%% file: fig11-b-DOF-at-hor-2.pdf_tex
%% Creator: Inkscape inkscape 0.91, www.inkscape.org
%% Accompanies image file 'fig11-b-DOF-at-hor-2.pdf' (pdf, eps, ps)
\begingroup%
  \makeatletter%
  \providecommand\color[2][]{%
    \errmessage{(Inkscape) Color is used for the text in Inkscape, but the package 'color.sty' is not loaded}%
    \renewcommand\color[2][]{}%
  }%
  \providecommand\transparent[1]{%
    \errmessage{(Inkscape) Transparency is used (non-zero) for the text in Inkscape, but the package 'transparent.sty' is not loaded}%
    \renewcommand\transparent[1]{}%
  }%
  \providecommand\rotatebox[2]{#2}%
  \ifx\svgwidth\undefined%
    \setlength{\unitlength}{384bp}%
    \ifx\svgscale\undefined%
      \relax%
    \else%
      \setlength{\unitlength}{\unitlength * \real{\svgscale}}%
    \fi%
  \else%
    \setlength{\unitlength}{\svgwidth}%
  \fi%
  \global\let\svgwidth\undefined%
  \global\let\svgscale\undefined%
  \makeatother%
  \begin{picture}(1,0.5859375)%
    \put(0,0){\includegraphics[width=\unitlength,page=1]{fig11-b-DOF-at-hor-2.pdf}}%
    \put(0.18387854,0.00495713){\color[rgb]{0,0,0}\makebox(0,0)[b]{\smash{$\tss{2000}$}}}%
    \put(0.31885757,0.00495713){\color[rgb]{0,0,0}\makebox(0,0)[b]{\smash{$\tss{4000}$}}}%
    \put(0.4538366,0.00495713){\color[rgb]{0,0,0}\makebox(0,0)[b]{\smash{$\tss{6000}$}}}%
    \put(0.58881565,0.00495713){\color[rgb]{0,0,0}\makebox(0,0)[b]{\smash{$\tss{8000}$}}}%
    \put(0.7237947,0.00495713){\color[rgb]{0,0,0}\makebox(0,0)[b]{\smash{$\tss{10000}$}}}%
    \put(0.85877371,0.00495713){\color[rgb]{0,0,0}\makebox(0,0)[b]{\smash{$\tss{12000}$}}}%
    \put(0.0654195,0.42609421){\color[rgb]{0,0,0}\makebox(0,0)[rb]{\smash{$\tss{10^{-5}}$}}}%
    \put(0.0654195,0.27687422){\color[rgb]{0,0,0}\makebox(0,0)[rb]{\smash{$\tss{10^{-8}}$}}}%
    \put(0.0654195,0.12765426){\color[rgb]{0,0,0}\makebox(0,0)[rb]{\smash{$\tss{10^{-11}}$}}}%
    \put(0.07821284,0.55163801){\color[rgb]{0,0,0}\makebox(0,0)[b]{\smash{$\tss{D}$}}}%
    \put(0.91774464,0.0446817){\color[rgb]{0,0,0}\makebox(0,0)[lb]{\smash{$\tss{2\pi T\,t}$}}}%
    \put(-0.17876059,0.49159164){\color[rgb]{0,0,0}\makebox(0,0)[lt]{\begin{minipage}{0.25945112\unitlength}\raggedleft \end{minipage}}}%
  \end{picture}%
\endgroup%

%% file: critical.tex
\subsection{Critical slowing down}
\label{sec::critExp}

In subsections \ref{sec::normalQNM} and \ref{sec::condQNM}, we saw 
that $\omega_{I}\rightarrow0$ 
as the phase transition at $\kappa=\kappa_{c}$ is approached. This implies that 
near the phase transition, the 
characteristic time scale $\tau=\omega_{I}^{-1}$ diverges, a well known 
phenomenon known as \textit{critical slowing 
down}. Specifically, the theory of dynamic critical phenomena (see e.g.~\cite{RevModPhys.49.435} and 
\cite{Maeda:2008hn,Maeda:2009wv,Natsuume:2010vb,Sonner:2014tca} for a holographic context) suggests a divergence of the form 
\begin{align}
 \tau\sim\left(\frac{T-T_c}{T_c}\right)^{-z\nu},
 \label{znu}
\end{align}
where $z$ and $\nu$ are \textit{critical exponents}. For holographic superconductors, it was found that 
\begin{align}
 z=2 \text{\ \ and\ \ }\nu=\frac{1}{2}
 \label{conductorexponents}
\end{align}
independently of the dimension
\cite{Maeda:2009wv,Natsuume:2010vb,Sonner:2014tca}\footnote{See also
  e.g.~\cite{PhysRevB.43.130} for a non-holographic study of critical
  exponents in superconductors.}. As the Kondo 
model under consideration resembles a holographic superconductor in $AdS_2$ space, we would naively expect it to fall into the same universality class, i.e.~to have the same critical exponents 
\eqref{conductorexponents}. Furthermore in \cite{Basu:2013soa}, the
exponents \eqref{conductorexponents} were found in a holographic model
involving a double-trace operator
in a background spacetime with  black hole horizon. 

We would hence naively expect that \eqref{conductorexponents} also applies to 
the Kondo model under investigation here. However, the boundary of $AdS_2$ has 
$d_s=0$  spatial dimensions, and as the 
definition of $\nu$ refers to the correlation length $\xi$ of the system, this 
critical exponent (as well as others) is not even well-defined in our model. 
Consequently, the often employed \textit{(hyper) scaling relations} between the 
critical exponents are not naively applicable in our model. This, of
course, is related to the fact that due to the well known Coleman-Mermin-Wagner theorem 
\cite{PhysRevLett.17.1133,Coleman1973}, phase transitions with spontaneous 
symmetry breaking do not 
occur in low dimensions at all. This theorem however does not apply in
the large $N$ limit where long-range fluctuations are suppressed \cite{coleman2015introduction}.
In the following we will show 
that the holographic Kondo model considered, which involves an $SU(N)$
spin in the large $N$ limit, shows a behaviour of the form \eqref{znu} with 
an exponent $z\nu=1$, as expected from \eqref{conductorexponents}.

We begin with the uncondensed phase, where equation \eqref{kappa} holds. Linearising 
this about $\omega=0$, we obtain
\begin{align}
 \kappa(\omega)&=\kappa_{c}+\kappa^{(1)} \omega+\mO(\omega^2).
 \label{kappaseries}
\end{align}
 Here, we have defined the constant
\begin{align}
 \kappa^{(1)}=\frac{i \psi 
^{(1)}\left(\frac{1}{2}-\frac{i}{2}\right)}{\left(H_{-\frac{1}{2}+\frac{i}{2}}+\gamma +\log (2)+\psi 
^{(0)}\left(\frac{1}{2}-\frac{i}{2}\right)\right){}^2}\approx -189.64 + 63.20 i,
\end{align}
where $\psi^{(n)}$ is the polygamma function and $\gamma$ is the Euler constant. For complex $\omega$ close to zero 
and real $\kappa-\kappa_{c}$, we hence see
\begin{align} 
\tau\sim\omega_{I}^{-1}=\frac{1}{\left(\kappa-\kappa_{c}\right)\Im\left(\frac
{1}{\kappa^{(1)}}\right)}.
\end{align}
Using
$
 \frac{T}{T_c}=e^{\frac{1}{\kappa_{c}}-\frac{1}{\kappa}},
$
this yields the result 
\begin{align}
 \tau \sim 
\frac{1}{\kappa_{c}\Im\left(\frac{1}{\kappa^{(1)}}\right)}\left(\frac{T-T_c}{T_c
}\right)^{-1},
\end{align}
consistent with the expectation based on equations \eqref{znu} and \eqref{conductorexponents}. This analytic argument 
described the critical slowing down for $T>T_c$, however the critical exponents $z$ and $\nu$ can be defined for both 
the normal and the condensed phase. In the latter case, where $T<T_c$, we do not have analytical results for the 
relaxation of the system, but as figure \ref{fig:QNMs-cond-2b} shows, our numerical results indicate that at 
$T\lesssim T_c$,
\begin{align}
 \tau\sim\omega_{I}^{-1}\sim \left(\frac{T-T_c}{T_c}\right)^{-1}.
\end{align}
This again agrees with the expectation from equations \eqref{znu} and \eqref{conductorexponents}.

%%%%%%%%%%%%%%%%%%%%%%%%%%%%%%%%%%%%%%%%%%%%%%%%%%%%%%%%%%%%%%%%%%%%%%%%%%%%%%%%
%%%%%%%%%%%%%%%%%%%%%%%%%%%%%%%%%%%%%%%%%%%%%%%%%%%%%%%%%%%%%%%%%%%%%%%%%%%%%%%%

\subsection{Power-law behaviour and discrete scale invariance at the critical point}
\label{sec::kappacrit}

Above we saw that due to the critical slowing down, the time 
scale governing the exponential 
decay of perturbations of the system diverges as $\kappa\rightarrow\kappa_{c}$. 
In the following, we investigate the 
evolution of the system as we quench the Kondo coupling $\kappa_1$ from the condensed phase $\kappa_1 < \kappa_c$ right 
onto the critical value $\kappa_1 = \kappa_c$. 

\begin{figure}[H]
\subfloat[][
 Log-linear plot of $\langle\mathcal{O}\rangle$ vs.~time.
 ]{
  \def\svgwidth{.45\textwidth}
\executeiffilenewer{fig12-re-and-im-beta-linear.svg}{fig12-re-and-im-beta-linear.pdf}%
{inkscape -z -D --file=fig12-re-and-im-beta-linear.svg %
--export-pdf=fig12-re-and-im-beta-linear.pdf --export-latex}%
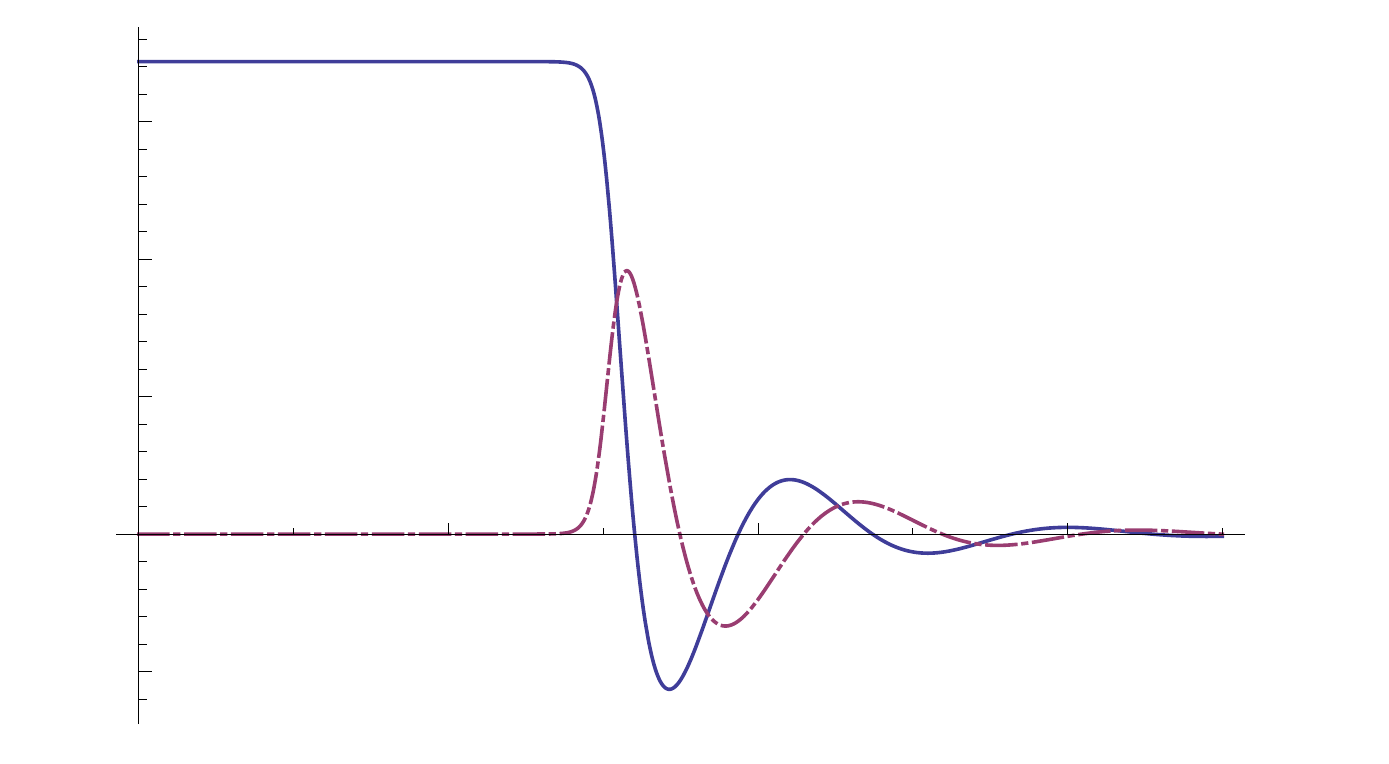%

}
 \hfill
\subfloat[][
 Log-log plot of $\langle\mathcal{O}\rangle$ vs.~time.
 ]{
  \def\svgwidth{.45\textwidth}
\executeiffilenewer{fig12-re-and-im-beta.svg}{fig12-re-and-im-beta.pdf}%
{inkscape -z -D --file=fig12-re-and-im-beta.svg %
--export-pdf=fig12-re-and-im-beta.pdf --export-latex}%
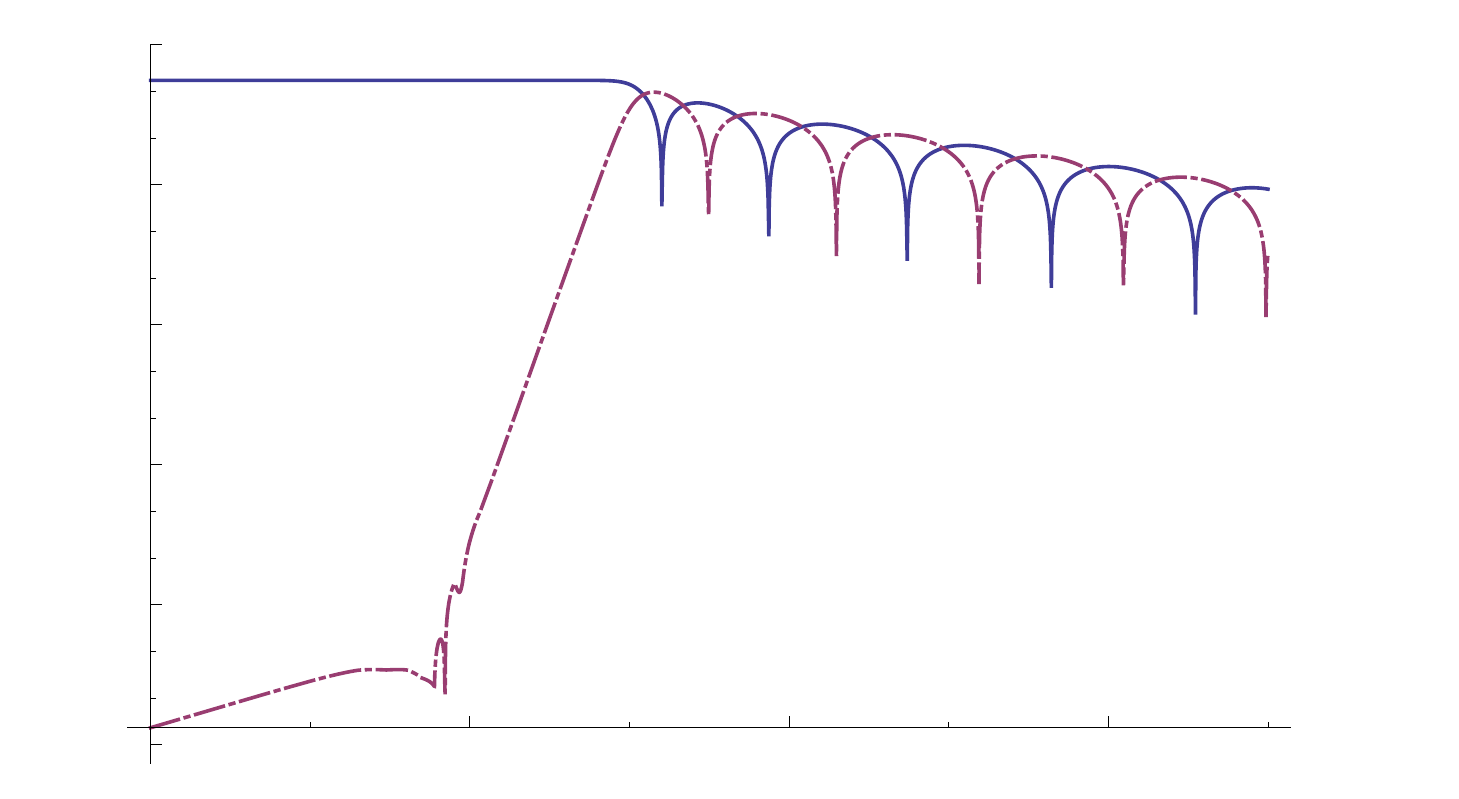%

}
\vfill
\subfloat[][
 Log-log plot of $|\langle\mathcal{O}\rangle|$ vs.~time.
 ]{
  \def\svgwidth{.45\textwidth}
\executeiffilenewer{fig12-abs-beta.svg}{fig12-abs-beta.pdf}%
{inkscape -z -D --file=fig12-abs-beta.svg %
--export-pdf=fig12-abs-beta.pdf --export-latex}%
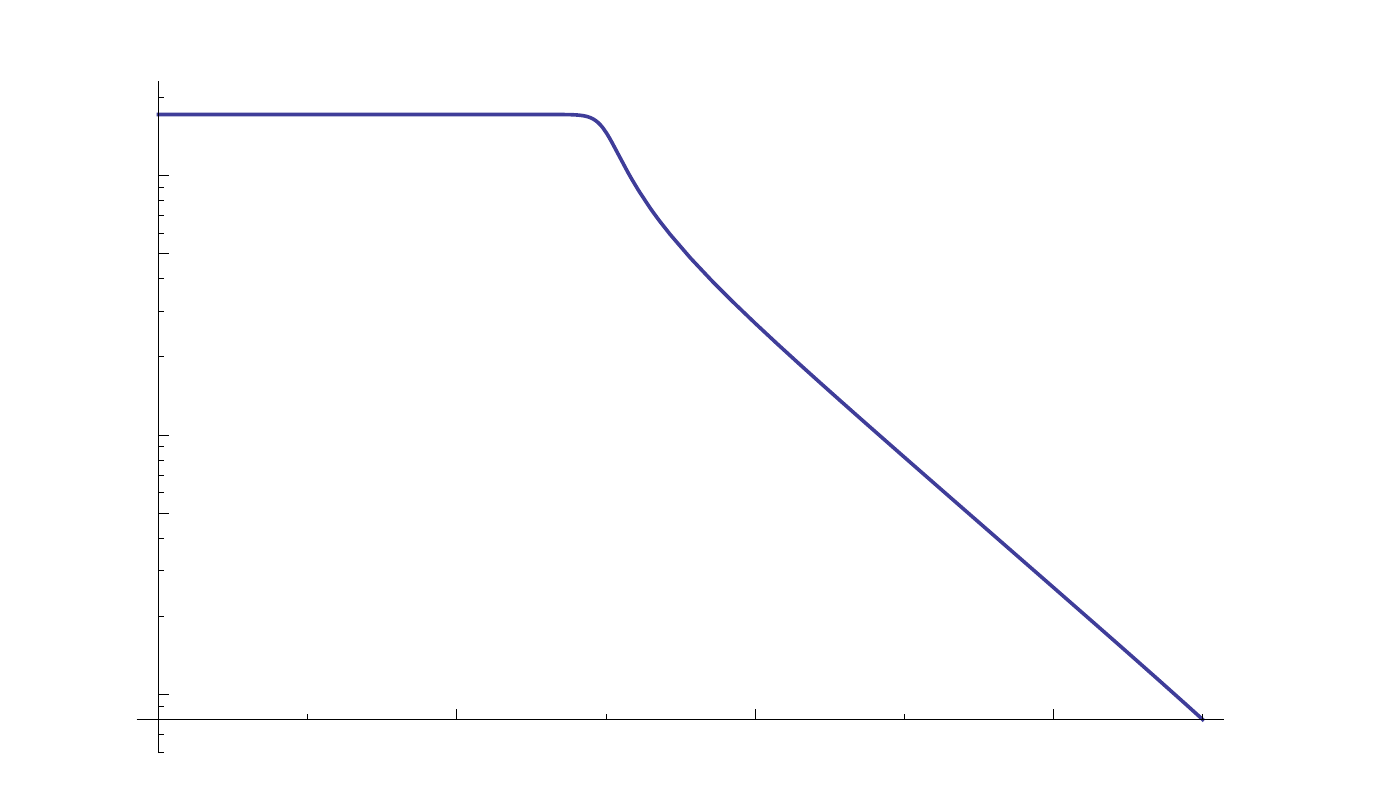%

}
 \hfill
\subfloat[][
 Log-log plot of $\Delta_t$ vs.~time.
 ]{
  \def\svgwidth{.45\textwidth}
\executeiffilenewer{fig12-Delta_t.svg}{fig12-Delta_t.pdf}%
{inkscape -z -D --file=fig12-Delta_t.svg %
--export-pdf=fig12-Delta_t.pdf --export-latex}%
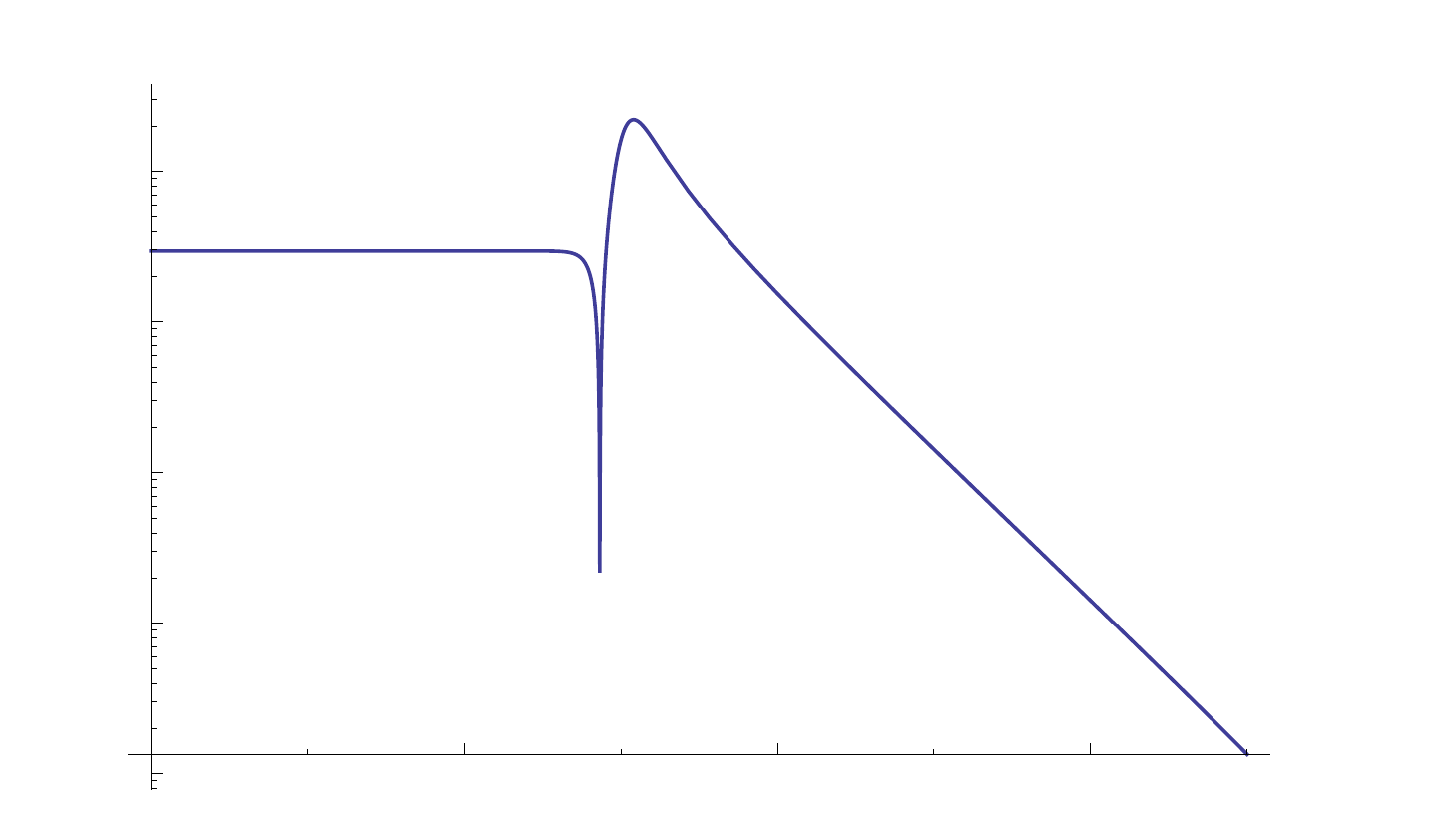%

}
\caption{A log-linear (a) and log-log (b) plot of 
$\alpha_1$ 
and $\alpha_2$ as functions of time. Below, we see log-log plots of the 
gauge-invariant quantities $\Delta_t = \mu - \pd \psi_0$ and $\beta = \langle 
\mathcal{O} \rangle / N$ which are not featuring any oscillations in 
(logarithmic) time.}
\label{fig:critical-1}
\end{figure}

The numerical results for a quench of this type are shown in figure
\ref{fig:critical-1}, where we show the time-dependent 
behaviour of $\beta_1(t)$ and $\beta_2(t)$ after the 
quench. We clearly see that the values relax to zero for late times, i.e.~that 
the systems settles to the solution $\Phi=0$ as appropriate for the onset of the normal phase. At first glance, the curves 
in figure \ref{fig:critical-1} appear to look qualitatively similar to the QNM 
depicted 
earlier, e.g.~in figure \ref{fig:generic-cond2norm-re-and-im}. However, there is a 
significant difference: Figure 
\ref{fig:generic-cond2norm-re-and-im} is a $\log$ plot and shows a behaviour 
$\sim \Re(e^{-i\omega t})$ with complex $\omega$, 
while in contrast figure \ref{fig:critical-1} is a $\log$-$\log$ plot and hence 
shows a behaviour \begin{align}
 \beta_1(t)&\sim \Re(e^{-i\upsilon\log(t)})=\Re(t^{-i 
\upsilon})=t^{\upsilon_{I}}\cos(\upsilon_{R}\log(t)),
 \label{powerlaw}
 \\
  \beta_2(t)&\sim \Im(e^{-i\upsilon \log(t)})=\Im(t^{-i 
\upsilon})=-t^{\upsilon_{I}}\sin(\upsilon_{R}\log(t)). 
\end{align}
Specifically, from the data of figure \ref{fig:critical-1} we may read off
\begin{align}
 \upsilon_{I}\approx -0.502 \text{\ \ and\ \ } \upsilon_{R}\approx 1.51\,,
\end{align}
both of which are less than $1\%$ off the  fractional values of 
$\upsilon_I = -1/2$ and $\upsilon_R = 3/2$.
When quenching the system to a final value $\kappa_{final}\lesssim\kappa_{c}$ 
or 
$\kappa_{final}\gtrsim\kappa_{c}$, the results of the earlier sections 
\ref{sec::normalQNM} and 
\ref{sec::condQNM} suggest that we should expect an exponential fall-off 
with decreasing exponent $\omega$ as $\kappa\rightarrow\kappa_{c}$. 
Our numerical results show that for $\kappa_{final}=\kappa_{c}$, the naively 
expected infinitely slow exponential 
decay gives way to a power law behaviour of the form \eqref{powerlaw}. 

In fact, late time \textit{power-law tails} are common in the study of QNMs. However, they are often associated 
with the QNMs of asymptotically \textit{flat} black holes, see e.g.~\cite{Konoplya:2011qq} for a review. In 
asymptotically AdS spaces in contrast, power-law tails of QNMs are usually absent \cite{PhysRevD.62.024027}. 
Furthermore, while in \cite{Konoplya:2011qq} a number of systems are mentioned 
that exhibit QNMs with power-law tails of the form
$t^{-a}\sin(\tilde{\mu} t)$, in \eqref{powerlaw}
 we observe a so-called \textit{log-periodic} (damped) 
oscillation. Here, the amplitude of an oscillating function decays as a power 
law ($\sim t^{-a}$), while the 
oscillation takes place in logarithmic time ($\sim \sin(b\log t)$). This
behaviour is known to be characteristic for systems exhibiting \textit{discrete 
scale invariance} and the associated \textit{complex critical exponents}, see 
\cite{Sornette1998239} for a review. In short, discrete scale invariance means 
that a theory is invariant under scale transformations only for specific scales, 
i.e.
\begin{align}
\mO_{DSI}(x)=\hat{\mu}\mO_{DSI}(\lambda x)
\label{dsi}
\end{align}
\textit{only} for specific scales $\lambda$. In general, the solution to \eqref{dsi} may 
then  take the form \cite{Sornette1998239}
\begin{align}
\mO_{DSI}(x)\propto x^\sigma,\ \ \sigma=-\frac{\log \hat{\mu}}{\log \lambda}+i\frac{2\pi n}{\log \lambda},
\end{align}
hence the connection with complex exponents and log-periodicity. This phenomenon 
has been observed\footnote{Note, however, that in many of these examples it is 
not the time variable $t$ in which log-periodic oscillations is observed.} 
in a wide range of physical systems, including stock markets and earthquakes 
(see \cite{Sornette1998239} and references therein for these two examples), but 
also quenches in condensed matter models \cite{PhysRevB.90.184202}, black hole 
formation 
\cite{PhysRevLett.70.9,PhysRevLett.70.2980,PhysRevD.51.4198,PhysRevD.52.5850} 
and even holographic models \cite{Liu:2009dm,Faulkner:2009wj,Hartnoll:2015rza}. 
See also \cite{Georgi:2016tjs} for a recent application of discrete scale 
invariance in QFT toy model building. We observe that in our model, similarly to 
what was found in \cite{PhysRevD.51.4198,PhysRevD.52.5850}, the discrete scale 
invariance, i.e.~the presence of a non-zero $\upsilon_{R}$ in 
\eqref{powerlaw}, manifests itself in form of a phase rotation
\begin{align}
t&\rightarrow \lambda t
\\
\frac{\alpha(t)}{|\alpha(t)|}&\rightarrow\lambda^{-i\upsilon_{R}}\frac{\alpha(t)
}{|\alpha(t)|}.
\end{align}
The quantities $|\alpha(t)|$ and $|\beta(t)|$ hence do not show any signatures 
of discrete scale invariance, only a power-law fall-off determined by 
$\upsilon_{I}$. The physical variable in the holographic Kondo model is of 
course the complex vev $\beta(t)\sim \left<\mO\right>$ \eqref{eq:vev-beta}. 
Hence while the modulus $|\left<\mO(t)\right>|$ decreases as a power-law, its 
complex phase rotates with $\sim\log t$. Equivalently, we see that the (bulk) 
gauge-invariant quantity $\Delta_t = \mu - \pd \psi_0$ falls off towards the 
limiting value $\Delta_t=1/2$ as $\sim 
t^{-1}$.

A definitive interpretation of how the discrete scale invariance arises in the 
model under investigation cannot be given just based on the numerical
results presented, and it is not possible to determine
what sets the corresponding scale. Starting from 
the observation \eqref{powerlaw}, we propose an ansatz of the form 
\begin{align} 
\phi_1(t,z)&=t^{\upsilon_{I,\phi}}\cos(\upsilon_{R}\log(t))\tilde{\phi}
(z)+\mO(t^x,x<\upsilon_{I,\phi}<0),
 \\ 
\phi_2(t,z)&=t^{\upsilon_{I,\phi}}\sin(-\upsilon_{R}\log(t))\tilde{\phi}
(z)+\mO(t^x,x<\upsilon_{I,\phi}<0),
 \label{powerlawansatz}
 \\
 a_t(t,z)&=\frac{Q}{z}+\mu+\mO(t^x,x<0)
\end{align}
and insert it into the equations of motion \eqref{eq:fulleom}-\eqref{eq:fulleom2}, 
using $-Q=\mu=\frac{1}{2}$ as throughout the paper. In this way, we obtain the 
lowest-order equation 
\begin{align}
4 z^2 h(z) \left(h'(z) \tilde{\phi }'(z)+h(z) \tilde{\phi }''(z)\right)+(z-1)^2 \tilde{\phi }(z)=0
 \label{tildeequation}
\end{align}
for $\tilde{\phi}(z)$, which has the two independent solutions
\begin{align}
\tilde{\phi}_{reg}=\sqrt{\frac{2 z}{z+1}}P_{-\frac{1}{2}+\frac{i}{2}}\left(\frac{4 
z}{z+1}-1\right)\text{\ \ 
and\ \ }
\tilde{\phi}_{irreg}=\sqrt{\frac{z}{z+1}} \, _2F_1\left(\frac{1}{2}-\frac{i}{2},\frac{1}{2}+\frac{i}{2};1;\frac{2 
z}{z+1}\right),
\end{align}
where $P_n(x)$ stands for the Legendre polynomial of the first kind and $_2F_1$ is the hypergeometric function. We find 
that $\tilde{\phi}_{irreg}$ diverges at the event horizon $z=z_H=1$, and consequently the physical 
solution to \eqref{tildeequation} is given by $const.\times\tilde{\phi}_{reg}$, with no further parameter to adjust 
boundary conditions. The boundary expansion of $\tilde{\phi}_{reg}$ reads
\begin{align}
 \tilde{\phi}_{reg}(z)\sim 
const.\times\sqrt{z}\left(
\log(z)+\left(H_{-\frac{1}{2}+\frac{i}{2}}+H_{-\frac{1}{2}-\frac{i}{2}}+\log(2)
\right)...
\right),
\end{align}
implying
\begin{align} 
\kappa_{1}=\frac{1}{H_{-\frac{1}{2}+\frac{i}{2}}+H_{-\frac{1}{2}-\frac{i}{2}}
+\log (2)}=\kappa_{c},
\end{align}
see equations \eqref{eq:alpha-is-kappa-beta} and  \eqref{eq:kappacrit}. We 
have hence proven that a power law ansatz such as \eqref{powerlawansatz} can 
\textit{only} solve the equations of motion if the boundary condition is fixed at the critical value 
$\kappa=\kappa_{c}$. However, the lowest-order 
equations obtained via the ansatz \eqref{powerlawansatz} do not fix the 
parameters $\upsilon_{I,\phi}$ and $\upsilon_{R,\phi}$. Presumably this 
would require the appropriate inclusion of higher orders into the ansatz.

The analytic study of power law tails of QNMs is an important subject, for 
example concerning the question whether these tails are intrinsically non-linear 
phenomena, see again \cite{Konoplya:2011qq} for an overview. There exist 
specialised analytical methods to treat these problems \cite{PhysRevD.52.2118}, 
and the methods employed in \cite{PhysRevD.51.4198,PhysRevD.52.5850} may also 
have applicability to our system. Due to the possible importance of 
higher-order effects, a full analytical analysis of the emerging log-periodic 
behaviour of our system is however beyond the scope of the present paper, and 
will be left to future investigations.

%% file: fig12-re-and-im-beta-linear.pdf_tex
%% Creator: Inkscape inkscape 0.91, www.inkscape.org
%% Accompanies image file 'fig12-re-and-im-beta-linear.pdf' (pdf, eps, ps)
\begingroup%
  \makeatletter%
  \providecommand\color[2][]{%
    \errmessage{(Inkscape) Color is used for the text in Inkscape, but the package 'color.sty' is not loaded}%
    \renewcommand\color[2][]{}%
  }%
  \providecommand\transparent[1]{%
    \errmessage{(Inkscape) Transparency is used (non-zero) for the text in Inkscape, but the package 'transparent.sty' is not loaded}%
    \renewcommand\transparent[1]{}%
  }%
  \providecommand\rotatebox[2]{#2}%
  \ifx\svgwidth\undefined%
    \setlength{\unitlength}{400bp}%
    \ifx\svgscale\undefined%
      \relax%
    \else%
      \setlength{\unitlength}{\unitlength * \real{\svgscale}}%
    \fi%
  \else%
    \setlength{\unitlength}{\svgwidth}%
  \fi%
  \global\let\svgwidth\undefined%
  \global\let\svgscale\undefined%
  \makeatother%
  \begin{picture}(1,0.55)%
    \put(0,0){\includegraphics[width=\unitlength,page=1]{fig12-re-and-im-beta-linear.pdf}}%
    \put(0.32322514,0.13172707){\color[rgb]{0,0,0}\makebox(0,0)[b]{\smash{$\tss{10^{2}}$}}}%
    \put(0.54614502,0.13172707){\color[rgb]{0,0,0}\makebox(0,0)[b]{\smash{$\tss{10^{4}}$}}}%
    \put(0.76906487,0.13172707){\color[rgb]{0,0,0}\makebox(0,0)[b]{\smash{$\tss{10^{6}}$}}}%
    \put(0.0751992,0.35353045){\color[rgb]{0,0,0}\makebox(0,0)[rb]{\smash{$\tss{10^{-3}}$}}}%
    \put(0.0751992,0.1568005){\color[rgb]{0,0,0}\makebox(0,0)[rb]{\smash{$\tss{0}$}}}%
    \put(0.9129628,0.16515831){\color[rgb]{0,0,0}\makebox(0,0)[lb]{\smash{$\tss{2\pi T \, t}$}}}%
    \put(0.44643944,0.48661246){\color[rgb]{0.24705882,0.23921569,0.6}\makebox(0,0)[lb]{\smash{$\tss{\Re\langle\mathcal{O}\rangle/N\sqrt{2\pi T}}$}}}%
    \put(0.47966724,0.32074211){\color[rgb]{0.6,0.23921569,0.44313725}\makebox(0,0)[lb]{\smash{$\tss{\Im\langle\mathcal{O}\rangle/N\sqrt{2\pi T}}$}}}%
  \end{picture}%
\endgroup%

%% file: fig12-re-and-im-beta.pdf_tex
%% Creator: Inkscape inkscape 0.91, www.inkscape.org
%% Accompanies image file 'fig12-re-and-im-beta.pdf' (pdf, eps, ps)
\begingroup%
  \makeatletter%
  \providecommand\color[2][]{%
    \errmessage{(Inkscape) Color is used for the text in Inkscape, but the package 'color.sty' is not loaded}%
    \renewcommand\color[2][]{}%
  }%
  \providecommand\transparent[1]{%
    \errmessage{(Inkscape) Transparency is used (non-zero) for the text in Inkscape, but the package 'transparent.sty' is not loaded}%
    \renewcommand\transparent[1]{}%
  }%
  \providecommand\rotatebox[2]{#2}%
  \ifx\svgwidth\undefined%
    \setlength{\unitlength}{420bp}%
    \ifx\svgscale\undefined%
      \relax%
    \else%
      \setlength{\unitlength}{\unitlength * \real{\svgscale}}%
    \fi%
  \else%
    \setlength{\unitlength}{\svgwidth}%
  \fi%
  \global\let\svgwidth\undefined%
  \global\let\svgscale\undefined%
  \makeatother%
  \begin{picture}(1,0.54761905)%
    \put(0,0){\includegraphics[width=\unitlength,page=1]{fig12-re-and-im-beta.pdf}}%
    \put(0.09012715,0.50868545){\color[rgb]{0,0,0}\makebox(0,0)[rb]{\smash{$\tss{10^{-2}}$}}}%
    \put(0.09012715,0.41183879){\color[rgb]{0,0,0}\makebox(0,0)[rb]{\smash{$\tss{10^{-5}}$}}}%
    \put(0.09012715,0.31499216){\color[rgb]{0,0,0}\makebox(0,0)[rb]{\smash{$\tss{10^{-8}}$}}}%
    \put(0.09012715,0.21814553){\color[rgb]{0,0,0}\makebox(0,0)[rb]{\smash{$\tss{10^{-11}}$}}}%
    \put(0.09012715,0.12129887){\color[rgb]{0,0,0}\makebox(0,0)[rb]{\smash{$\tss{10^{-14}}$}}}%
    \put(0.32472098,0.01034317){\color[rgb]{0,0,0}\makebox(0,0)[b]{\smash{$\tss{10^{2}}$}}}%
    \put(0.54269751,0.01034317){\color[rgb]{0,0,0}\makebox(0,0)[b]{\smash{$\tss{10^{4}}$}}}%
    \put(0.76067403,0.01034317){\color[rgb]{0,0,0}\makebox(0,0)[b]{\smash{$\tss{10^{6}}$}}}%
    \put(0.90603253,0.05075455){\color[rgb]{0,0,0}\makebox(0,0)[lb]{\smash{$\tss{2\pi T \, t}$}}}%
    \put(0.13853014,0.51058705){\color[rgb]{0.24705882,0.23921569,0.6}\makebox(0,0)[lb]{\smash{$\tss{\Re\langle\mathcal{O}\rangle/N\sqrt{2\pi T}}$}}}%
    \put(0.3514146,0.18729604){\color[rgb]{0.6,0.23921569,0.44313725}\makebox(0,0)[lb]{\smash{$\tss{\Im\langle\mathcal{O}\rangle/N\sqrt{2\pi T}}$}}}%
  \end{picture}%
\endgroup%

%% file: fig12-abs-beta.pdf_tex
%% Creator: Inkscape inkscape 0.48.4, www.inkscape.org
%% Accompanies image file 'fig12-abs-beta.pdf' (pdf, eps, ps)
\begingroup%
  \makeatletter%
  \providecommand\color[2][]{%
    \errmessage{(Inkscape) Color is used for the text in Inkscape, but the package 'color.sty' is not loaded}%
    \renewcommand\color[2][]{}%
  }%
  \providecommand\transparent[1]{%
    \errmessage{(Inkscape) Transparency is used (non-zero) for the text in Inkscape, but the package 'transparent.sty' is not loaded}%
    \renewcommand\transparent[1]{}%
  }%
  \providecommand\rotatebox[2]{#2}%
  \ifx\svgwidth\undefined%
    \setlength{\unitlength}{400bp}%
    \ifx\svgscale\undefined%
      \relax%
    \else%
      \setlength{\unitlength}{\unitlength * \real{\svgscale}}%
    \fi%
  \else%
    \setlength{\unitlength}{\svgwidth}%
  \fi%
  \global\let\svgwidth\undefined%
  \global\let\svgscale\undefined%
  \makeatother%
  \begin{picture}(1,0.575)%
    \put(0,0){\includegraphics[width=\unitlength]{fig12-abs-beta.pdf}}%
    \put(0.09722191,0.44015983){\color[rgb]{0,0,0}\makebox(0,0)[rb]{\smash{$\tss{10^{-3}}$}}}%
    \put(0.09722191,0.25346439){\color[rgb]{0,0,0}\makebox(0,0)[rb]{\smash{$\tss{10^{-4}}$}}}%
    \put(0.09722191,0.06851711){\color[rgb]{0,0,0}\makebox(0,0)[rb]{\smash{$\tss{10^{-5}}$}}}%
    \put(0.32768452,0.01600349){\color[rgb]{0,0,0}\makebox(0,0)[b]{\smash{$\tss{10^{2}}$}}}%
    \put(0.54340401,0.01600349){\color[rgb]{0,0,0}\makebox(0,0)[b]{\smash{$\tss{10^{4}}$}}}%
    \put(0.75912346,0.01600349){\color[rgb]{0,0,0}\makebox(0,0)[b]{\smash{$\tss{10^{6}}$}}}%
    \put(0.89518951,0.05457802){\color[rgb]{0,0,0}\makebox(0,0)[lb]{\smash{$\tss{2\pi T \, t}$}}}%
    \put(0.11194396,0.53933095){\color[rgb]{0,0,0}\makebox(0,0)[b]{\smash{$\tss{\left|\langle\mathcal{O}\rangle\right|/N \sqrt{2\pi T}}$}}}%
  \end{picture}%
\endgroup%

%% file: fig12-Delta_t.pdf_tex
%% Creator: Inkscape inkscape 0.91, www.inkscape.org
%% Accompanies image file 'fig12-Delta_t.pdf' (pdf, eps, ps)
\begingroup%
  \makeatletter%
  \providecommand\color[2][]{%
    \errmessage{(Inkscape) Color is used for the text in Inkscape, but the package 'color.sty' is not loaded}%
    \renewcommand\color[2][]{}%
  }%
  \providecommand\transparent[1]{%
    \errmessage{(Inkscape) Transparency is used (non-zero) for the text in Inkscape, but the package 'transparent.sty' is not loaded}%
    \renewcommand\transparent[1]{}%
  }%
  \providecommand\rotatebox[2]{#2}%
  \ifx\svgwidth\undefined%
    \setlength{\unitlength}{420bp}%
    \ifx\svgscale\undefined%
      \relax%
    \else%
      \setlength{\unitlength}{\unitlength * \real{\svgscale}}%
    \fi%
  \else%
    \setlength{\unitlength}{\svgwidth}%
  \fi%
  \global\let\svgwidth\undefined%
  \global\let\svgscale\undefined%
  \makeatother%
  \begin{picture}(1,0.57142857)%
    \put(0,0){\includegraphics[width=\unitlength,page=1]{fig12-Delta_t.pdf}}%
    \put(0.08212773,0.44618646){\color[rgb]{0,0,0}\makebox(0,0)[rb]{\smash{$\tss{10^{-3}}$}}}%
    \put(0.08212773,0.34299629){\color[rgb]{0,0,0}\makebox(0,0)[rb]{\smash{$\tss{10^{-4}}$}}}%
    \put(0.08212773,0.23980615){\color[rgb]{0,0,0}\makebox(0,0)[rb]{\smash{$\tss{10^{-5}}$}}}%
    \put(0.08212773,0.13661601){\color[rgb]{0,0,0}\makebox(0,0)[rb]{\smash{$\tss{10^{-6}}$}}}%
    \put(0.12214934,0.52886784){\color[rgb]{0,0,0}\makebox(0,0)[lb]{\smash{$\tss{\frac{\left|\Delta_t\right|}{2\pi T} - \frac{1}{2}}$}}}%
    \put(0.8867388,0.05218542){\color[rgb]{0,0,0}\makebox(0,0)[lb]{\smash{$\tss{2\pi T \, t}$}}}%
    \put(0.31860653,0.01158574){\color[rgb]{0,0,0}\makebox(0,0)[b]{\smash{$\tss{10^{2}}$}}}%
    \put(0.53323804,0.01158574){\color[rgb]{0,0,0}\makebox(0,0)[b]{\smash{$\tss{10^{4}}$}}}%
    \put(0.74786951,0.01158574){\color[rgb]{0,0,0}\makebox(0,0)[b]{\smash{$\tss{10^{6}}$}}}%
  \end{picture}%
\endgroup%

%% file: conclusion.tex
\subsection{Summary}
\label{sec::summary}

We have studied quantum quenches in a holographic Kondo model. We 
solved 
the full time-dependent dynamics of our model numerically using spectral and 
finite difference methods, and we studied the relaxation of the system under 
various quench protocols. We found that the relaxation is determined by the 
lowest QNM of the initial state of the system, which describes an 
exponential decay, with an additional oscillatory profile for quenches
to the normal phase. The lowest QNM provides an 
excellent description of the relaxation of the system not only at late times,
as one expects close to equilibrium, but also almost immediately as the system
begin to relax after the quench of the Kondo coupling: there does not appear to be
any appreciable region in the onset of the relaxation that is not described 
by the lowest QNM.  
This seems to be a generic feature in holographic systems modelling strong 
dynamics. Most importantly, we found that in the condensed (screened)
phase, the leading QNM is purely imaginary, which corresponds to
over-damping. In a temperature region below $T_c$, we have $\omega
\propto -i \langle {\mathcal{O}} \rangle^2$. 
This is consistent with expectations from the behaviour of the Kondo resonance
\cite{Erdmenger:2016jjg}. At low temperatures, we see a deviation from mean-field behaviour, $\omega\propto-i\log(\langle {\mathcal{O}} \rangle)$.

The time-dependence of the flux through $AdS_2$ at the black
hole horizon, which is dual to the size of the impurity representation,  describes the 
decrease of degrees of freedom after a quench to the condensed
phase. This corresponds to a measure for the formation of the Kondo
cloud at the site of the impurity. We found the decrease of degrees of freedom to be
exponential. 

In section \ref{sec::critExp} we studied in more detail the \textit{critical 
slowing down} of the system near the phase transition at $T=T_c$, and showed the 
corresponding (combination of) critical exponents to be $z\nu=1$, just as 
expected from the similar holographic models 
\cite{Basu:2013soa,Maeda:2008hn,Maeda:2009wv,Natsuume:2010vb,Sonner:2014tca}. However, we 
also pointed out that due to the low dimensionality of our model, great care is 
required when interpreting critical exponents and their (hyper) scaling 
relations. Interestingly, the critical exponents of a large $N$ Kondo-Heisenberg 
lattice near a quantum critical point have been studied in  
\cite{PhysRevLett.98.026402,PhysRevB.78.035109}, with the result $z=3$.

Section \ref{sec::kappacrit} was then denoted to the study of quenches that 
lead directly to the critical point, $\kappa\rightarrow\kappa_{c}$. 
Our numerical results for quenches of this type, displayed in 
figure \ref{fig:critical-1}, show a damped \textit{log-periodic} behaviour of 
the 
form \eqref{powerlaw}. This log-periodic behaviour is known to be a telltale 
sign of \textit{discrete scale invariance}, reviewed in \cite{Sornette1998239} 
and observed in holographic models already in 
\cite{Liu:2009dm,Faulkner:2009wj,Hartnoll:2015rza}. Unfortunately our numerics 
alone do not offer insight into the underlying mechanism causing the emergence 
of this phenomenon, and a lowest-order ansatz for the late time solutions of 
the 
equations of motion does not fix the involved complex critical exponent. A full 
analytical treatment of this interesting problem is hence left for future 
study. We note that
in \cite{Liu:2009dm,Faulkner:2009wj}, it was speculated that the emerging 
discrete scale invariance is connected to the physics of a near-horizon $AdS_2$ 
region. Similarly, in our model, we are effectively working with an asymptotic 
$AdS_2$ spacetime. This seems to imply that an $AdS_2$ 
structure is 
advantageous to the emergence of discrete scale invariance in holographic 
models, which  may have interesting implications for the Sachdev-Ye-Kitaev model 
\cite{Sachdev:1992fk,Sachdev:2010um,Maldacena:2016hyu}.
 
\subsection{Outlook}
\label{sec::outlook}
 
One obvious generalisation of our approach
will be to investigate quenches and other time-dependent phenomena in the holographic
two-impurity Kondo model of \cite{O'Bannon:2015gwa}.  Moreover, it
will be instructive to study the
$T=0$ limit in further detail. This requires to add a quartic term to
the scalar potential in order to ensure a well-defined IR fixed point
and a finite condensate for $T\rightarrow 0$  (see also \cite{Erdmenger:2015spo}).

Also, as pointed out above, 
the emergence of \textit{discrete scale invariance} for critical quenches 
$\kappa\rightarrow\kappa_{c}$ deserves further study. In fact, we see from 
figure \ref{fig:QNMs-cond-2b} that in the condensed phase 
$\omega_{I}\rightarrow0$ not only for $T/T_c\rightarrow1$, but also for 
$T/T_c\rightarrow0$. This seems to indicate that in the holographic Kondo model, 
\textit{critical slowing down} does not only occur near the phase transition at 
$T=T_c$, but also at $T=0$. Whether this critical slowing down at zero 
temperature will be accompanied by similar $\log$-periodic oscillations to those 
that we found at $T=T_c$ is however not clear. This requires a further
study of the $T\rightarrow 0$ limit in a model which ensures stability
of the IR fixed point, as described above.

One interesting further direction may be to combine the study of time-dependent 
phenomena in the holographic Kondo model of 
\cite{Erdmenger:2013dpa,O'Bannon:2015gwa} that was carried out in this work with 
the study of backreaction and entanglement entropy done in 
\cite{Erdmenger:2014xya,Erdmenger:2015spo,Erdmenger:2015xpq}. In particular, in 
\cite{Erdmenger:2015spo,Erdmenger:2015xpq}, it was shown that in the backreacted 
holographic Kondo model, there is a natural geometric length scale that takes 
the role of the Kondo scale $\xi_K$. A study of a holographic model
allowing for both time-dependence and 
backreaction may hence allow the study of the characteristic time and length 
scales 
involved in the formation of the Kondo cloud in a similar way to what was done 
in \cite{2014arXiv1409.0646N} on the field theory side. Futhermore, calculations 
of the time-dependence of entanglement entropy after a quench in the holographic 
Kondo model may be compared to the results of field theory 
studies such as \cite{PhysRevB.90.115101}.
 
Finally, recently there appeared  models of Kondo physics in the context of QCD~\cite{Ozaki:2015sya}
and Lorentz violation~\cite{Bazeia:2016pra}. By adapting the methods
of this paper, we expect that quenches in these
models can also be studied holographically.

\section*{Acknowledgements}

We are very grateful to
Paul Chesler and to Julian Sonner
for sharing their insights about numerical analysis of time-dependent 
problems in holography. Furthermore we would like to thank
Carlos Hoyos, Nina Miekley, Andy O'Bannon, Ioannis Papadimitriou and Jonas Probst 
for illuminating discussions.

%% file: appendix.tex
\section{Gauge fixing, coordinates and field redefinition}
\label{sec:ansatzEF}
We apply the radial gauge for both the gauge field $a$ on the 
$(1+1)$-dim.~defect 
manifold as well as the Chern-Simons field in the $(2+1)$-dim.~bulk. Regularity 
of the Chern-Simons field implies $\left. A_t \right|_{z_H} = 0$ which prolongs 
to $A_t = 0$ throughout the bulk after imposing the equations of motion.
Hence only $A_x$ is non-trivial.
As it is orthogonal to the defect manifold, it does not appear in the 
equations of motion for neither the scalar nor the gauge field.
Thus we may neglect the Chern-Simons field in our analysis.

As already mentioned around eq.~\eqref{eq:boundary-exp}, the asymptotic behaviour 
of the scalar field in Schwarzschild-like coordinates is given by
\begin{gather}
 \phi_i (t,z) \approx \sqrt{z} \, \left(\alpha_i(t) \log(z) + \beta_i(t) 
	      \right) + O(z^{3/2})
\end{gather}
as we approach the AdS boundary at $z=0$.
The difficulty in dealing with the equations numerically is this non-analytic 
boundary behaviour. 

We address this issue with a field redefinition and a change of 
coordinates. 
The change of variables can get rid of all (non-analytic at $z=0$) 
$\sqrt{z}$ terms in the boundary expansion if we let $y^2 = z$. With this 
definition the boundary and horizon are at $y=0$ and $y=\sqrt{z_H}$, 
respectively. 
This redefinition only changes the log terms by a factor of two, however, so 
the fields are still non-analytic at $y=0$. 
In terms of the old coordinates, we define the new differentials to be
\begin{equation}
\totd t = \totd v + \frac{2 \,y \,\totd y}{h(y)} \, , \quad \totd z = 2\,y\, 
\totd y \, .
\end{equation}
The metric~\eqref{eq:fullmetric} becomes
\begin{equation}
\label{eq:metricEF}
ds^2 = \frac{1}{y^4}\left(-h(y)\,\totd v^2 - 4 \,y \,\totd v\, \totd y + 
\totd x^2\right) \,,
\end{equation}
where $h(y) = 1- y^4$.
Note that $v$ and $t$ have the same level sets at the boundary, $y=0$, so that 
we can simply replace them when analysing boundary properties.
The radial gauge $a_z = 0$ in Schwarzschild-like coordinates translates into 
$a_y = 2\,y\, a_v / h(y)$
in our adapted EF-like coordinates.
We decompose the scalar field into its real and imaginary part 
$\Phi = \phi_1 + i \phi_2$. 
Applying this ansatz, the relevant 
equations of motion~\eqref{eq:fulleom}-\eqref{eq:fulleom2} are explicitly 
given 
by
\begin{align}
0 &= 
\pd_y^{\,2}\phi_{1} 
- \frac{4\,\pd_y\pd_v\phi_{1}}{h(y)}
- \frac{(1+3 y^4) \,\pd_y \phi_{1}}{y\,h(y)}
- \frac{2 y \,\left( \phi_{2} \,\pd_y a_v + 2 a_v \,\pd_y 
\phi_{2} \right)}{h(y)} \, ,
\\
0 &= 
\pd_y^{\,2}\phi_{2} 
- \frac{4\,\pd_y\pd_v\phi_{2}}{h(y)}
- \frac{(1+3 y^4) \,\pd_y \phi_{2}}{y\,h(y)}
+ \frac{2 y \,\left( \phi_{1} \,\pd_y a_v + 2 a_v \,\pd_y 
\phi_{1} \right)}{h(y)} \, ,
\\
0 &=
\pd_y^{\,2} a_v +\frac{3}{y}\,\pd_y a_v - 
\frac{4}{y^3}\left(\phi_2\,\pd_y\phi_1 - \phi_1\,\pd_y\phi_2\right) \, ,
\\
0 &= 
\pd_y\pd_v a_v 
+ \frac{4}{y^3} a_v \left(\phi_1^2+\phi_2^2\right) \nonumber 
\\
&\hphantom{=} 
- \frac{2}{y^4}\,\left(\left[h(y)\,\pd_y\phi_1+2\,y\,\pd_v\phi_1\right]\phi_2
-\phi_1\left[h(y)\,\pd_y\phi_2+2\,y\,\pd_v\phi_2\right]\right) \, ,
\label{kondo:eq:EFphifull}
\end{align}
where, as anticipated, the Chern-Simons field does not enter the equations of 
motion of the fields restricted to the defect.

The next step is a field redefinition where we subtract a number of 
dominant log terms in the boundary expansion. If we subtract enough terms, the 
redefined fields  will have regular second derivatives at $y=0$. This also 
means that the non-analytic contributions from the log terms only appear at 
higher order.
By applying \eqref{eq:alpha-is-kappa-beta}, the $v$-dependent boundary 
expansion of the remaining fields looks like
\begin{align}
 \phi_i(v,y) &\approx \beta_i(v) y \left( 1 + 2\,\kappa_i(v) \log y \right) 
 \nonumber
 \\
 &\hphantom{=}+ y^3 \left(a_i^{(4)}(v) \log^4 y + \dots + a_i^{(1)}(v) \log y
 + a_i^{(0)}(v) \right) + \dots,
 \label{eq:full-exp-phi}
 \\
%  \phi_2(v,y) &\approx y^3 \left( b_1(v) \log y + b_0(v) \right) + \dots,
%  \\
 a_v(v,y) &\approx -\frac{1}{2\,y^2} + \mu(v) 
 + c^{(3)}(v) \log^3 y + \dots + c^{(1)}(v) \log y 
 \nonumber
 \\
 &\hphantom{=}+ y^2 \left(d^{(4)}(v) \log^4 y + \dots + d^{(1)}(v) \log y
 + d^{(0)}(v) \right) + \dots,
 \label{eq:full-exp-av}
\end{align}
where $a_i^{(k)}(v)$, $c^{(k)}(v)$ and $d^{(k)}(v)$ are functions of 
$\beta_i(v)$, 
$\kappa_i(v)$, $\mu(v)$ and their derivatives. We choose to subtract all terms 
containing a $\log y$ and divergent terms up to $\mO(y^3)$. In other words, 
we define
\begin{align}
 \tilde{\phi}_i(v, y) &= \frac{1}{y} \left( 
  \phi_i(v,y) - s^{(\phi_i)}
 \right),
 \\
%  \tilde{\phi}_2(v,y) &= \phi_2(v,y) - s_b,
%  \\
 \tilde{a}_v(v,y) &= a_v(v,y) - s^{(a_v)} \, ,
\end{align}
where
\begin{align}
 s^{(\phi_i)} &=  2\,y \,\beta_i(v)\,\kappa_i(v) \,\log y + y^3 
\left(a_i^{(4)}(v) 
\log^4 y + 
\dots + a_i^{(1)}(v) \log y \right),
 \\
%  s_b &= y^3  b_1(v) \log y,
%  \\
 s^{(a_v)} &= -\frac{1}{2\,y^2} 
 + c^{(3)}(v) \log^3 y + \dots + c^{(1)}(v) \log y, 
 \\
 &\hphantom{=}+ y^2 \left(d^{(4)}(v) \log^4 y + \dots + d^{(1)}(v) \log y
  \right) + \dots
\end{align}

These tilded fields have regular second $y$-derivatives on the domain 
$0\le y \le 1$. Moreover, the boundary value of $\tilde{\phi}_i$ is $\beta_i$ 
and the boundary value of $\tilde{a}_v$ is $\mu$.
With these redefinitions the equations of motion become too long to reproduce 
on the page here. However, they involve only fields and derivatives that are 
regular, which provides some numerical stability.

%%%%%%%%%%%%%%%%%%%%%%%%%%%%%%%%%%%%%%%%%%%%%%%%%%%%%%%%%%%%%%%%%%%%%%%%%%%%%%%%
%%%%%%%%%%%%%%%%%%%%%%%%%%%%%%%%%%%%%%%%%%%%%%%%%%%%%%%%%%%%%%%%%%%%%%%%%%%%%%%%

\section{Numerical evolution scheme}
\label{sec:evolutionscheme}

Our goal is to find solutions in which we give $\kappa_1$ a $v$-dependent 
profile.
We always start from a static solution, so the assumption is that $\kappa_1$ is 
constant for all $v<0$.
At $v = 0$, $\kappa_1(v)$ becomes a time--dependent function which smoothly 
connects to the initial constant value. 
This function could be a Gaussian or a hyperbolic tangent, for example.
By solving the equations of motion, we find the resulting motion of the fields 
and, consequently, the coefficients $\beta_1(v)$, $\beta_2(v)$ and $\mu(v)$ in 
\eqref{eq:full-exp-av} and \eqref{eq:full-exp-phi}.
We use an implicit Crank--Nicholson integration method in $v$ and 
pseudospectral methods in $y$.

The first step is to find the static solution to the equations of motion. 
For that, we get an initial guess for the solutions by using the shooting 
method: we choose the initial value of $\kappa_1$, then adjust $\beta_1$, 
$\beta_2$ and $\mu$ until integration of the solutions from boundary to horizon 
yields solutions that are regular everywhere. 
After that, we discretise the equations of motion and solve the 
resulting non-linear matrix equations with the output of the shooting method as 
initial guess. 
More precisely, the equations are discretised on Chebyshev-Lobatto collocation 
points, starting at $y = \epsilon$ and ending at $y=1-\epsilon$. 
Differential operators are replaced by pseudospectral differentiation matrices.
For $N$ collocation points, the fields 
are discretised to $(\tilde{\phi}_1)_1, \dots, (\tilde{\phi}_1)_N$ 
and similarly for $\tilde{\phi}_2$ and $\tilde{a}_v$. 

Since we know the boundary expansion for each of the fields, the first 
component of each can be replaced by that boundary expansion up to a given 
order, which will be made up of terms containing $\beta_1$, $\beta_2$, 
$\kappa_1$ and $\mu$, only. 
We apply a numerical algorithm to solve for the $3N$ components
$(\tilde{\phi}_1)_2, \dots, (\tilde{\phi}_1)_N$, $(\tilde{\phi}_2)_2, \dots, 
(\tilde{\phi}_2)_N$, $(\tilde{a}_v)_2, \dots, (\tilde{a}_v)_N$, $\beta_1$, 
$\beta_2$ and $\mu$.

For our purposes, we found that it is sufficient to consider $N=50$ 
points, $\epsilon = 10^{-3}$ and to cut the boundary expansion at next-to-leading 
order. 

Starting with the static solution found in the previous section, we can use a 
time-marching method to evolve the solutions in $v$. 
We discretise the spacetime in the $v$-direction, using an evenly-spaced grid 
of step size $\Delta v$. 
The equations are stiff, so we use implicit methods. We use Crank--Nicholson, 
which has 
an error that is second order in the $\Delta v$.
We start with the static solution at $v_0$ and earlier, so that the fields and 
coefficients at $v_i$ equal the fields and coefficients at $v_0$ for $i \le 0$. 
At each step we then calculate the new fields and coefficients at $v_{i+1}$ 
from the values on the previous time slices. 

So far this is general. The Crank--Nicholson method specifies precisely how we 
calculate values on $v$-slice $v_{i+1}$. We use the equations of motion and 
discretise them in the following way. For 
$f \in \left\{
\tilde{\phi}_1(\cdot, y), \tilde{\phi}_2(\cdot, y), \tilde{a}_v(\cdot, y),
\beta_1, \beta_2, \mu
\right\}$ we make the replacements
\begin{align}
 f(v) &\rightarrow f(v_{i+1}) 
 \nonumber \\
 \partial_v f(v) &\rightarrow 2\frac{f(v_{i+1}) - f(v_i)}{\Delta v} 
  - \partial_v f(v_{i})
 \nonumber \\
  \partial_v^2 f(v) &\rightarrow 2\frac{
   2\frac{f(v_{i+1}) - f(v_i)}{\Delta v} - \partial_v f(v_{i})
  }{\Delta v} 
  - \partial_v^2 f(v_{i})
 \label{eq:CrankNicholsonReplacements}\\
 &\dots \nonumber
\end{align}
where for each $v$-derivative at slice $v_{i+1}$ we use the substitution rules 
to make a replacement until the equations contain no $v$-derivatives evaluated 
at $v_{i+1}$. Note that even though the original equations were first order in 
$v$, the equations for the tilded fields contain higher derivatives of the 
coefficients $\beta_1$, $\beta_2$, $\kappa$ and $\mu$.

The discretised equations of motion are used as follows. We know the values of 
the fields and their derivatives at $v$-slice $i$. Initially we use the static 
solution and set all the $v$-derivatives to zero. If we substitute in these 
numerical values, the discretised equations of motion are then algebraic in 
values of $f$ at $v_{i+1}$. We solve for these new 
values, store them, and also use
equations~\eqref{eq:CrankNicholsonReplacements} to find the $v$-derivatives at 
$v_{i+1}$.

Solving for the field values at $v_{i+1}$ is similar to the 
static case. We first discretise the equations in the $z$-direction as well, 
getting $3N$ equations. We then replace $(\tilde{\phi}_1(v_{i+1})_1$, 
$(\tilde{\phi}_2(v_{i+1}))_1$, and $(\tilde{a}_v(v_{i+1}))_1$ with their 
($v$-dependent) boundary expansions and solve for the 
$3N$ 
components
$(\tilde{\phi}_1(v_{i+1}))_2, \dots, (\tilde{\phi}_1(v_{i+1}))_N$,  
$(\tilde{\phi}_2(v_{i+1}))_2, \dots, 
(\tilde{\phi}_2(v_{i+1}))_N$, $(\tilde{a}_v(v_{i+1}))_2, \dots, 
(\tilde{a}_v(v_{i+1}))_N$, $\beta_1(v_{i+1})$, $\beta_2(v_{i+1})$ and 
$\mu(v_{i+1})$. 
We adjust $\Delta v$ for each quench such that the shortest time scale 
appearing in the respective run is well resolved. All of the discussed 
algorithms were implemented and computed in \textsc{Mathematica}.